\documentclass[twocolumn,pre,english]{revtex4-1}
\usepackage[T1]{fontenc}
\usepackage{babel}
\usepackage{array}
\usepackage{float}
\usepackage{booktabs}
\usepackage{textcomp}
\usepackage{amsmath}
\usepackage{amsthm}
\usepackage{amssymb}
\usepackage{graphicx}
\usepackage{esint}
\usepackage{xcolor}
\usepackage{color}
\usepackage[unicode=true,pdfusetitle,
 bookmarks=true,bookmarksnumbered=false,bookmarksopen=false,
 breaklinks=false,pdfborder={0 0 1},backref=false,colorlinks=false]
 {hyperref}
 
\usepackage[sort&compress]{natbib}

\usepackage{silence}

\makeatletter

\providecommand{\tabularnewline}{\\}

\makeatother

\begin{document}

\title{Signatures of Spectral Crossovers in the Short- and Long-range Spectral Correlations of a Disordered Spin-chain with Kramers Degeneracy}
\author{Debojyoti Kundu}
\email{debojyoti.kundu.physics@gmail.com}

\author{Santosh Kumar}
\email{Corresponding Author : skumar.physics@gmail.com}

\author{Subhra Sen Gupta}
\email{Corresponding Author : subhro.sengupta@gmail.com}

\affiliation{\textrm{\textit{Department of Physics, Shiv Nadar Institution of Eminence (SNIoE), Gautam
Buddha Nagar, Uttar Pradesh 201314, India}}}

\begin{abstract}
We investigate several distinct spectral crossovers amongst various
integrable (Poissonian) and quantum-chaotic (Wigner-Dyson) limits
of a $1\mathrm{D}$ disordered quantum spin ($\mathrm{S}=1/2$) model,
by tuning the relative amplitudes of various Hamiltonian parameters
to retain or break relevant unitary and anti-unitary symmetries. The
spin model consists of an isotropic and deterministic Heisenberg term,
a random Ising term, an anisotropic and antisymmetric, but deterministic
Dzyaloshinskii-Moriya (DM) term, and a Zeeman coupling to a random,
inhomogeneous magnetic field. Since we are specially interested in
crossovers involving a Gaussian Symplectic Ensemble (GSE) limit, we
carry out all our calculations with an odd number of lattice sites
(spins) that naturally results in eigenspectra with Kramers degeneracies
(KD's). The  various crossovers (viz.,the re-entrant Poissonian-to-GSE-to-Poissonian, Poissonian-to-GUE, GSE-to-GUE and the re-entrant Poissonian-to-GOE-to-Poissonian crossovers) are investigated via detailed studies of both
short-range (Nearest Neighbour Spacings Distribution, NNSD) and long-range
{[}spectral rigidity, $\Delta_{3}(L)$ and number variance, $\Sigma^{2}(L)${]}
spectral correlations, where $L$ is the \textit{spectral interval}
over which the long-range statistic is examined. The short-range studies
show excellent agreement with RMT predictions. One of the highlights
of this study is the systematic investigation of the consequences
of retaining both eigenvalues corresponding to every Kramers doublet,
in a crossover involving the GSE limit, and see how it evolves to
a limit where the KD is naturally lifted. This is seen most clearly
in the NNSD study of the GSE-to-GUE (Gaussian Unitary Ensemble) transition, achieved
by gradually lifting the KD, using the random magnetic field. The NNSD plot in the GSE limit here exhibits a Dirac delta peak at zero splitting and a
renormalized GSE hump at finite splitting, whose general analytical
form and its asymptotic limit are derived. With an increasing symmetry
breaking magnetic field the NNSD shows an interesting, dynamic two-peaked
structure that finally converges to the standard GUE lineshape. We
explain this trend in terms of a competition between the splittings
amongst distinct Kramers doublets (related to unitary symmetries)
and the Zeeman-like splittings induced by a breaking of the anti-unitary time-reversal symmetry (TRS). This is investigated via the NNSD, the Marginal
Spectral Density (MSD) and the Densities of States (DOS) for both
spin models and RMT crossover matrix models. The first and the final short-range studies involve re-entrant Poissonian-to-GOE(GSE)-to-Poissonian crossovers,
where the final Poissonian is obtained by a many-body localization
of states in the strongly disordered limit, whereas the initial Poissonian regime involve much more delocalized eigenstates. In the long-range spectral
correlation studies, we shed light on the extent of agreement between
our physical spin systems and RMT predictions. We find that the spin
systems depart from the ideal RMT predictions for relatively finite
$L\sim10-15$ at least, for the \textit{spectral rigidity} and a much
smaller $L\sim2-4$ for the \textit{number variance}. It is further
seen that the departure is usually sooner at the uncorrelated (Poissonian)
upper end compared to the correlated (Wigner-Dyson) lower end. We
carry out a detailed comparison between the \textit{local} and the
\textit{global} crossover points, associated with the short-range
and the long-range statistics respectively, and find that in most cases they seem to agree reasonably well, but for a few exceptions. Our studies also show that
the long-range correlations may serve to distinguish between the two
Poissonian limits (\textit{non-localized} and \textit{localized})
in the re-entrant crossovers, which the short-range correlations fail
to distinguish.
\end{abstract}
\maketitle

\section{Introduction}

The microscopic many-body quantum interactions in a solid are very
complex, due to presence of strongly correlated electrons and ions.
It is quite impossible to exactly solve the dynamics of such complex
systems, but one can certainly gain some insights through some relatively
simplified models, such as the Hubbard models and its
generalizations \cite{Hubbard_original,Fazekas's_book,Lieb_Hubbard_model,Mahan_many_particle_physics,essler_frahm_1D_Hubbard_model}, obtained after several levels of simplifying assumptions, starting with the Born-Oppenheimer approximation that effectively integrates out the lattice degrees of freedom leading to purely electronic Hamiltonians. In the presence of large onsite Coulomb and Exchange interactions,
the inter-site charge fluctuations and the intra-site orbital and spin fluctuations become irrelevant
and these fermionic models can be further reduced to a variety
of \textit{spin}-only Hamiltonians. The \textit{isotropic Heisenberg
model}, the \textit{XX model}, the \textit{XY model}, the \textit{anisotropic XXZ} and \textit{XYZ models}, the \textit{Ising model}, etc. \cite{Heisenberg_original_paper,Lenz_Ising_model_1920,Ising_original_paper,Bethe_ansatz_XXX_model,Baxter_XYZ_model,Lieb_XY_model}, are among the most common ones. Even after so many levels of simplification, most of these interacting
spin-models cannot be solved analytically for exact solutions, in arbitrary dimensions. The presence of spin-orbit coupling or magnetic field induced anisotropies further complicate the situation, by introducing terms like the 2-spin anisotropic Dzyaloshinskii-Moriya (DM) exchange interaction, or the 3-spin scalar-spin-chiral interaction, that leads to even richer phase diagrams with exotic spin phases  \cite{Moriya_DM_1,Moriya_DM_2,Dzyaloshinskii_DM,Spin_chirality_Wen_Wilczek,Spin_chirality_Rokhsar,Spin_chirality_Freericks,Diptiman's-paper}.
So one resorts to numerical solutions. 

The presence of disorder in solids can further increase the complexity
of correlated electron systems. Anderson and co-workers addressed
the physics of real (amorphous) solids with impurities but in the absence of interactions
\cite{Anderson's_impurity,Scaling_theory_impurity_Abrahams_Anderson-1,Scaling_theory_impurity_Abrahams_Anderson-2,Fifty_years_of_Anderson_localization,Lee_Ramakrishnan_Disorder_review}.
But only limited success has been achieved while dealing with the
generic disorder problem in correlated electron systems \cite{Efros_disordered_system_1975,Romer_Punnoose,Punnoose_Finkel'stein-1,Punnoose_Finkel'stein-2}. Coming back to spin Hamiltonians, it is encouraging to note that in 1D, several of the above models are integrable and often lead to exact solutions. However, this integrability and related predictability via analytical solutions quickly fades away as soon as disorder is introduced, either in form of site- or bond-disorder, or via coupling to an inhomogeneous and stochastic external magnetic field \cite{Debojyoti_paper_1_2022,Avishai-Richert-PRB,Modak's-paper_2014}. Owing to the complexity of correlated electronic and spin systems and unavailability of exact solutions, one seeks some statistical techniques which can capture some crucial features of the eigenspectra. In this regard, random matrix theory (RMT) naturally appears as a powerful formalism. Originally introduced in the context of studying neutron scattering spectra from heavy nuclei \cite{Wigner-before-RMT-1951,Wigner-Conference,Wigner-book-1957,Wigner-RMT-1,Wigner-RMT-2,Wigner-RMT-3,Wigner-RMT-4,Dyson-RMT,Mehta-Book}, it has gradually found its place in the investigation of spectral properties of large complex atoms, molecules, quantum chaotic systems, quantum many-body systems with disorder, quantum dots \cite{Mehta-1960,Mehta-Book,Haake-Chaos_book,Reichl_chaos_book,Stockmann_chaos_book,T.GuhrReview,Beenakker_RMT_in_Quantum_Transport,Alhassid_quantum_dots,Shklovskii_crossover_physical_system}, etc. RMT is used to model the relevant operators in a given problem exhibiting complexity and, {\em inter alia}, can predict universal features of the associated spectra. 

Disorder in physical systems in general, and in quantum spin models, in particular, results in a transition to a non-integrable regime, the spectral statistics of the corresponding Hamiltonians being linked to the three Canonical ensembles (GOE, GUE and GSE) of RMT via the Bohigas, Giannoni and Schmit (BGS) Conjecture \cite{BGS-conjecture}, as summarized succinctly in the next section. In contrast, the spectral statistics of the former Integrable Hamiltonians are well described by Poissonian ststistics, {\em a-la} the Berry-Tabor Conjecture \cite{Berry-Tabor_conjecture_paper_1977.0140}. The various physical symmetries of the Hamiltonians are intimately related to the RMT (Wigner-Dyson) spectral classes mentioned above, which is detailed in the next section for ready reference. Even though most of these non-integrable quantum spin Hamiltonians have no classically chaotic counterpart, the very emergence of Wigner-Dyson spectral statistics is now well accepted to be the quantum signatures of Chaos, and consequently these are often referred to as the Quantum-Chaotic regime. Researchers have examined both short-range \cite{Wigner-surmise-for-high-orde-Abul-Magd,Avishai-Richert-PRB,Collura_Bose_Hubbard_2,Chavda-Kota-PLA-1,Iyer_Oganesyan_quasi-periodic_system,Modak's-paper_2014,Shklovskii_crossover_physical_system,Debojyoti_paper_1_2022} and long-range \cite{Santhanam_long_range,Long_range_Pandey_Puri,long_range_Sarika_Jalan,Magd_Unfolding_long_range,Bertrand_spin_chain_ratio_1} spectral fluctuation behavior of spectra of physical systems using various spectral measures to assess the extent of agreement with RMT and to unveil integrability or non-integrability aspects.

In our previous work \cite{Debojyoti_paper_1_2022}, we considered a Heisenberg spin-1/2 chain in the presence of a Zeeman coupling to a spatially inhomogeneous and random magnetic field  and a scalar spin-chirality term. Our main focus there was to explore the short-range (Nearest Neighbor Spacing Distribution, NNSD, and Ratio Distribution, RD) spectral properties of the above spin-chain and spectral crossovers amongst the integrable (Poissonian) and two Wigner-Dyson ensembles (GOE and GUE). In the present work, we consider a spin-1/2 quantum spin-chain model involving an isotropic  nearest-neighbor Heisenberg coupling, in the simultaneous presence of random Ising interactions, anti-symmetric DM interactions, and Zeeman-coupled to a random, inhomogeneous magnetic field (details in Sec. \ref{sec:Spin-Hamiltonians}). By tuning the relative amplitudes of the various terms in this complex Hamiltonian, we realize a plethora of RMT spectral crossovers, as detailed later and summarized in Table \ref{tab:Crossovers_between_various_ Symmetry_ Classes_and_their_criteria} (see below). These crossovers are examined via the studies of both short-range (NNSD) as well as long-range (Spectral Rigidity, $\Delta_3(L)$ and Number Variance, $\Sigma^2(L)$) spectral correlations. In the process we also carry out a comparison between the level of correspondence between the short-range correlation dictated {\em Local} and the long-range correlation dictated {\em Global} crossover points.

A highlight of this work is a detailed study of some crossovers involving the GSE regime {\em retaining the inherent Kramers degeneracy (KD)} due to an odd number of lattice sites hosting an odd number of spin-1/2's \footnote{An even number of lattice sites hosting an even number of spin-1/2’s does not give rise to the desired Kramers degenerate situation, essential for observing the GSE distribution; see e.g. Ref. \cite{Debojyoti_paper_1_2022}.}. This is in contrast to more conventional studies where the KD is artificially removed before studying the spectral statistics. This gives rise, for example in the GSE-to-GUE crossover NNSD, to a dynamic double-peak structure in the crossover regime, which we analyze in detail in terms of the Marginal Spectral Density (MSD) as well as the full Densities of States (DOS) in the context of both spin Hamiltonians as well as RMT matrix models. In the GSE limit, we also present a derivation of an analytical expression for the NNSD in the presence of KD, dubbed as the modified or diluted GSE distribution. This will also be seen to manifest in the long-range studies in terms of a reduced spectral correlation, not only with respect to the standard GSE, but also the standard GUE.

Moreover, in the course of our studies, we have examined two cases of {\em re-entrant} transitions in the form of the Poissonian-GSE-Poissonian and the Poissonian-GOE-Poissonian crossovers, both of which exhibit two distinct Poissonian regimes. In either case we encounter, in the beginning, a Poissonian regime where several eigenstates are fairly delocalized and end with a Poissonian regime where the eigenstates are strongly localized due to a very strong Ising anisotropy. Although these distinct Poissonian regimes are not distinguishable via the NNSD studies, we demonstrate that they may be distinguished via their long-range spectral fluctuations.

The rest of the paper is organized as follows. In Sec. \ref{sec:SPATIO-TEMPORAL-SYMMETRY-REQUIREMENTS},
we discuss various spatio-temporal symmetries of the physical Hamiltonians
that are required for them to belong to a specific Symmetry Class.
Next, in Sec. \ref{sec:Spin-Hamiltonians}, we describe the spin-chain
Hamiltonian used in this study and its various competing terms, and
the various unitary or anti-unitary symmetries that they individually
preserve or violate, and how their joint action classifies the full
Hamiltonian into one Wigner-Dyson class or another. In Sec. \ref{sec:Random-Matrix-Theory},
we review various RMT key concepts, like the short-range and long-range
spectral correlation properties, and summarize the analytical RMT
results used in our analysis. Next, in Sec. \ref{sec:Calculations-and-Results},
we present the details of our calculations and showcase the results.
It also includes the analysis and discussions of our results. In Conclusion,
we summarize our findings in Sec. \ref{sec:Conclusion}. Details of some of the derivations etc. are presented in four separate appendices.

\section{Classification of A Hamiltonian into Integrable and Non-integrable
Classes: Spatio-Temporal Symmetry Requirements\label{sec:SPATIO-TEMPORAL-SYMMETRY-REQUIREMENTS}}

In this short section, we summarize the various physical symmetry
requirements on a physical Hamiltonian, in order for its short-range
and long-range spectral statistics to be classified as that of an
Integrable (Poissionian) or Non-integrable/Quantum-Chaotic (Wigner-Dyson
classes) system. Physical symmetries are usually classified as the
more common Unitary class (e.g. rotations, parity, translations, time-translations
etc.) or the more exotic Anti-unitary class (time-reversal symmetries, complex conjugation symmetry, particle-hole or charge conjugation symmetry etc.) \cite{Mehta-Book,Haake-Chaos_book,Dyson-RMT,Mehta_Dyson_paper,Beenakker_review}. For our purposes in this paper, we will
limit ourselves to various rotation operations as relevant Unitary
symmetries and $conventional$ and $unconventional$ time-reversal
operations as relevant anti-unitary symmetries, as shown in Table
\ref{tab:Conditions_for_RMT_ensembles}. As also seen from Table \ref{tab:Conditions_for_RMT_ensembles},
this limits us to the Integrable (Poissonian) and the three classic
Wigner-Dyson classes (GOE, GUE, GSE). This has been mainly compiled
on the basis of the Refs. \cite{Mehta-Book} and \cite{Haake-Chaos_book}.

The spatial symmetries relevant in this context are represented by
the unitary operators $R_{\alpha}=e^{i\pi\mathrm{\mathcal{J^{\alpha}}/\hbar}}$
($\alpha=\mathrm{x},\mathrm{y}$), where ${\cal J^{\alpha}}$ is the
$\alpha$-th component of the relevant angular momentum of the system
under consideration. $R_{\alpha}$ is seen to be the rotation
operator about the $\ensuremath{\hat{\alpha}}$-axis by an angle $\pi$,
and for this reason it is referred to as a \textit{parity }operator
in some references \cite{Haake-Chaos_book}. For a spin-$\frac{1}{2}$
system with no other spatial degrees of freedom, the generator of
rotations about the axis $\hat{\alpha}$ is the \textit{Pauli} operator,
$\sigma^{\alpha}=\boldsymbol{\sigma}\cdot\hat{\alpha}$ related to
the spin operator via $\mathrm{S^{\alpha}}=\frac{\hbar}{2}\sigma^{\alpha}$. 

The \textit{conventional} time-reversal symmetry operator is defined
by the usual relation $T_{0}=e^{i\pi\mathrm{S^{y}/\hbar}}K$, whereas the \textit{unconventional}
time-reversal symmetry operator is given by $T=e^{i\pi\mathrm{S^{x}/\hbar}}T_{0}$
\cite{Haake-Chaos_book,Avishai-Richert-PRB,Modak's-paper_2014},
where $\mathit{\mathrm{S^{y}}}$($\mathit{\mathrm{S^{x}}}$) is the
$\mathrm{y}$($\mathrm{x}$)-component of the spin operator and $K$
is the \textit{complex conjugation operator}, with the property $K^{2}=1$.
Acting on a spin $\mathrm{\mathbf{S}}$, $T_{0}$ reverses the sign
of $all$ its components, while $T$ reverses the sign of only the
$\mathrm{x}$-component. For a system of spin-$\frac{1}{2}$ particles,
$T_{0}^{2}=+1$ for an \textit{even} number of spins, and $T_{0}^{2}=-1$
for an \textit{odd} number of spins. On the other hand, $T^{2}=+1$ irrespective of the number of spins involved. 

It is reasonably well-established by now, based on the \textit{Berry-Tabor
Conjecture} \cite{Berry-Tabor_conjecture_paper_1977.0140}, that
for \textit{Integrable} Systems the eigenvalue fluctuation statistics follow
the Poissonian distribution in the sense that its eigenvalues seem
to behave like a sequence of uncorrelated random variables, with no level-repulsion. This is usually associated with the Dyson-index $\beta=0$. This seems
to be true for many Physical Hamiltonians without Disorder but also
effectively true for Hamiltonians with large diagonal-disorder and
relatively negligible off-diagonal elements correlating them (for
example, in Ref. \cite{Debojyoti_paper_1_2022}, the purely Heisenberg
case or the high magnetic field case, respectively). On the other hand, the Bohigas,
Giannoni and Schmit (BGS) Conjecture \cite{BGS-conjecture} seems
to ensure that the emergence of Wigner-Dyson Statistics (with the degree of level-repulsion, signified by the Dyson-index $\beta=1,2$ and $4$ respectively, for the GOE, GUE and GSE distributions) for level-spacings
is a hallmark of \textit{non-integrable} or \textit{Quantum-Chaotic} systems, even for
many-body quantum Hamiltonians which do not have a classically chaotic
counterpart. The spin-chain Hamiltonian used in our study and its
various limits, which give rise to distinct symmetry classes, are
described in the following section.

\begin{table*}
\caption{Conditions under which a quantum system (e.g. the spin-chain model
in our case) achieves different RMT symmetry classes, mathematical
nature of the corresponding Hamiltonian matrices, along with the information
about integrability of the system. Here the anti-unitary operator
$\mathcal{T}$ represents the genereic anti-unitary time-reversal
symmetry operator ($T_{0}=e^{i\pi\mathrm{S^{y}/\hbar}}K$, or $T=e^{i\pi\mathrm{S^{x}/\hbar}}T_{0}$),
and the unitary operator $R_{\alpha}$ represents the general spatial
(or parity) symmetry (e.g. $R_{\alpha}=e^{i\pi\mathrm{\mathcal{J^{\alpha}}/\hbar}}$, $\alpha=\mathrm{x}$ or $\mathrm{y}$).
KD stands for \textit{Kramers Degeneracy, }$N_{s}$ is the number
of spin-$\frac{1}{2}$'s in the system\textit{.} \label{tab:Conditions_for_RMT_ensembles}}

\centering{}%
\begin{tabular}[t]{>{\raggedright}p{4cm}>{\raggedright}p{7.5cm}>{\centering}p{2.3cm}>{\centering}p{3.5cm}}
\hline 
\centering{}Mathematical Nature of the Hamiltonian matrix & $\;$$\;\;\;$$\;$$\;$Physical Symmetry (and other) Requirements & Symmetry Class & Integrability of the System\tabularnewline
\hline 
\begin{raggedright}
I. Any diagonalizable
\par\end{raggedright}
{[}Random or non-random (see second column){]} & \begin{raggedright}
$\;$$\;$$\;$$\;$No specific symmetry requirements
\par\end{raggedright}
\begin{enumerate}
\item No disorder or randomness; OR
\item Localized states with large diagonal disorder (and no or small off-diagonal
correlations).
\end{enumerate}
 & 
\centering{}Poissonian ($\beta=0$)

 & Integrable\tabularnewline
\hline 
\begin{raggedright}
II. Real-Symmetric
\par\end{raggedright}
(Usually random with finite off-diagonal correlations) & \begin{enumerate}
\item $\left[H,\mathcal{T}\right]=0$, $\mathcal{T}^{2}=+1$ (No KD present;
$N_{s}$ even) and no specific spatial symmetry requirements; OR
\item $\left[H,\mathcal{T}\right]=0$, $\mathcal{T}^{2}=-1$ (KD present;
$N_{s}$ odd), and at least \textit{two} spatial symmetries ($R_{\alpha}$,
$\alpha=1,2$), along with the conditions: $\left[H,R_{\alpha}\right]=0=\left[\mathcal{T},R_{\alpha}\right],$
$R_{\alpha}^{2}=-1$, $\left\{ R_{1},R_{2}\right\} =0$ (anti-commutation).
\end{enumerate}
 & 
\centering{}GOE $\;$$\;$$\;$$\;$$\;$($\beta=1$)

 & Non-integrable (Quantum-chaotic)\tabularnewline
\hline 
\begin{raggedright}
III. Complex-Hermitian
\par\end{raggedright}
(Usually random with finite off-diagonal correlations) & \begin{enumerate}
\item $\left[H,\mathcal{T}\right]\neq0$ ($N_{s}$ even or odd) and no specific
spatial symmetry requirements; OR
\item $\left[H,\mathcal{T}\right]=0$, $\mathcal{T}^{2}=-1$ (KD present;
$N_{s}$ odd), and \textit{one} spatial symmetry ($R_{\alpha}$),
along with the conditions: $\left[H,R_{\alpha}\right]=0=\left[\mathcal{T},R_{\alpha}\right],$
$R_{\alpha}^{2}=-1$.
\end{enumerate}
 & 
\centering{}GUE $\;$$\;$$\;$$\;$$\;$($\beta=2$)

 & Non-integrable (Quantum-chaotic)\tabularnewline
\hline 
\begin{raggedright}
IV. Quaternionic Self-dual
\par\end{raggedright}
(Usually random with finite off-diagonal correlations) & \noindent $\;$$\;$$\;$$\;\;\;$$\;$$\;$$\left[H,\mathcal{T}\right]=0$,
$\mathcal{T}^{2}=-1$ (KD present; $N_{s}$ odd),

\noindent $\;$$\;$$\;$$\;\;\;$$\;$$\;$\textit{no} spatial symmetries. & 
\centering{}$\;$ GSE $\;$$\;$$\;$$\;$$\;$($\beta=4$)

 & Non-integrable (Quantum-chaotic)\tabularnewline
\hline 
\end{tabular}
\end{table*}

\section{Methodology : The Spin Hamiltonian, Its Symmetries And Choice of
Basis\label{sec:Spin-Hamiltonians}}

Our one-dimensional spin Hamiltonian, $H$, has $N$ lattice sites
with one spin-$\frac{1}{2}$ per site, and is given by,
\begin{align}
\nonumber
H & =H_{h}+H_{ir}+H_{r}+H_{DM} \\
\nonumber
 & =\sum_{j=1}^{N-1}J\boldsymbol{\mathrm{S}}_{j}.\boldsymbol{\mathrm{S}}_{j+1}+\sum_{j=1}^{N-1}J\epsilon_{j}\mathrm{S^{z}}_{j}\mathrm{S^{z}}_{j+1}\\
 & +\sum_{j=1}^{N}h_{j}\mathrm{S^{z}}_{j}+\sum_{j=1}^{N-1}\boldsymbol{\boldsymbol{D}}\cdot[\mathbf{S}_{j}\times\mathbf{S}_{j+1}].
\label{eq:totalhamil}
\end{align}
This spin-chain Hamiltonian consists of four terms. The \textit{first}
term, $H_{h}$, is the usual isotropic spin-$\frac{1}{2}$\textit{
Heisenberg} term, where $\boldsymbol{\mathrm{S}}_{j}$ is the spin
operator at site $j$ (and $\mathrm{S_{\mathit{j}}^{z}}$ its $\mathrm{z}$-component),
with $J$ as the nearest-neighbor exchange interaction. The \textit{second}
term, $H_{ir}$, is a random \textit{Ising }term, where the exchange
interaction is randomized by multiplying $\ensuremath{J}$ with the
dimensionless random parameter $\epsilon_{j}$, which follows a Gaussian
distribution having zero mean and variance $\epsilon^{2}$. The \textit{third}
term, $H_{r}$, couples the spin system to a spatially inhomogeneous
and random magnetic field. The parameters, $h_{j}$, characterizing
the random, inhomogeneous site magnetic fields, follow a Gaussian
distribution, having zero mean and variance $h^{2}$ \cite{Avishai-Richert-PRB,Modak's-paper_2014,Debojyoti_paper_1_2022}.
The \textit{fourth} term, $H_{DM}$, is the anti-symmetric \textit{Dzyaloshinskii-Moriya}
(DM) interaction \cite{Moriya_DM_1,Moriya_DM_2,Dzyaloshinskii_DM,Hamazaki_DM_RMT_paper,Vahedi_Quantum_chaos_DM}.
$H_{DM}$ describes the anisotropic effective spin-spin coupling between
neighboring spins, induced in second-order perturbation theory via
the on-site spin-orbit coupling terms and the intra-site exchange
interaction between the relevant sites, after integrating out the
orbital degrees of freedom, while retaining spin as an operator. The
vector coupling constant $\boldsymbol{\boldsymbol{D}}$ in $H_{DM}$,
carries the orbital contribution and the inter-site exchange interaction,
while the rest is the anti-symmetric spin part ($\mathbf{S}_{j}\times\mathbf{S}_{j+1}$).
This often leads to \textit{canted spin arrangements} \cite{Yosida_Theory_of_Magnetism}
in real materials, while competing with the Heisenberg term. A detailed
expansion of this term in terms of the site $(\mathrm{S_{\mathit{j}}^{z}},\mathrm{S}_{\mathit{j}}^{+},\mathrm{S}_{\mathit{j}}^{-})$
operators is given in Appendix \ref{sec:simplification-of-DM-term},
for ready reference. 

\begin{table*}
\caption{Invariance of the different Hamiltonian terms under various Unitary
and Anti-unitary symmetry operations, and the conserved ${\bf S}$
component. Here $\mathcal{R}(\hat{e},\theta)=e^{i\theta\boldsymbol{\mathcal{J}}\cdotp\hat{e}/\hbar}$,
represents the general rotation operator about an axis $\hat{e}$
and by an angle $\theta$, generated by the relevant angular momentum
operator, $\boldsymbol{\mathcal{J}}$. \label{tab:Conditions_for_Hamiltonian_terms-1}}

\centering{}%
\begin{tabular}{>{\centering}m{1.9cm}>{\centering}m{1cm}>{\centering}m{1cm}>{\centering}m{1.5cm}>{\centering}m{3cm}>{\centering}m{2.5cm}}
\hline 
\centering{}Hamiltonian Term & \centering{}$T_{0}$ & \centering{}$T$ & \centering{}$\mathcal{R}(\hat{e},\theta)$ &  Rotational Invariance Axis & \centering{}Conserved ${\bf S}$ Component\tabularnewline
\hline 
\centering{}$H_{h}$ & \centering{}$\surd$ & \centering{}$\surd$ & \centering{}$\surd$  & Any axis & \centering{}$\ensuremath{\mathrm{S}^{\mathrm{z}}}$\tabularnewline
\hline
$H_{ir}$ & $\surd$ & $\surd$ & $\surd$  & $\mathrm{z}$-axis and $\pi$-rotation about any $\hat{e}$ in $\mathrm{xy}$-plane & $\ensuremath{\mathrm{S}^{\mathrm{z}}}$\tabularnewline
\hline
\centering{}$H_{r}$ & \centering{}$\times$ & \centering{}$\surd$ & \centering{}$\surd$  & \centering{}$\mathrm{z}$-axis & \centering{}$\ensuremath{\mathrm{S}^{\mathrm{z}}}$\tabularnewline
\hline
\centering{}$H_{DM}$ & \centering{}$\surd$ & \centering{}$\times$ & \centering{}$\surd$  & \centering{}$\hat{D}$-axis & \centering{}$\ensuremath{{\bf S}\cdot\hat{D}}$\tabularnewline
\hline 
\end{tabular}
\end{table*}

As already pointed out in the Introduction, the eigenvalue fluctuation
statistics of the quantum systems, are guided by the preservation
or breaking of various unitary and anti-unitary symmetries by the
different terms in the Hamiltonian. \textit{Note that for the full
Hamiltonian to obey a certain symmetry, all individual terms must
abide by it. On the other hand, if even one term violates a certain
symmetry, the whole Hamiltonian does not respect that symmetry anymore}.
With this in mind, we tabulate in Table \ref{tab:Conditions_for_Hamiltonian_terms-1},
the invariance/non-invariance of each term of our Hamiltonian with
respect to the two anti-unitary discrete symmetries $T$ and $T_{0}$,
and any possible unitary rotational symmetries, $\mathcal{R}(\hat{e},\theta)=e^{i\theta\boldsymbol{\mathcal{J}}\cdotp\hat{e}/\hbar}$,
and the last column summarizes if the total $\mathrm{S^{z}}$ (total
$\mathrm{z}$-component of spin for the lattice; $\mathrm{S^{z}}=\mathrm{\sum_{\mathit{j}=1}^{\mathit{N}}}\mathrm{S}_{j}^{\mathrm{z}}$)
is a conserved quantum number.

Any angular momentum operator $\boldsymbol{\mathcal{J}}$ is odd under
the action of the conventional time-reversal symmetry operator, i.e.,
$T_{0}\mathcal{\boldsymbol{\mathcal{J}}}T_{0}^{-1}=-\mathcal{\boldsymbol{\mathcal{J}}}$.
The spin angular momentum is thus odd under the time-reversal symmetry
and the number of spin operators involved in a Hamiltonian term, decides
its \textit{evenness} or \textit{oddness} under the $T_{0}$ operation.
For this reason, $H_{h}$, $H_{ir}$, and $H_{DM}$ are (conventional)
time-reversal symmetry invariant ($T_{0}H_{h}T_{0}^{-1}=H_{h}$, similarly
for $H_{ir}$ and $H_{DM}$ ), but $H_{r}$ is not. Now, $H_{h}$,
$H_{ir}$, and $H_{r}$ are even and $H_{DM}$ is odd under the \textit{unconventional
time-reversal} symmetry, which is represented by the anti-unitary
operator $T$. Again, all the Hamiltonian terms other than $H_{h}$,
break the full rotational invariance (isotropy), but they may be invariant
under certain special rotation operations, as elaborated in Table
\ref{tab:Conditions_for_Hamiltonian_terms-1}. However, if different
terms of the full Hamiltonian are invariant under rotations about
different axes, then in general the Hamiltonian may lack any rotational
symmetry at all. 

To construct the Hamiltonian in a matrix form, we consider a site-spin
direct product basis with a spin-$\frac{1}{2}$ at each lattice site.
An up-spin ($\ensuremath{\mathrm{m_{\mathit{j}}^{z}}=\frac{1}{2}\equiv}\ensuremath{\uparrow}$)
or a down-spin ($\ensuremath{\mathrm{m_{\mathit{j}}^{z}}=-\frac{1}{2}\equiv}\ensuremath{\downarrow}$)
can occupy each of the $N$ lattice sites of the system, where $\mathrm{\mathrm{m_{\mathit{j}}^{z}}}$
is the eigenvalue of $\mathrm{S_{\mathit{j}}^{z}}$, so we have $2^{N}$
number of basis states \{$\left|\mathrm{m_{1}^{z}m_{2}^{z}m_{3}^{z}\mathbf{....}m_{\mathit{N}}^{z}}\right\rangle $\}
\cite{Debojyoti_paper_1_2022}. From Table \ref{tab:Conditions_for_Hamiltonian_terms-1},
we notice that the Hamiltonian terms $H_{h}$, $H_{ir}$, and $H_{r}$
commute with $\mathrm{S^{z}}$ but $H_{DM}$ does not. While for $H_{h}$,
$\mathrm{S}^{\pm}$ always appear in pairs, $H_{ir}$ and $H_{r}$
only involve site $\mathrm{S}^{\mathrm{z}}$ operators, and hence
these three terms can never change the total $\mathrm{S}^{\mathrm{z}}$.
On the other hand, as is apparent from Eq. (\ref{eq:DM_appendix_A_4})
of Appendix \ref{sec:simplification-of-DM-term}, which shows the
full decomposition of the DM term, only the z-component conserves
total $\mathrm{S^{z}}$, while the x- and y-components are combinations
of terms that change the total $\mathrm{S^{z}}$ by $\pm1$. Thus,
in the absence of the DM term, the different total $\mathrm{S^{z}}$
symmetry sectors are irreducible blocks of the Hamiltonian and hence
the eigenvalues between the different sectors are uncorrelated, while
only those within a given sector are correlated. On the other hand,
the presence of the DM term introduces off-diagonal terms between
these irreducible blocks causing all eigenvalues of the full Hamiltonian
to become correlated. As a result, in order to observe Wigner-Dyson
distributions, we must consider a fixed $\mathrm{S^{z}}$ restricted
subspace when $H_{DM}=0$, while we are $not$ permitted to do a similar
symmetry adaptation when $H_{DM}$ is finite \cite{Mehta-Book,Santos_Gubin_Inverted_spin},
and the $full$ basis must be considered. In our calculations with
this spin-chain model, we have considered systems where $N$ is \textit{odd},
so that $T^{2}=-1$ and Kramers degeneracy is imposed. For simplicity,
we keep the $J=1$ (antiferromagnetic) in our calculations. We need
the entire energy spectrum for our spectral correlation studies, so
we use the full exact diagonalization methods to obtain the energy
eigenvalues. As a result, the system sizes we can access are limited
to some extent. 

We now explore the conditions under which our spin-chain system achieves
different RMT symmetry classes in light of the prior discussions surrounding
Tables \ref{tab:Conditions_for_RMT_ensembles} and \ref{tab:Conditions_for_Hamiltonian_terms-1}.
In the presence of only the Heisenberg term ($H_{h}$) in Eq. (\ref{eq:totalhamil}),
the Hamiltonian is preserved under all unitary and anti-unitary symmetries,
discussed above. Also, there is no disorder in the system, so the
fluctuation statistics of the eigenvalues are expected follow the
Poissonian distribution. For $H_{1}=H_{h}+H_{ir}$, randomness is
introduced in the spin-chain system along $\mathrm{z}$, without breaking
either of the time-reversal symmetries. $H_{1}$ is also real-symmetric
and as a result of all this, is expected to belong to the Gaussian
Orthogonal Class. If we consider  a $\mathit{fixed}$ $\mathrm{S^{z}}$
$\mathit{subspace}$, the fluctuation statistics of the eigenvalues
is then expected to follow the GOE distribution. Now, for the $random$
Hamiltonian $H_{2}$ ($=H_{h}+H_{r}+H_{DM}$), both anti-unitary symmetries
are broken ($H_{DM}$ breaks the $T$ symmetry) and the matrix representation
becomes complex-Hermitian (the DM term induces the complex nature,
as is clearly seen from Eq. (\ref{eq:DM_appendix_A_4}) of Appendix
\ref{sec:simplification-of-DM-term}, where the $\mathrm{x}$- and
$\mathrm{z}$- components of the DM term are \textit{pure imaginary}
and \textit{off-diagonal} and add on to the real terms from the other
parts of $H_{2}$). The quantum system represented by $H_{2}$, thus
belongs to the Gaussian Unitary Class. For a \textit{full basis} calculation,
the spectral fluctuation statistics is now expected to follow the
GUE distribution. Lastly, the $random$ Hamiltonian $H_{3}$ ($=H_{h}+H_{ir}+H_{DM}$)
preserves the $T_{0}$ symmetry (breaks $T$ symmetry due to the $H_{DM}$
term) and breaks all the unitary\textit{ spin rotational} (or spatial)
symmetries, for the chosen direction of $\boldsymbol{\boldsymbol{D}}$.
For the system represented by this Hamiltonian, for which conventional
time-reversal symmetry is the only remaining symmetry, and with an
\textit{odd} number of sites ($N$) and hence spins, we encounter
Kramers degeneracy in the full basis calculation as the only systematic
degeneracy left, and the system belongs to the Gaussian Symplectic
Class. The quantum system can then be represented by a quaternionic
self-dual matrix and the spectral fluctuation statistics is now expected
to follow the GSE distribution. Variations of the relative amplitudes
of the various terms in $H$, lead to spectral crossovers amongst
the Poissonian and the various Wigner-Dyson distributions. The form
of these distributions are tabulated for ready reference in Table
\ref{tab:NNSD formulas}.

Table \ref{tab:Crossovers_between_various_ Symmetry_ Classes_and_their_criteria}
shows the crossover criteria for various Symmetry Classes in the spin-chain
systems ($H_{1}$, $H_{2}$, $H_{3}$, and $H$). We also list the
parameters (in $H$) that remained fixed  during a crossover and those
that need to be tuned on in order to break a symmetry and undergo
a crossover. 

\begin{table*}
\caption{Crossovers between various Symmetry Classes and their criteria (as
defined in Table \ref{tab:Conditions_for_RMT_ensembles}). Various
relevant Hamiltonian parameter values, including the value of the
tuning parameter at the NNSD crossover, are also included. \label{tab:Crossovers_between_various_ Symmetry_ Classes_and_their_criteria}}

\centering{}%
\begin{tabular}{>{\raggedright}p{3.7cm}>{\centering}p{3.5cm}>{\centering}p{3.5cm}>{\centering}p{4.5cm}>{\raggedright}p{2.5cm}}
\hline 
\begin{centering}
Hamiltonian
\par\end{centering}
\centering{}($N$; Basis Type) & \centering{}Crossover From (Symmetry Criterion from Table \ref{tab:Conditions_for_RMT_ensembles}) & \centering{}Crossover To (Symmetry Criterion from Table \ref{tab:Conditions_for_RMT_ensembles}) & \centering{}Fixed Parameters & \begin{centering}
Tuning Parameter 
\par\end{centering}
\centering{}(Value at NNSD Crossover)\tabularnewline
\hline 
\raggedright{}$H_{3}$ (13; Full basis) & \centering{}Poissonian (I.1) & \centering{}GSE (IV) & \centering{}$J=1.0;D=0.2$ & $\;\;\;\;$$\epsilon$ ($0.6$)\tabularnewline
\raggedright{}$H_{3}$ (13; Full basis) & \centering{}GSE (IV) & \centering{}Poissonian (I.2) & \centering{}$J=1.0;D=0.2$ & $\;\;\;\;$$\epsilon$ ($20.0$)\tabularnewline
\raggedright{}$H_{2}$ (13; Full basis) & \centering{}Poissonian (I.1) & \centering{}GUE (III.1) & \centering{}$J=1.0;D=0.2$ & $\;\;\;\;$$h$ ($0.15$)\tabularnewline
\raggedright{}$H$ (13; Full basis) & \centering{}GSE (IV) & \centering{}GUE (III.1) & \centering{}$J=1.0;\epsilon=0.6;D=0.2$ & $\;\;\;\;$$h$ ($0.015$)\tabularnewline
$H_{1}$ (13; $\mathrm{S}^{\mathrm{z}}=1/2$ sector) & \centering{}Poissonian (I.1) & \centering{}GOE (II.2) & $J=1.0$ & $\;\;\;\;$$\epsilon$ ($0.5$)\tabularnewline
$H_{1}$ (15; $\mathrm{S}^{\mathrm{z}}=1/2$ sector) & \centering{}Poissonian (I.1) & \centering{}GOE (II.2) & $J=1.0$ & $\;\;\;\;$$\epsilon$ ($0.4$)\tabularnewline
$H_{1}$ (13; $\mathrm{S}^{\mathrm{z}}=1/2$ sector) & \centering{}GOE (II.2) & \centering{}Poissonian (I.2) & $J=1.0$ & $\;\;\;\;$$\epsilon$ ($20.0$)\tabularnewline
$H_{1}$ (15; $\mathrm{S}^{\mathrm{z}}=1/2$ sector) & \centering{}GOE (II.2) & \centering{}Poissonian (I.2) & $J=1.0$ & $\;\;\;\;$$\epsilon$ ($15.0$)\tabularnewline
\hline 
\end{tabular}
\end{table*}

As is customary for random systems, we need to consider the process
of \textit{configuration averaging} by diagonalizing an ensemble of
$\mathcal{M}$ matrices. Each configuration is a matrix representation
of the system with parameters generated at random from Gaussian distributions
with fixed widths (standard deviations), $h$ and $\epsilon$ \cite{Debojyoti_paper_1_2022},
respectively, for the two random terms, as relevant. Similar averaging
is then repeated for each value of $h$ or $\epsilon$. In Sec. \ref{sec:Calculations-and-Results},
we specify the number of configurations used in an ensemble, for each
lattice size. Usually, the larger the Hamiltonian matrix dimension,
the smaller the number of configurations over which averaging is required
to be performed, in agreement with the principle of $spectral$ $ergodicity$
\cite{thermalization-1_Lebowitz_Penrose,Penrose_1979,thermalization_Giovanni}.
While it is customary to remove one of the Kramers degenerate partners
from the eigenvalues before studying spectral correlations and obtaining
the standard GSE statistics, we also explore spectral correlations retaining
the Kramers degeneracies in the spectrum.

\section{Random Matrix Theory (RMT)\label{sec:Random-Matrix-Theory}}

In this Section, we describe the measures of the short-range and long-range
spectral correlations in RMT studies used in this work, and include
the RMT analytical results for them. Density of states (DOS) of a
physical system is nonuniform, so to compare spectral correlations
between different systems, one needs to remove the system-dependent
level density from the eigenspectrum, and scale it in terms of the
mean level spacing. For this, we need to implement the \textit{unfolding}
procedure before comparing our calculated results with the standard
RMT results. In our calculations, calculated distribution of states
is fitted using polynomials and the fitted polynomial is used to unfold
the eigenspectra \cite{Mehta-Book,Haake-Chaos_book,Avishai-Richert-PRB,Debojyoti_paper_1_2022}.
For an ordered sequence $\mathop{\varepsilon_{1}<\cdots<\varepsilon_{n}}$,
of $n$ energy eigenvalues, the unfolded eigenvalues are calculated
using $\tilde{\varepsilon}_{j}=\int_{\varepsilon_{1}}^{\varepsilon_{j}}\rho(\varepsilon')d\varepsilon'$,
where $\rho(\varepsilon)=d\mathcal{N}(\varepsilon)/d\varepsilon$
is the fitted DOS, and $\mathcal{N}(\varepsilon)$ is the cumulative
DOS (or, the \textit{spectral staircase function}).

\subsection{Short-range level correlation statistics\label{subsec:Short-range-level-correlation}}

In RMT, it is standard practice to study the short-range level correlations
via the nearest neighbor spacing distribution (NNSD). It quantifies
the local fluctuations of energy eigenvalues of a given system \cite{Mehta-Book,Haake-Chaos_book,T.GuhrReview,Brody_Random_Matrix_Physics_review}. The
nearest-neighbor level spacing of the unfolded eigenvalues is defined
as $\ensuremath{s_{j}=\tilde{\varepsilon}_{j+1}-\tilde{\varepsilon}_{j}}$.
The corresponding probability density function, $P(s)$, can be compared
with the analytical RMT results. The Wigner surmise formulae for the
three Dyson symmetry classes along with the Poisson distribution are
compiled in Table \ref{tab:NNSD formulas}, for this purpose. \begin{table}
\caption{Probability distributions of nearest-neighbor spacings for unfolded eigenvalues \cite{Mehta-Book,Haake-Chaos_book}.\label{tab:NNSD formulas}}
\begin{tabular}{cc}
\hline  
{\footnotesize{}Type of Distribution} & {\footnotesize{}NNSD Probability Density}\tabularnewline
\hline 
{\footnotesize{} Poissonian} & $P_{Poi}(s)=\exp(-s)$\tabularnewline
{\footnotesize{}GOE} & {\footnotesize{}$\mathop{P_{GOE}(s)=(\pi s/2)\exp(-\pi s^{2}/4)}$}\tabularnewline
{\footnotesize{}GUE} & {\footnotesize{}$\mathop{P_{GUE}(s)=(32s^{2}/\pi^{2})\exp(-4s^{2}/\pi)}$}\tabularnewline
{\footnotesize{}GSE} & {\footnotesize{}$\mathop{P_{GSE}(s)=(2^{18}s^{4}/3^{6}\pi^{3})\exp(-64s^{2}/9\pi)}$}\tabularnewline
\hline 
\end{tabular}
\end{table}

In this paper, one of our interests is to study the Poissonian-to-GOE,
Poissonian-to-GUE, and Poissonian-to-GSE crossovers in NNSD. It is
also compelling to study the GSE-to-GUE crossover with and without
removing the Kramers degeneracy from the spectra, using the \textit{crossover
matrix model} and the spin-chain model. Within RMT, these crossovers
can be modeled using the Pandey-Mehta Hamiltonian \cite{PandeyMehta1983,MehtaPandey1983,Lenz_Haake_crossover_matrix},
\begin{equation}
\mathcal{H}=(1-\alpha)\mathcal{H}_{0}+\alpha\mathcal{H}_{1},\label{eq:crossover_matrix_model}
\end{equation}
 where at $\alpha=0$ \footnote{The $\alpha$ in the following expression should not be confused with the
component index in Table \ref{tab:Conditions_for_RMT_ensembles} and in Sec. \ref{sec:SPATIO-TEMPORAL-SYMMETRY-REQUIREMENTS}}, the matrix model is governed by the symmetry
of $\mathcal{H}_{0}$, and the finite $\alpha$ ($0<\alpha<1$) introduces
perturbation through $\mathcal{H}_{1}$. At $\alpha=1$, the other extreme
is achieved, where the matrix model is governed by the symmetry of
$\mathcal{H}_{1}$. By varying $\alpha$ between $0$ and 1, we can
study the crossover between two distinct symmetry classes in RMT.

As previously discussed, the Gaussian Symplectic Class possesses Kramers
degeneracy. After removing one of the identical eigenvalues from each
of the Kramers doublets, the spectral correlation statistics are usually
examined. In this paper, we want to look at the spectral fluctuation
for the GSE class \textit{without} eliminating the KD from the eigenspectra.
Since, in the absence of any spatial symmetries, we are only left
with a series of Kramers doublets, one may guess that this will lead
to a GSE-like distribution along with a singular peak at $s=0$. But
in view of this, the whole distribution needs to be renormalized.
Below, we present such a modified NNSD formula for an eigenspectrum of $n$ levels, a detailed derivation
of which is provided in Appendix \ref{sec:NNSD-of-the-GSE-Class-with-KD}.

\begin{align}
\nonumber
\mathcal{P}_{GSE}^{n}(s) & =\ensuremath{\mathop{\left[\frac{2^{12}}{3^{6}\pi^{3}}\left(\frac{n-2}{n-1}\right)^{6}s^{4}\right]\exp\left[-\frac{16}{9\pi}\left(\frac{n-2}{n-1}\right)^{2}s^{2}\right]}}\\
& +\left(\frac{n}{n-1}\right)\delta(s).
\label{eq:NNSD_GSE_Without_removing_KD}
\end{align}
 where the \textit{Dirac Delta,} $\delta(s)$, appears because of
the presence of KD in the spectra. Since level-spacings cannot be
negative by definition ($s\ge0$), we need to consider along with
Eq. (\ref{eq:NNSD_GSE_Without_removing_KD}), the definition $\int_{0}^{\infty}\delta(s)ds:=\frac{1}{2}$.
In the large $n$ (number of levels) limit, the Eq. (\ref{eq:NNSD_GSE_Without_removing_KD})
becomes, 
\begin{equation}
\mathcal{P}_{GSE}(s)=\ensuremath{\mathop{\left(\frac{2^{12}}{3^{6}\pi^{3}}s^{4}\right)\exp\left(-\frac{16}{9\pi}s^{2}\right)}}+\delta(s).\label{eq:NNSD_GSE_Without_removing_KD_large_n}
\end{equation}
 Starting from GSE, it would be interesting to observe how the initial
delta function peak broadens in the NNSD, when one transitions to
an another symmetry class.

\subsection{Long-range level correlation statistics\label{subsec:Long-range-level-correlation}}

As discussed earlier in the Introduction, the long-range eigenvalue
fluctuation studies are required to ascertain the extent of universal
RMT behavior in a physical system. The two most popular RMT measures
to study the long-range spectral properties are the \textit{spectral
rigidity} ($\Delta_{3}$-statistic) and the \textit{number variance} ($\Sigma^{2}$-statistic) \cite{Mehta-Book,Haake-Chaos_book,T.GuhrReview,Brody_Random_Matrix_Physics_review}.

For an unfolded eigenspectrum, the \textit{spectral staircase function},
$\mathcal{N}(\tilde{\varepsilon})$, denotes the number of levels
having energy between $0$ and $\tilde{\varepsilon}$. This can be
thought of as the cumulative or integrated DOS, $\mathcal{N}(\tilde{\varepsilon})=\int_{0}^{\tilde{\varepsilon}}\rho(\tilde{\varepsilon}')d\tilde{\varepsilon}'$.
The \textit{least-square} deviation of $\mathcal{N}(\tilde{\varepsilon})$
from the \textit{best fit} straight line ($a\tilde{\varepsilon}+b$,
where $a$ and $b$ are obtained from the fit), is defined as the
\textit{spectral rigidity }{[}$\Delta_{3}(L)${]}\textit{,} for a
finite interval $L$ of the eigenspectrum. It is given by the expression:
\begin{equation}
\Delta_{3}(L)=\left\langle \frac{1}{L}\underset{a,b}{\mathrm{min}}\left(\int_{E}^{E+L}[\mathcal{N}(\tilde{\varepsilon})-a\tilde{\varepsilon}-b]^{2}d\tilde{\varepsilon}\right)\right\rangle ,\label{eq:spectral_rigidity}
\end{equation}
 where $E$ is the starting position and $\left\langle \cdots\right\rangle $
denotes the average over several choices of $E$ (\textit{spectral}
average) and also over several disordered configurations \cite{Santhanam_long_range,Mehta_Dyson_paper,Mehta-Book,Stockmann_chaos_book}.
The latter \textit{ensemble averaging} over several disordered configurations
is performed to mainly obtain statistically smooth data for finite
lattice sizes, in our numerical calculations. The analytical RMT formulae
for the Poissonian and the Wigner-Dyson ensemble statistics, are given
in the Table \ref{tab:Spectral_rigidity_formulae}. These analytical
expressions for the Wigner-Dyson ensembles are approximate results
in the large $L$ limit. In our studies, we use the full exact integral
expressions involving the two-level $\mathit{cluster}$ $\mathit{functions}$
\cite{Mehta-Book,Mehta_Dyson_paper}, discussed in the Appendix
\ref{sec:Full-Integral-Expressions-for-Delta-and-Sigma}. 
\begin{table}
\centering{}\caption{Spectral rigidity ($\Delta_{3}$-statistic) expressions for the Poissonian
and the Wigner-Dyson ensembles, approximated for large $L$. Here
$\gamma$ is Euler's constant. The full integral expressions (any
$L$) of $\Delta_{3}(L)$ for the Poissonian and the Wigner-Dyson
ensembles can be found in the Appendix \ref{sec:Full-Integral-Expressions-for-Delta-and-Sigma},
and are the ones used in our analysis throughout. \label{tab:Spectral_rigidity_formulae}}
\begin{tabular}{>{\centering}p{1.75cm}>{\raggedright}p{7.2cm}}
\hline 
Type of Ensemble & \centering{}Spectral Rigidity\tabularnewline
\hline 
Poissonian & {\footnotesize{}$\left[\Delta_{3}(L)\right]_{Poi}=\frac{L}{15}$}\tabularnewline
GOE & {\footnotesize{}$\left[\Delta_{3}(L)\right]_{GOE}=\frac{1}{\pi^{2}}\left(\ln(2\pi L)+\gamma-\frac{5}{4}-\frac{\pi^{2}}{8}\right)+\mathcal{O}\left(L^{-1}\right)$}\tabularnewline
GUE & {\footnotesize{}$\left[\Delta_{3}(L)\right]_{GUE}=\frac{1}{2\pi^{2}}\left(\ln(2\pi L)+\gamma-\frac{5}{4}\right)+\mathcal{O}\left(L^{-1}\right)$}\tabularnewline
GSE & {\footnotesize{}$\left[\Delta_{3}(L)\right]_{GSE}=\frac{1}{4\pi^{2}}\left(\ln(4\pi L)+\gamma-\frac{5}{4}+\frac{\pi^{2}}{8}\right)+\mathcal{O}\left(L^{-1}\right)$}\tabularnewline
\hline 
\end{tabular}
\end{table}

For the \textit{number variance statistic}, given an unfolded eigenspectrum,
one examines the variation in the \textit{number of energy levels},
$\mathfrak{n}(E,L)$, defined as $\mathfrak{n}(E,L)=\int_{E}^{E+L}\rho(\tilde{\varepsilon})d\tilde{\varepsilon}$,
in an energy interval of given length $L$ and as a function of the
starting energy $E$. The \textit{number variance statistic} is then
defined as \cite{Santhanam_long_range,Mehta_Dyson_paper,Mehta-Book,Stockmann_chaos_book}:
\begin{align}
\nonumber
\Sigma^{2}(L) & =\left\langle \mathfrak{n}(E,L)^{2}\right\rangle -\left\langle \mathfrak{n}(E,L)\right\rangle ^{2}\\
& =\left\langle \mathfrak{n}(E,L)^{2}\right\rangle -L^{2},
\label{eq:number_variance}
\end{align}
where the average of $\mathfrak{n}(E,L)$ becomes $L$, which is easy to see because the average spectral density for an unfolded spectrum is unity.
Here also, as in the case of spectral rigidity, we perform both \textit{spectral}
and \textit{ensemble} averages in our relevant numerical calculations.
The analytical RMT formulae of $\Sigma^{2}(L)$-statistic, for the
Poissonian and Wigner-Dyson ensembles, are given in the Table \ref{tab:number_variance_formulae}.
See the Appendix \ref{sec:Full-Integral-Expressions-for-Delta-and-Sigma}
for the full integral expressions involving the two-level $\mathit{cluster}$
$\mathit{functions}$ \cite{Mehta-Book,Mehta_Dyson_paper}, which
are being used in our studies.

In concluding this section, we also note that the $\Sigma^{2}$-statistic
exhibits more fluctuations or oscillations, on the average, compared
to the $\Delta_{3}$-statistic. This will also be seen in our studies
of the physical spin models. In the context of RMT, this may be understood
from the fact that the $\Delta_{3}$-statistic can be represented as an
integral transform involving the $\Sigma^{2}$-statistic \cite{T.GuhrReview,Vyas_long_range_comparison,Brody_Random_Matrix_Physics_review},
given by the Eq. (\ref{eq:appendix_C_integral_transform_between_Sigma_and_Delta3}),
leading to the smoother nature of the \textit{spectral rigidity} compared
to the \textit{number variance}. 

\begin{table}
\caption{Number Variance ($\Sigma^{2}$-statistic) expressions for the Poissonian
and the Wigner-Dyson ensembles, approximated for the large $L$. Here
$\gamma$ is Euler's constant. The full integral expressions (any
$L$) of $\Sigma^{2}(L)$ for the Poissonian and the Wigner-Dyson
ensembles can be found in the Appendix \ref{sec:Full-Integral-Expressions-for-Delta-and-Sigma},
and are the ones used in our analysis throughout. \label{tab:number_variance_formulae}}

\centering{}%
\begin{tabular}{>{\centering}p{1.75cm}>{\raggedright}p{7.2cm}}
\hline 
Type of Ensemble & \centering{}Number Variance\tabularnewline
\hline 
Poissonian & {\footnotesize{}$\left[\Sigma^{2}(L)\right]_{Poi}=L$}\tabularnewline
GOE & {\footnotesize{}$\left[\Sigma^{2}(L)\right]_{GOE}=\frac{2}{\pi^{2}}\left(\ln(2\pi L)+1+\gamma-\frac{\pi^{2}}{8}\right)+\mathcal{O}\left(L^{-1}\right)$}\tabularnewline
GUE & {\footnotesize{}$\left[\Sigma^{2}(L)\right]_{GUE}=\frac{1}{\pi^{2}}\left(\ln(2\pi L)+1+\gamma\right)+\mathcal{O}\left(L^{-1}\right)$}\tabularnewline
GSE & {\footnotesize{}$\left[\Sigma^{2}(L)\right]_{GSE}=\frac{1}{2\pi^{2}}\left(\ln(4\pi L)+1+\gamma+\frac{\pi^{2}}{8}\right)+\mathcal{O}\left(L^{-1}\right)$}\tabularnewline
\hline 
\end{tabular}
\end{table}

One of the main objectives of this paper is to investigate the correlations
between far-off eigenvalues of our spin-chain systems and see how
closely they resemble the universal RMT behavior outlined in this
Section. We further investigate how closely the spectral crossovers,
characterized via changes in the short-range and the long-range spectral
statistics, correspond with each other.

\section{Calculations and Results\label{sec:Calculations-and-Results}}

In this section, we discuss the details of our short- and long-range
level correlation calculations, as well as an analysis of the densities
of states (DOS) associated with our spin Hamiltonians vis-a-vis RMT
matrix models for some of the relevant symmetry crossovers.  

\subsection{Nearest-Neighbor Spacing Distributions (NNSD)\label{subsec:NNSD_results}}

In this section, we report the results from the study of the nearest-neighbor
spacings of energy eigenvalues computed from the Hamiltonian $H$,
for an odd number of lattice sites. Here, we consider the lattice
size $N=13$, which has a matrix dimension $n=8192$ ($2^{N}$). We
carry out numerical exact-diagonalization calculations with an ensemble
size of $\mathcal{M\mathrm{=}\mathrm{15}}$ configurations, while
noting that systems with lattice sizes less than $N=13$, do not follow
the standard RMT ensemble results, closely enough. Also, since the
DM term connects the various total $\mathrm{S}^{\mathrm{z}}$ sectors
(see Appendix \ref{sec:simplification-of-DM-term}), spin-symmetry
adaptation is not feasible in several of these calculations, and the
full basis must be used. Hence performing exact-diagonalization calculations
for larger lattice sizes having $N=15$ ($n=32768$) or more, is computationally
very expensive, in view also of the configuration averaging required
in all calculations, and so we do not attempt it. As discussed in
Sec. \ref{sec:Spin-Hamiltonians}, for odd $N$, we get \textit{Kramers
Degeneracy, }i.e.,\textit{ }each eigenstate is doubly degenerate,
in the case when there is no spatial symmetry left in the system,
as in the GSE limit. Here the standard practice is to systematically
remove one of the Kramers degenerate partners by hand before studying
the spectral correlation properties of the model. While we have done
this, in this work we have also done calculations retaining the Kramers
degeneracy (KD) and compared the results with the standard case where
KD has been removed, as discussed below. This often leads to very
interesting multi-peak distributions that smoothly evolve from the
GSE to other limits where the KD is absent. We consider open boundary
conditions (OBC) throughout our calculations, as extra degeneracies
might occur in the spectra with periodic boundary conditions (PBC).
As mentioned in Sec. \ref{sec:Random-Matrix-Theory}, we have unfolded
the spectra using polynomial fits to make the average level spacing
equal to unity. 

\begin{figure}
\begin{centering}
\includegraphics[scale=0.4]{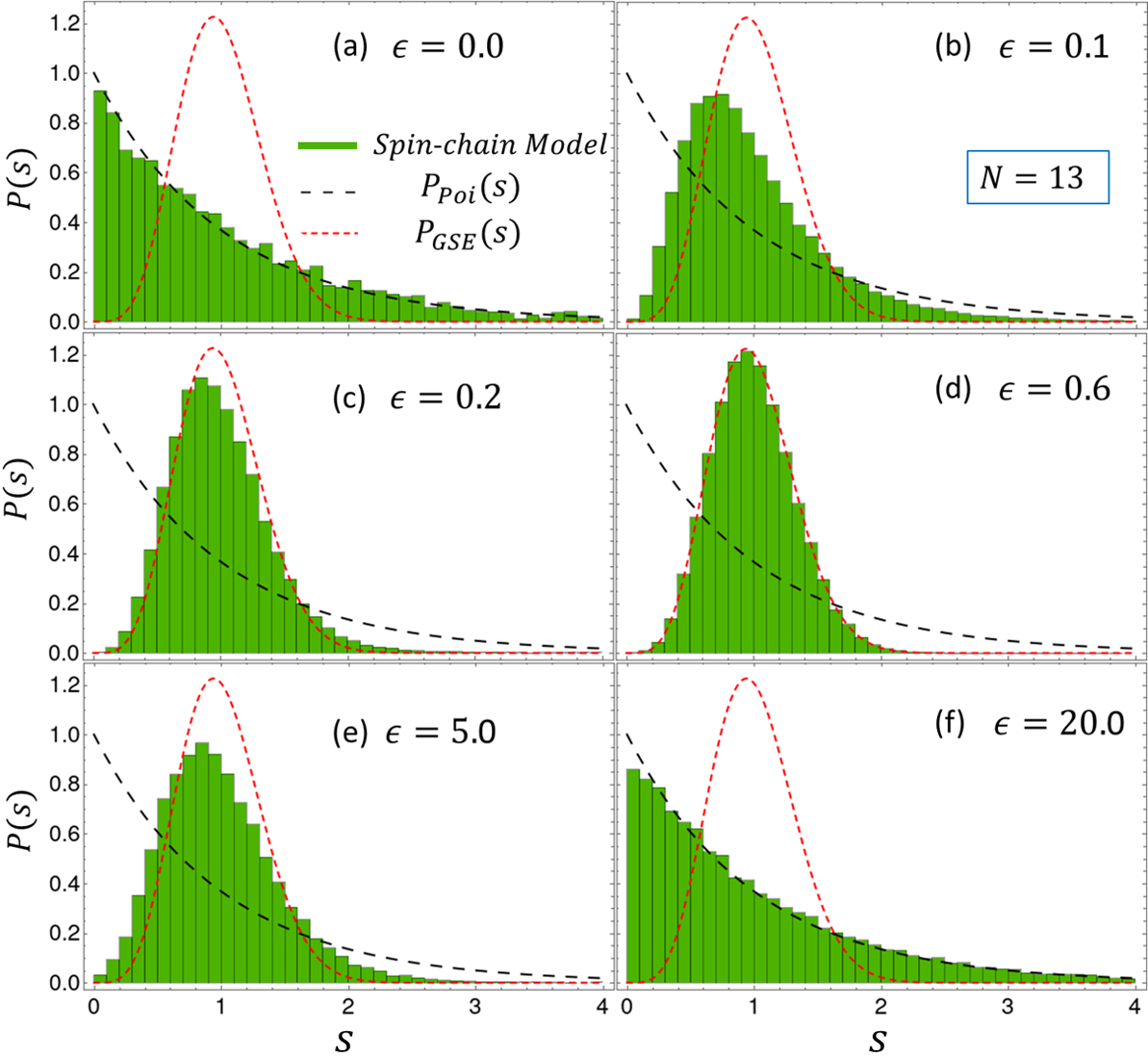}
\par\end{centering}
\caption{NNSD for $N=13$ (basis size $n=8192$ and the number of configurations
$\mathcal{M\mathrm{=}\mathrm{15}}$) in the {\em re-entrant} Poissonian-to-GSE-to-Poissonian crossover
with increasing $\epsilon$, fixed $D=0.2$. (a) and (d) show the
two limiting cases, namely the Poissonian and the GSE respectively,
whereas (b) and (c) show two of the intermediate cases (see also Table
\ref{tab:Crossovers_between_various_ Symmetry_ Classes_and_their_criteria}). A further increase of $\epsilon$ leads to a re-entrant crossover into the Poissonian regime via intermediate stages, as shown in panels (e) and (f). This new Poissonian regime is characterized by a strong many-body localization of its eigenstates, compared to the Poissonian at $\epsilon=0$.
\label{fig:1_Poi_GSE_NNSD}}
\end{figure}

\begin{figure}
\begin{centering}
\includegraphics[scale=0.33]{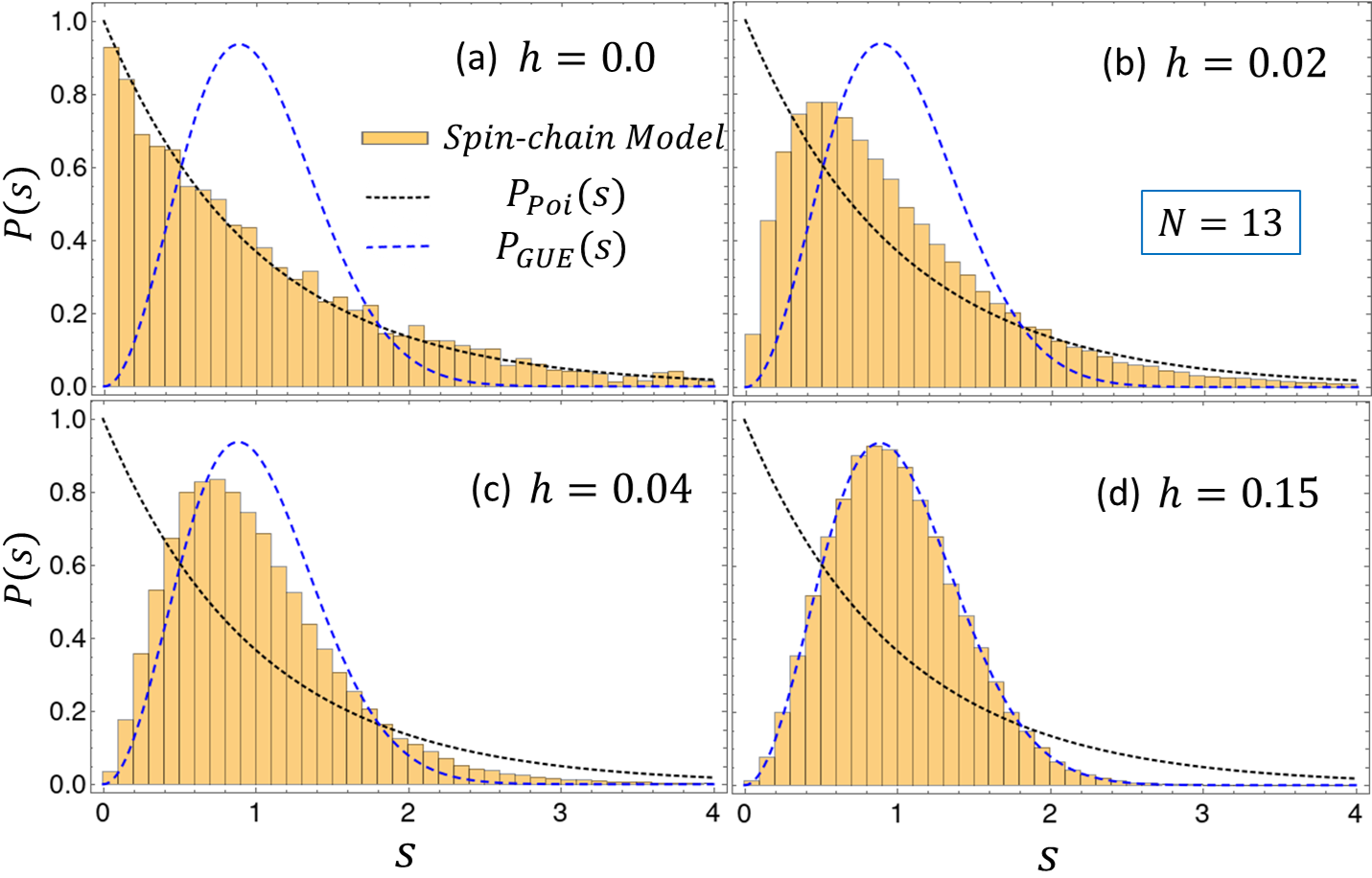}
\par\end{centering}
\caption{NNSD for $N=13$ ($n=8192$ and $\mathcal{M\mathrm{=}\mathrm{15}}$) in the Poissonian-to-GUE crossover with increasing $h$, fixed $D=0.2$.
(a) and (d) show the two limiting cases, namely the Poissonian and
the GUE respectively, whereas (b) and (c) show two of the intermediate
cases (see also Table \ref{tab:Crossovers_between_various_ Symmetry_ Classes_and_their_criteria}).
\label{fig:2_Poi_GUE_NNSD}}
\end{figure}

In Ref. \cite{Debojyoti_paper_1_2022}, we explored the short-range
spectral correlations and crossovers amongst the Poissonian, GOE,
and GUE distributions, for an \textit{even} number of sites using
a spin-chain model. In this paper, the primary motive behind working
with a spin-chain model having an \textit{odd} number of sites, is
to achieve the Gaussian Symplectic class (GSE) distribution, and study
crossovers from/to other distributions in the integrable (Poissonian)
or other quantum-chaotic (Wigner-Dyson) limits. To this end we first
take up the Poissonian-to-GSE crossover. We tune the relative strengths
of the various interactions in the Hamiltonian $H_{3}$ to obtain
this crossover (see Table \ref{tab:Crossovers_between_various_ Symmetry_ Classes_and_their_criteria}).
In the presence of the Heisenberg term (with $J=1.0$ always, so that
all interactions are measured in units of $J$) and the DM term, and
no disorder (randomness) the Poissonian distribution is obtained.
To go over to a GSE distribution one needs a complete breaking of
any spatial (rotational) symmetry in the system, and this is obtained
by the joint action of the $H_{ir}$ and the $H_{DM}$ terms. The $H_{ir}$ term possesses a 2-fold spin-rotational symmetry about any direction $\hat{e}$ in the $\mathrm{xy}$-plane, and reduces the spherical symmetry of $H_{h}$ to cylindrical symmetry about the $\mathrm{z}$-axis, so that only
$\mathrm{S^{\mathrm{z}}}$ is conserved. $H_{DM}$ alone results in
a cylindrical symmetry only about the direction of $\hat{D}$ so that
only $\mathbf{S}\cdot\hat{D}$ is conserved (see Table \ref{tab:Conditions_for_Hamiltonian_terms-1}
and Appendix \ref{sec:Commutation_Between_H_DM_and_S_dot_D_cap}, for further details). Evidently for
any direction of $\boldsymbol{D}$ distinct from the Ising axis and any axis in the $\mathrm{xy}$-plane, there
is no rotational symmetry left at all. With this understanding, we
make the choice $\boldsymbol{D}=(D_{\mathrm{x}},D_{\mathrm{y}},D_{\mathrm{z}})=(D,D,D)$.
$H_{DM}$ also preserves $T_{0}$ (but breaks the $T$-symmetry),
which seems sufficient to yield a GSE.

It is however found that \textit{sufficient} breaking of the rotational
symmetry of $H_{h}$ by $H_{DM}$, to effect a crossover to the GSE
distribution, is obtained only at about $D=0.2$. Hence we fix the
magnitude of $D$ at this value at the Poissonian end, and then gradually
turn on $H_{ir}$ ($\epsilon=0\rightarrow\mathrm{finite}$), which
generates \textit{diagonal disorder}, to finally cross over to the
GSE distribution at $\epsilon\sim0.6$. Here we have followed the standard practice of removing
one Kramers degenerate partner from each KD. Also the \textit{full
basis} has been retained due to the DM term, as discussed earlier. The results of these calculations
are shown in Figs. \ref{fig:1_Poi_GSE_NNSD}(a)-\ref{fig:1_Poi_GSE_NNSD}(d).
It compares the results of the spin-chain model (histograms) with
the canonical Poissonian and GSE distributions. As seen, fixing $D=0.2$,
we obtain the Poissonian distribution at $\epsilon=0.0$ {[}see panel
(a){]} the pure GSE distribution at $\epsilon=0.6$ {[}see panel (d){]}
and these show an exceptional agreement with the ideal distributions
(broken lines). For intermediate values of $\epsilon$, one obtains
hybrid distributions which match with neither limiting distribution,
examples of which are seen in panels (b) and (c). It is to be noted in this context, that in the absence of any time-reversal breaking anti-unitary symmetry, the Kramers degeneracy persists throughout this transition. Hence retaining the Kramers doublets, like we do in the GSE-to-GUE transition where TRS is progressively broken, is of no interest here, as it leads to a monotonous singular Dirac delta like peak at $s=0$ throughout the transition, essentially decoupled from the other peaks. Hence we discard one of the Kramers degenerate eigenvalues systematically before plotting the NNSD. This is in stark contrast to the GSE-to-GUE transition where this Dirac delta like peak moves and merges with the GSE-like peak, also transferring spectral weight in the process, throughout the transition, as will be seen presently.

A further increase of $\epsilon$ leads to a re-entrant crossover into the Poissonian regime via intermediate distributions that are neither Poissonian nor GSE, a representative case being shown in panel (e) of Fig. \ref{fig:1_Poi_GSE_NNSD}. In this large $\epsilon$ limit the model tends to a random-coupling, nearest neighbor 1D Ising model (used often to model spin-glasses), where the quantum fluctuations due to the Heisenberg and the DM terms are strongly suppressed by a large uniaxial exchange anisotropy, leading to eigenstates that are many-body localized in this new Poissonian regime. We will again encounter a somewhat similar situation in the context of the Poissonian-to-GOE-to-Poissonian crossover, to be discussed later in the paper.

Next, we study the Poissonian-to-GUE crossover by tuning the relative
strengths of the various terms in the Hamiltonian $H_{2}$ (see Table
\ref{tab:Crossovers_between_various_ Symmetry_ Classes_and_their_criteria}).
We start again with the non-random part of $H_{2}$ consisting of
$H_{h}$ and $H_{DM}$ and slowly turn on the random magnetic field
($H_{r}$) to cross over into the GUE regime. The details of the parameter
values and criteria used etc. are summarized in Table \ref{tab:Crossovers_between_various_ Symmetry_ Classes_and_their_criteria},
and the results of our calculations are shown in Fig. \ref{fig:2_Poi_GUE_NNSD}.
When $h=0$, the level-spacing statistics of the ordered model follows
the Poissonian distribution, as shown in Fig. \ref{fig:2_Poi_GUE_NNSD}(a).
Considering the spin-chain model $H_{2}$, at $h=0.15$ and $D=0.2$,
both conventional and unconventional time-reversal symmetries are
now significantly broken (in turn, the KD is lifted), and the NNSD
follows the GUE distribution {[}Fig. \ref{fig:2_Poi_GUE_NNSD}(d){]}.
The model $H_{2}$ exhibits the Poissonian-to-GUE crossover by varying
the random magnetic field $h$ between $0$ and $0.15$, with $D$
kept fixed at $0.2$ (see Table \ref{tab:Crossovers_between_various_ Symmetry_ Classes_and_their_criteria}).
While the calculated end members for the spin-chain model (histograms)
exhibit excellent agreement with the canonical Poissonian and GUE
distributions (broken lines), as seen from Figs. \ref{fig:2_Poi_GUE_NNSD}(a)
and \ref{fig:2_Poi_GUE_NNSD}(d), the intermediate range hybrid distributions
are shown in Figs. \ref{fig:2_Poi_GUE_NNSD}(b) and \ref{fig:2_Poi_GUE_NNSD}(c).
In the Poissonian limit, one of Kramers degenerate partners is discarded
from each Kramers doublet present, as the calculations were carried
out using the \textit{full basis}. 

\begin{figure}
\begin{centering}
\includegraphics[scale=0.4]{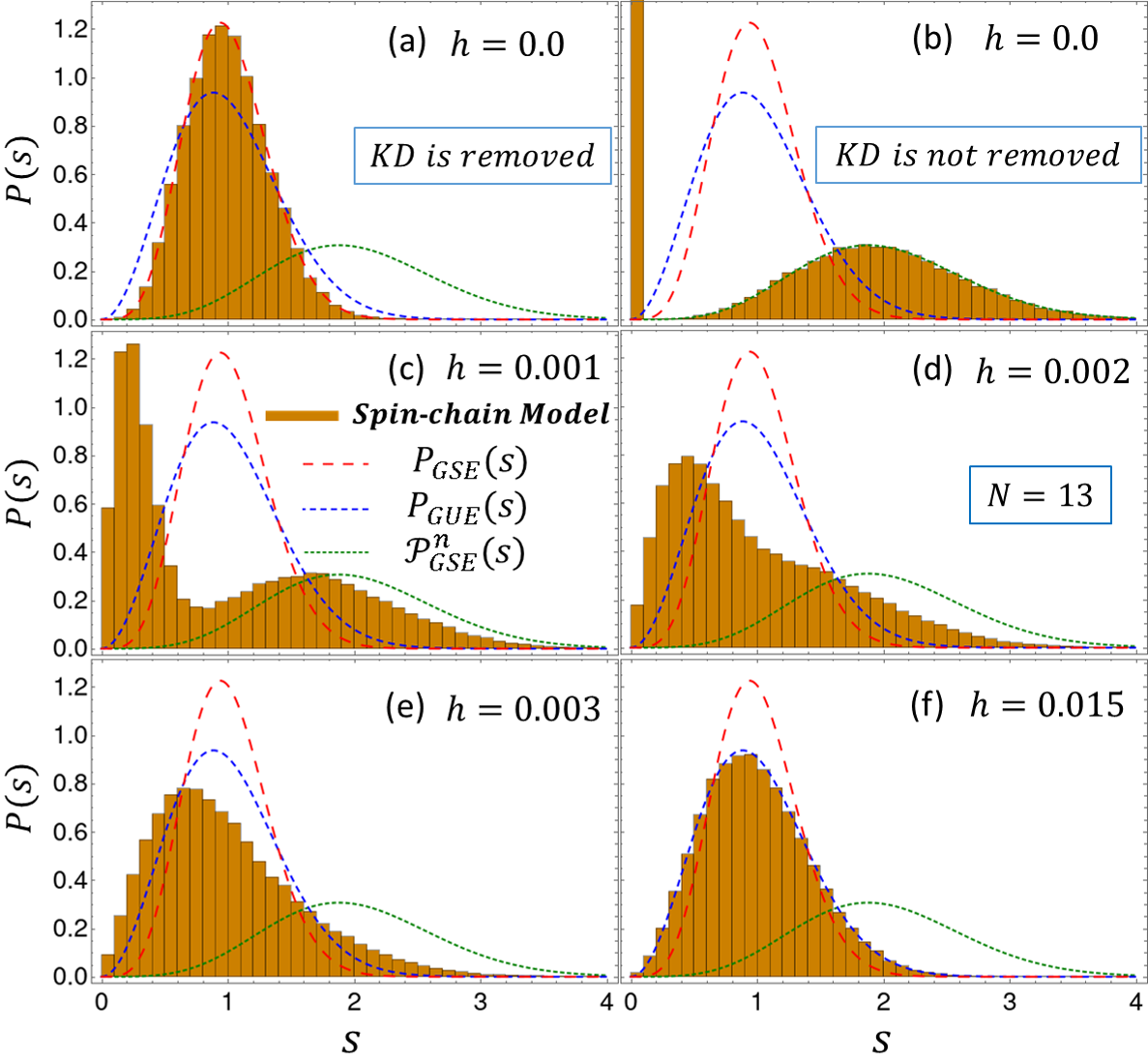}
\par\end{centering}
\caption{NNSD for $N=13$ ($n=8192$ and $\mathcal{M\mathrm{=}\mathrm{15}}$).
GSE-to-GUE crossover with increasing $h$, fixed $D=0.2$ and $\epsilon=0.6$
(see also Table \ref{tab:Crossovers_between_various_ Symmetry_ Classes_and_their_criteria}).
KD is present at $h=0$. (a) shows the NNSD after removing KD, and
it follows $P_{GSE}(s)$, as expected. In (b), the NNSD is plotted
without removing KD, which follows the analytical distribution $\mathcal{P}_{GSE}^{n}(s)$,
derived in Appendix \ref{sec:NNSD-of-the-GSE-Class-with-KD} and as
predicted by RMT. The agreement is excellent. KD is lifted for finite
$h$, and the NNSD is plotted in (c)-(f). $\mathcal{P}_{GSE}^{n}(s)$
to $P_{GUE}(s)$ crossover with increasing $h$ is observed. (f) shows
the limiting case, which coincides with the NNSD of GUE, at a relatively
modest value of $h=0.015$. An interesting double-peaked structure
is observed in the crossover regime, as seen in (c) and (d). \label{fig:3_Spin_model_GSE_GUE_NNSD}}
\end{figure}

\begin{figure}
\begin{centering}
\includegraphics[scale=0.3]{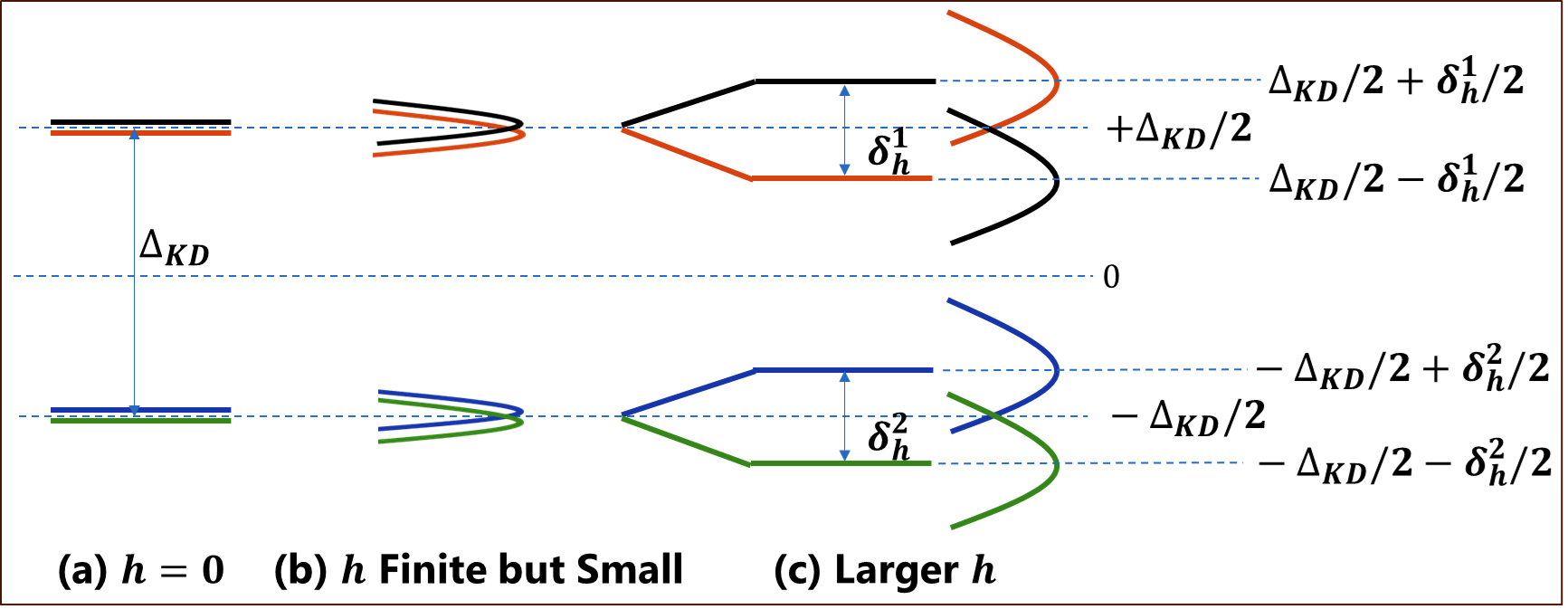}
\par\end{centering}
\caption{Evolution of the Marginal Spectral Densities (MSD) and level-spacings with increasing $h$. Panel (a) represents the case for $h=0$, where the Kramers doublets are truly degenerate, and one has only two distinct level-spacings, $s=0$ and $s=\Delta_{KD}$. The width due to the already present random distribution of the Ising $\epsilon$ is ignored here to focus on the synamics due to $h$ alone. The presence of $\epsilon$ will only contribute further to all broadenings, in addition to also broadening the discrete levels in panel (a). In panel (b), $h$ is small but finite and the Kramers doublets split, and the MSD now gain a narrow width, centered about the slightly split Kramers doublets. In panel (c), a larger finite value of $h$ further splits each of the Kramers doublets, but in general by different amounts, $\delta^{1}_{h}$ and $\delta^{2}_{h}$. The MSD evolves from a 2-peak structure in (b) to a 4-peak structure in (c), leading also to three finite nearest neighbor level-spacings.  \label{fig:4_Schematic_lifting_of_KD}}
\end{figure}

Next, \textit{without} removing the Kramers degeneracy from the spectrum,
we present an interesting study of the NNSD crossover between the
GSE and the GUE distributions, while also comparing the results with
the standard case where the KD was removed. As already discussed above, unlike in the Poissonian-to-GSE crossover where the TRS is never broken (and the KD's never lifted), the retaining of the Kramers doublets here, in the GSE-to-GUE crossover, is expected to show an interesting and dynamical evolution of the spectral shape across the crossover. We discuss this in detail now. The information about the
parameter values and the basis used are summarized in Table \ref{tab:Crossovers_between_various_ Symmetry_ Classes_and_their_criteria}
again. In Figs. \ref{fig:3_Spin_model_GSE_GUE_NNSD}(a)-\ref{fig:3_Spin_model_GSE_GUE_NNSD}(f),
we show the NNSD crossover between the GSE and the GUE distributions
for the spin-chain model $H$ (see Table \ref{tab:Crossovers_between_various_ Symmetry_ Classes_and_their_criteria}),
by varying the magnetic field $h$ ($D=0.2$ and $\epsilon=0.6$ are
fixed), which breaks the $T_{0}$ symmetry of the system, in the absence
of any spatial (rotational) symmetry throughout. In Fig. \ref{fig:3_Spin_model_GSE_GUE_NNSD}(a),
we show the results for the spin-chain model calculations (histogram)
and observe that it faithfully follows the standard GSE NNSD {[}$P_{GSE}(s)${]}
at $h=0.0$, where the calculation is carried out after discarding
the Kramers degeneracy from the spectrum. However, when KD is not removed
from the spectrum {[}Fig. \ref{fig:3_Spin_model_GSE_GUE_NNSD}(b){]}, the NNSD from the spin-chain
model follows the derived analytical result $\mathcal{P}_{GSE}^{n}(s)$
(Eq. \ref{eq:NNSD_GSE_Without_removing_KD}).
We can observe the \textit{Dirac delta} peak of $\mathcal{P}_{GSE}^{n}(s)$
at $s=0$, which originates from the \textit{zero} spacings between
the various Karamers doublets. The non-degenerate eigenvalues generate
the broad hump of the modified GSE-like distribution part of ${\cal P}_{GSE}^{n}(s)$,
at finite $s$ {[}$\sim(3/2)\sqrt{\pi/2}${]}, between the distinct Kramers doublets. The width of this hump at $h=0$ is contributed partly by the multitude of distinct splittings between various Kramers doublets and also largely by the width of the distribution of the random Ising term ($\epsilon=0.6$). A detailed
derivation of this modified GSE distribution {[}${\cal P}_{GSE}^{n}(s)${]}
retaining the Kramers doublets and for a general $n$, is presented
in Appendix \ref{sec:NNSD-of-the-GSE-Class-with-KD}, and its large $n$ limit is also discussed.
As seen, it consists of a Dirac delta peak at $s=0$ and a broad hump
at finite $s$ which, as we will see, is a variant of the original
GSE distribution. As we increase $h$, the Kramers degeneracy is lifted,
and the Dirac delta peak broadens and moves away from $s=0$. But,
the non-zero spacings are now reduced and overall hump at finite $s$
now moves towards a lower value of $s$. This trend and related spectral
weight transfer, begins to show up in Fig. \ref{fig:3_Spin_model_GSE_GUE_NNSD}(c)
and the two peaks continue to move towards each other in Fig. \ref{fig:3_Spin_model_GSE_GUE_NNSD}(d),
before finally merging into a single peak in the GUE limit (at $s\sim\sqrt{\pi}/2$), as from
Figs. \ref{fig:3_Spin_model_GSE_GUE_NNSD}(e) and \ref{fig:3_Spin_model_GSE_GUE_NNSD}(f).
It is also observed that a very small symmetry breaking field, $h=0.015$,
is enough to lead the spin system ($H$) into the GUE regime, and
the NNSD finally achieves the $P_{GUE}(s)$ {[}Fig. \ref{fig:3_Spin_model_GSE_GUE_NNSD}(f){]}.

\begin{figure}
\begin{centering}
\includegraphics[scale=0.4]{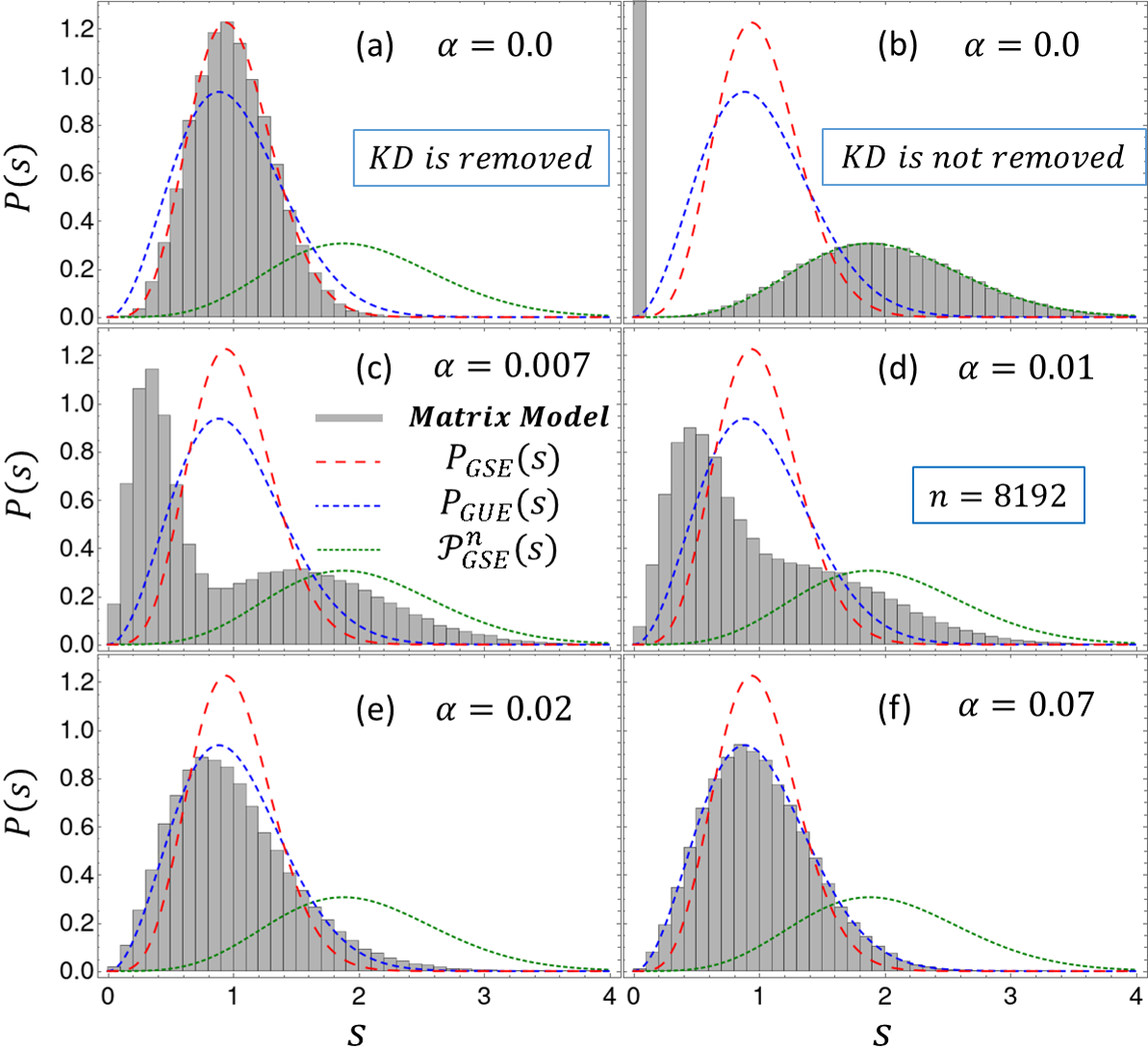}
\par\end{centering}
\caption{NNSD of unfolded eigenvalues of the matrix model {[}Eq. (\ref{eq:GSE_GUE_crossover_matrix_model}){]}
with $n=8192$, considering an ensemble of $\mathcal{M}=15$ matrices, for the GSE-to-GUE crossover. KD is present at $\alpha=0$. (a) shows NNSD after removing the KD,
and it follows $P_{GSE}(s)$, as expected. In (b), the NNSD is plotted
without removing the KD, which follows the analytical RMT distribution
$\mathcal{P}_{GSE}^{n}(s)$, as derived in Appendix \ref{sec:NNSD-of-the-GSE-Class-with-KD}.
KD is lifted for finite $\alpha$, and the NNSD is plotted in (c)-(f).
$\mathcal{P}_{GSE}^{n}(s)$ to $P_{GUE}(s)$ crossover with increasing
$\alpha$ is observed. (f) shows the limiting case ($\alpha=0.07$)
where the NNSD matches the GUE distribution. This crossover RMT parameter
value of $0.07$ is quite small, in agreement with the small value
of $h$ required for the physical crossover in Fig. \ref{fig:3_Spin_model_GSE_GUE_NNSD}.
Just as in the spin model, the unusual double-peak structure is seen
in the crossover regime, as shown in (c) and (d). \label{fig:5_Matrix_model_GSE_GUE_NNSD}}
\end{figure}

The understanding of this is especially intuitive within the physical
spin model. With the introduction of a finite $h$, the degeneracy
of all the Kramers doublets are simultaneously lifted, due to the breaking of
time-reversal symmetry. This Zeeman splitting between the various Kramers doublets now
constitute the new peak at a small but finite $s$, that replaces the
Dirac delta peak at $s=0$. It also gains a finite width due to the
fact that the Zeeman splitting of the different Kramers doublets are, in general
different depending on the value of $\mathrm{S^{z}}$ that appears
in the Kramers doublet (as states with $\pm\mathrm{S^{z}}$, the splitting is proportional
to $\mathrm{S^{z}}$). A second important contribution to the broadening, comes from
the width of the distribution for the $h_{j}$'s themselves, represented
by the value of $h$. Since the center-of-gravity of the individual
Kramers doublets remain intact when the magnetic field splits the Kramers doublets,
it is evident that as the Zeeman splittings of the Kramers doublets increase,
the nearest neighbour separation between the \textit{lower} Zeeman-split
partner of a certain Kramers doublet and the \textit{upper} Zeeman-split partnet
of its immediately lower Kramers doublet will reduce. This causes the higher-$s$
hump to move towards a lower $s$ value in Fig. \ref{fig:3_Spin_model_GSE_GUE_NNSD}(c), along with a concurrent movement of the Dirac delta derived small-$s$ peak towards higher energies, due to an overall increase in the Zeeman splittings of all Kramers doublets.
With a further increase in the magnetic field, the Zeeman splittings between the
Kramers doublets further increase, which concurrently reduces the splittings between
distinct Kramers doublet split-states, and the aforesaid movements of the peaks
continue, as seen from Fig. \ref{fig:3_Spin_model_GSE_GUE_NNSD}(d).

This is also schematically depicted in Fig. \ref{fig:4_Schematic_lifting_of_KD} for a simplified model with 2 Kramers doublets, and ignoring the effect of the finite $\epsilon$ for now, remembering that it also contributes to the widths of the various peaks seen in our actual calculations. For $h=0$, as shown in panel (a), the Kramers doublets are truly degenerate, so that one has two distinct level-spacings only, $s=0$ {\em within} the two Kramers doublets and a finite $s=\Delta_{KD}$ {\em between} the pair of Kramers doublets. With the introduction of a small but finite $h$ the Kramers doublets split, but now also gain a width due to the spread in $h$. This is shown in panel (b), by the narrow and slightly displaced distributions centered about the original discrete levels. With a further increase in $h$ to a larger finite value, each of the Kramers doublets split even more, but in general by different amounts given by the values $\delta^{1}_{h}$ and $\delta^{2}_{h}$ here, as shown in panel (c). The corresponding widths of the Marginal Spectral Densities (MSD), centered about these states, are also seen to increase, as $h$ increases. Now one has four peaks in the MSD, and three different level-spacings ($s$), in general, as seen from Fig. \ref{fig:4_Schematic_lifting_of_KD}. The bare discrete level energies are marked in the figure, that yields the values $s=\delta^{1}_{h},\; \delta^{2}_{h}$, and [$\Delta_{KD} - \frac{1}{2}(\delta^{1}_{h} + \delta^{2}_{h})$] for the three bare level-spacings. So even this simple model shows how the MSD may evolve from a 2-peak to a 4-peak structure, and the NNSD may evolve from a 2-peak to a 3-peak structure, for example, as $h$ is increased. This further clarifies the origin of the trends seen in our physical spin model, as depicted in Fig. \ref{fig:3_Spin_model_GSE_GUE_NNSD} above.

Although the above discussion on the origin of the double peak structure
within the spin-chain model is enlightening, it should not give the
reader the impression that the qualitative nature of this behavior
is specific to spin models alone. In fact we will show below that
this behavior is generic of any GSE-to-GUE crossover, whenever the
Kramers degeneracy is retained, rather than weeded out. Although this
is already borne out by the analytical calculation for ${\cal P}_{GSE}^{n}(s)$
in the GSE limit, an analytical calculation for the intermediate regime
may be a daunting task. The best way to then demonstrate the robustness
of this behavior, across the entire crossover, would be to repeat
the calculation for a \textit{crossover matrix model} numerically within the Pandey-Mehta
approach, which has no direct connection with any specific physical
model. To facilitate a reasonably detailed comparison with our calculation,
in terms of the shape of the spectral distribution etc., we do this
for the exact same matrix dimension as the spin-chain model, viz.
$n=8192$. For the GSE-to-GUE crossover, the \textit{crossover matrix
model} of Eq. (\ref{eq:crossover_matrix_model}) becomes, 
\begin{equation}
\mathcal{H}=(1-\alpha)\mathcal{H}_{GSE}+\alpha\mathcal{H}_{GUE}.\label{eq:GSE_GUE_crossover_matrix_model}
\end{equation}
We also keep the ensemble size for configuration averaging same as
that of the spin model (${\cal M}=15$). At $\alpha=0$, the NNSD
of the matrix model yields $P_{GSE}(s)$ as expected as expected {[}Fig.
\ref{fig:5_Matrix_model_GSE_GUE_NNSD}(a){]}, when KD is removed from
the spectrum. In the Figs. \ref{fig:5_Matrix_model_GSE_GUE_NNSD}(b)-\ref{fig:5_Matrix_model_GSE_GUE_NNSD}(f),
the NNSDs are plotted without removing the KD. Similar to the spin-chain
model, we again observe the \textit{Dirac delta} peak of $\mathcal{P}_{GSE}^{n}(s)$
at $s=0$. The Figs. \ref{fig:5_Matrix_model_GSE_GUE_NNSD}(b)-\ref{fig:5_Matrix_model_GSE_GUE_NNSD}(f),
represent the NNSD crossover between $\mathcal{P}_{GSE}^{n}(s)$ to
$P_{GUE}(s)$, with increasing $\alpha$. The intermediate distributions
{[}Figs. \ref{fig:5_Matrix_model_GSE_GUE_NNSD}(c) and \ref{fig:5_Matrix_model_GSE_GUE_NNSD}(d){]}
also show a similar two-peaked structure as observed for the physical
system {[}Figs. \ref{fig:3_Spin_model_GSE_GUE_NNSD}(c) and \ref{fig:3_Spin_model_GSE_GUE_NNSD}(d){]}.
We observe that the NNSD converges to the GUE distribution at $\alpha=0.07$,
which is rather small compared to the analytical RMT requirement ($\alpha=1.0$),
and seems to track the time-reversal symmetry breaking rather faithfully
and with minimal lag. It is also consistent with the extremely small
magnetic field required in Fig. \ref{fig:3_Spin_model_GSE_GUE_NNSD}(f),
to reach the GUE limit. This behavior is expected from such a large
dimensional matrix model, which is consistent with the discussions
in Refs. \cite{KumarPandey2011b,PandeyMehta1983,Brody_Random_Matrix_Physics_review,Pandey1981,Debojyoti_paper_1_2022},
regarding the rate of the crossovers with the matrix-dimensions. We
conclude from this RMT matrix model study that similar qualitative
NNSD crossover behavior between the GSE and the GUE limits should
be achieved with any relevant many-body quantum system, provided it
satisfies the relevant symmetry requirements as listed in Table \ref{tab:Conditions_for_RMT_ensembles}.

\begin{figure}
\begin{centering}
\includegraphics[scale=0.4]{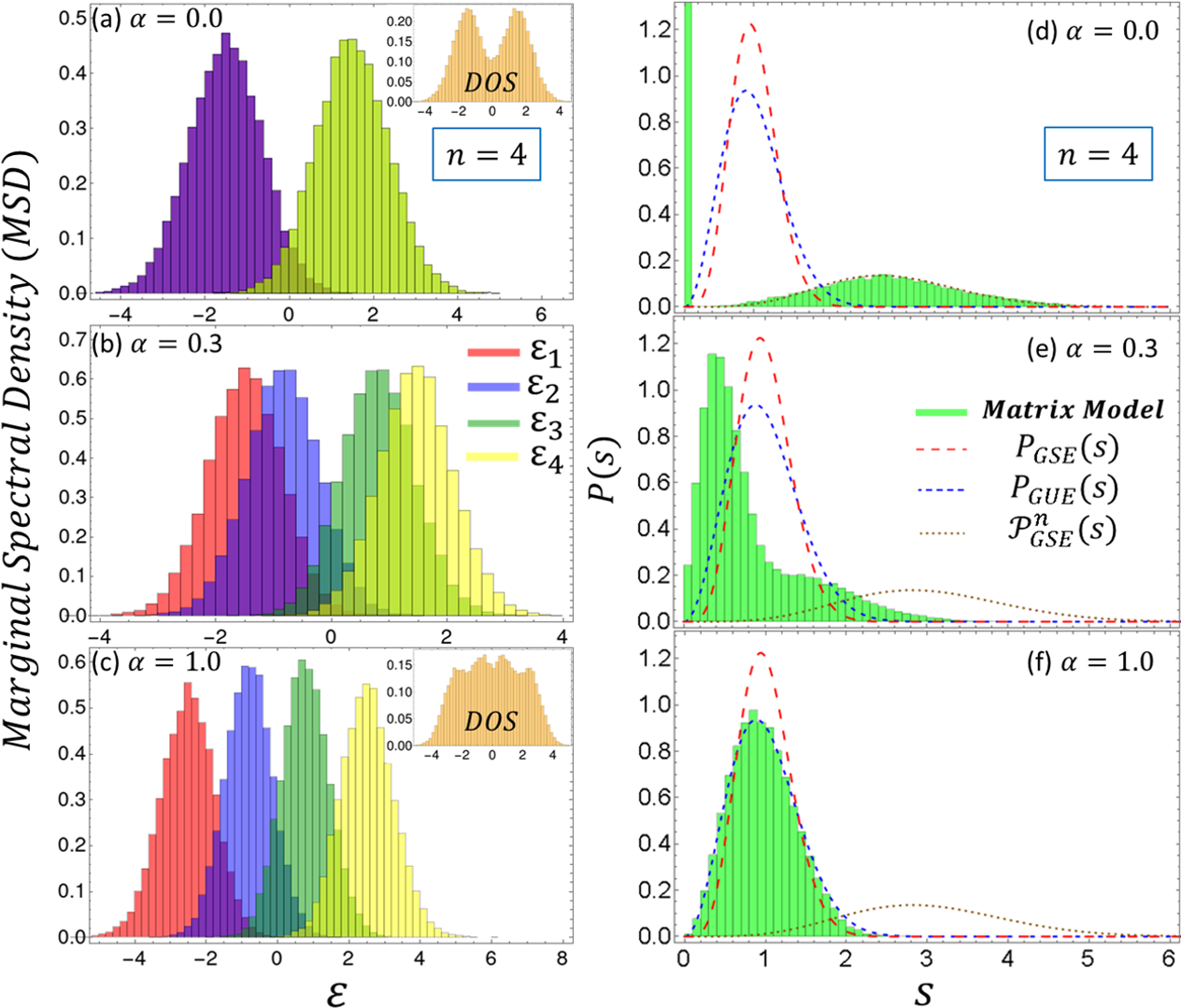}
\par\end{centering}
\caption{Marginal Spectral Density (MSD) of the four individual eigenvalues
($\varepsilon_{1},\varepsilon_{2},\varepsilon_{3},$ and $\varepsilon_{4}$)
of the matrix model {[}Eq. (\ref{eq:GSE_GUE_crossover_matrix_model}){]}
with $n=4$ and a configuration of $\mathcal{M}=$20,000 matrices.
It shows how KD is lifted for the GSE-to-GUE crossover with increasing
$\alpha$. (a) represents the MSD for the GSE matrix model ($\alpha=0$)
which exhibits KD, due to which the MSD of $\varepsilon_{1}$ and
$\varepsilon_{2}$ ($\varepsilon_{3}$ and $\varepsilon_{4}$) are
superimposed on one another. At finite $\alpha$, the splitting between
($\varepsilon_{1}$,$\varepsilon_{2}$) and ($\varepsilon_{3}$,$\varepsilon_{4}$)
increases, which can be seen in (b) and (c). (c) shows the limiting
case ($\alpha=1.0$) where the matrix model represents the GUE class,
and the splitting is maximum for the eigenvalues. The two insets in
(a) and (c) represent the combined DOS of the four eigenvalues, which
exhibit interesting two-peak and four-peak structures. (d)-(f) show
the corresponding NNSD plots. \label{fig:6_DOS_Matrix_model_GSE_GUE}}
\end{figure}

Though the large matrix model is useful in establishing the robustness
of the double-peak structure and its evolution across the crossover,
a smaller matrix model is more transparent in analyzing this behavior
and reinforcing our conjecture regarding the origin of this behavior,
made in the context of the spin model. Hence, in addition to the NNSD,
we also look at the evolution of the MSD across the crossover, in such a model, to see how the states themselves
evolve with $\alpha$. To this end, we consider the crossover matrix
model {[}Eq. (\ref{eq:GSE_GUE_crossover_matrix_model}){]} with dimension
$n=4$ and plot (Fig. \ref{fig:6_DOS_Matrix_model_GSE_GUE}) the MSD
of the individual four eigenvalues ($\varepsilon_{1},\varepsilon_{2},\varepsilon_{3},$
and $\varepsilon_{4}$), considering an ensemble of $\mathcal{M}=$20,000
matrices. It clearly shows how the Kramers degeneracy is lifted for
the GSE-to-GUE crossover with increasing $\alpha$. In Fig. \ref{fig:6_DOS_Matrix_model_GSE_GUE}(a),
we show MSD of the matrix model in the pure GSE limit ($\alpha=0$),
which exhibits the KD, as a result, the distribution of $\varepsilon_{1}$
and $\varepsilon_{2}$ ($\varepsilon_{3}$ and $\varepsilon_{4}$)
are superimposed on one another. With the increase of $\alpha$, the
splittings between ($\epsilon_{1},\epsilon_{2}$) and ($\epsilon_{3},\epsilon_{4}$)
increase, which can be seen from the displacements of their MSD peaks,
in Figs. \ref{fig:6_DOS_Matrix_model_GSE_GUE}(b) and \ref{fig:6_DOS_Matrix_model_GSE_GUE}(c).
In Fig. \ref{fig:6_DOS_Matrix_model_GSE_GUE}(b), the two peaks in
Fig. \ref{fig:6_DOS_Matrix_model_GSE_GUE}(a) have each split into
two Kramers partners. The \textit{violet} peak has separated into
a \textit{red} and a \textit{blue} one, while the \textit{yellowish-green}
peak splits into a \textit{pure green} and a \textit{pure yellow}
one. These splittings keep increasing with increasing $\alpha$. Fig.
\ref{fig:6_DOS_Matrix_model_GSE_GUE}(c) shows the limiting case ($\alpha=1.0$),
where the matrix model represents pure GUE, and the separations between
the MSD peaks are maximum. The two \textit{insets} in the Figs. \ref{fig:6_DOS_Matrix_model_GSE_GUE}(a)
and \ref{fig:6_DOS_Matrix_model_GSE_GUE}(c), represent the combined
DOS of the four eigenvalues, and show how the original two-peak structure
in \textit{inset} of the Fig. \ref{fig:6_DOS_Matrix_model_GSE_GUE}(a)
evolves into a interesting four-peak structure in \textit{inset} of the
Fig. \ref{fig:6_DOS_Matrix_model_GSE_GUE}(c). The above description
is well in line with our proposed conjecture above. The corresponding
NNSD plots are represented in the Figs. \ref{fig:6_DOS_Matrix_model_GSE_GUE}(d)-\ref{fig:6_DOS_Matrix_model_GSE_GUE}(f),
which are consistent with the higher dimensional behavior as seen
both for the spin-system {[}Figs. \ref{fig:3_Spin_model_GSE_GUE_NNSD}(b)-\ref{fig:3_Spin_model_GSE_GUE_NNSD}(f){]}
as also for the larger matrix model {[}Figs. \ref{fig:5_Matrix_model_GSE_GUE_NNSD}(b)-\ref{fig:5_Matrix_model_GSE_GUE_NNSD}(f){]}. 

\begin{figure}
\begin{centering}
\includegraphics[scale=0.33]{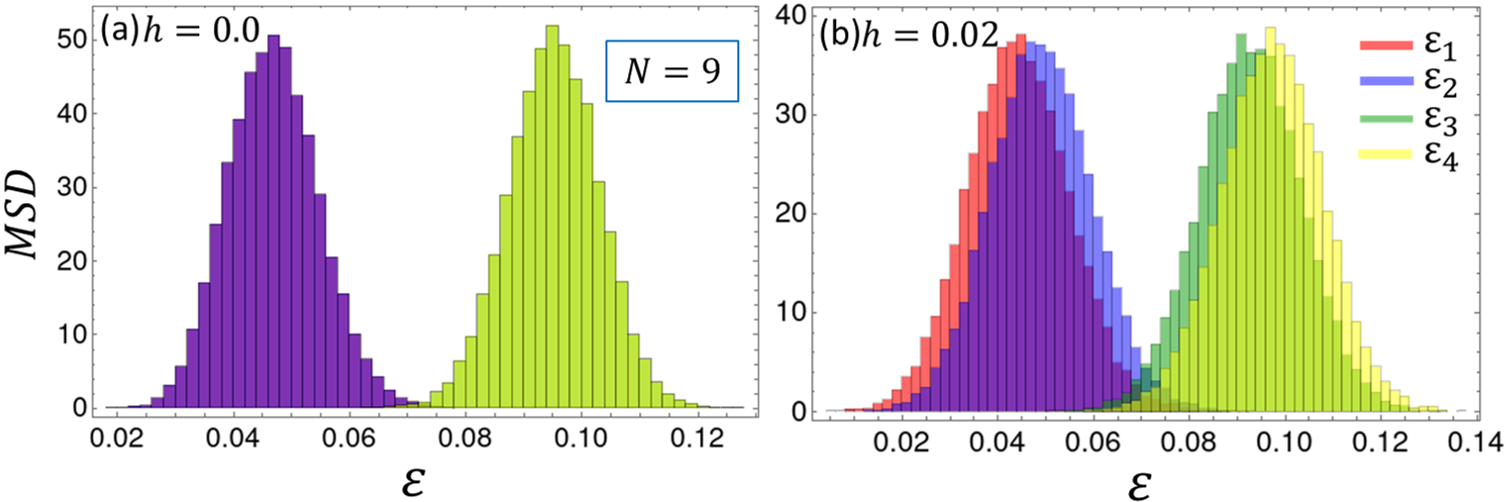}
\par\end{centering}
\caption{Marginal Spectral Density (MSD) of the four individual eigenvalues
($\varepsilon_{1},\varepsilon_{2},\varepsilon_{3},$ and $\varepsilon_{4}$)
(selected from the central peak region of the spectra), of the spin-chain
model $H$ with $N=9$ and a configuration of $\mathcal{M}=$20,000
matrices. It shows how the KD is lifted in a physical system for the
GSE-to-GUE crossover with increasing symmetry breaking field $h$
(fixed $\epsilon=0.6$, $D=0.2$). (a) represents the case of GSE
($h=0$) which exhibits the KD, due to which the MSD of $\varepsilon_{1}$
and $\varepsilon_{2}$ ($\varepsilon_{3}$ and $\varepsilon_{4}$)
are superimposed on one another. With an increase of $h$, the splittings
between ($\varepsilon_{1}$,$\varepsilon_{2}$) and ($\varepsilon_{3}$,$\varepsilon_{4}$)
increase, and (b) shows the splitting between MSD of the eigenspectrum,
in the GUE limit ($h=0.02$). \label{fig:7_DOS_Spin_model_GSE_GUE}}
\end{figure}

Finally, we consider four individual eigenvalues from a spin-chain
calculation and look at the evolution of their MSD with increasing
$h$, to demonstrate explicitly that similar physics is operative
as in the $n=4$ matrix model above. In Figs. \ref{fig:7_DOS_Spin_model_GSE_GUE}(a)
and \ref{fig:7_DOS_Spin_model_GSE_GUE}(b), we present the MSD of
the spin-chain model $H$, with lattice size $N=9$. We examine two
sets of the Kramers degenerate eigenvalues (keeping a separation of
$10$ eigenvalues) from the middle of the spectra, considering an
ensemble of $\mathcal{M}=$20,000 configurations, to obtain smooth
MSD plots.\footnote{To obtain a smooth MSD of only four individual eigenvalues, we need to consider a large number of configurations, and hence a $N=13$ calculation is computationally very expensive for this purpose.}
A small magnetic field, $h=0.02$ (fixed $\epsilon=0.6$, $D=0.2$),
is enough to lift the KD and drive the system from the GSE class to
the GUE class. Here, in Fig. \ref{fig:7_DOS_Spin_model_GSE_GUE}(b),
the two peaks in Fig. \ref{fig:7_DOS_Spin_model_GSE_GUE}(a) have
each split into two Kramers partners and four distinct peaks are observed,
similar to what we observe for the matrix model in Figs. \ref{fig:6_DOS_Matrix_model_GSE_GUE}(a)-\ref{fig:6_DOS_Matrix_model_GSE_GUE}(c). 

\begin{figure}
\begin{centering}
\includegraphics[scale=0.42]{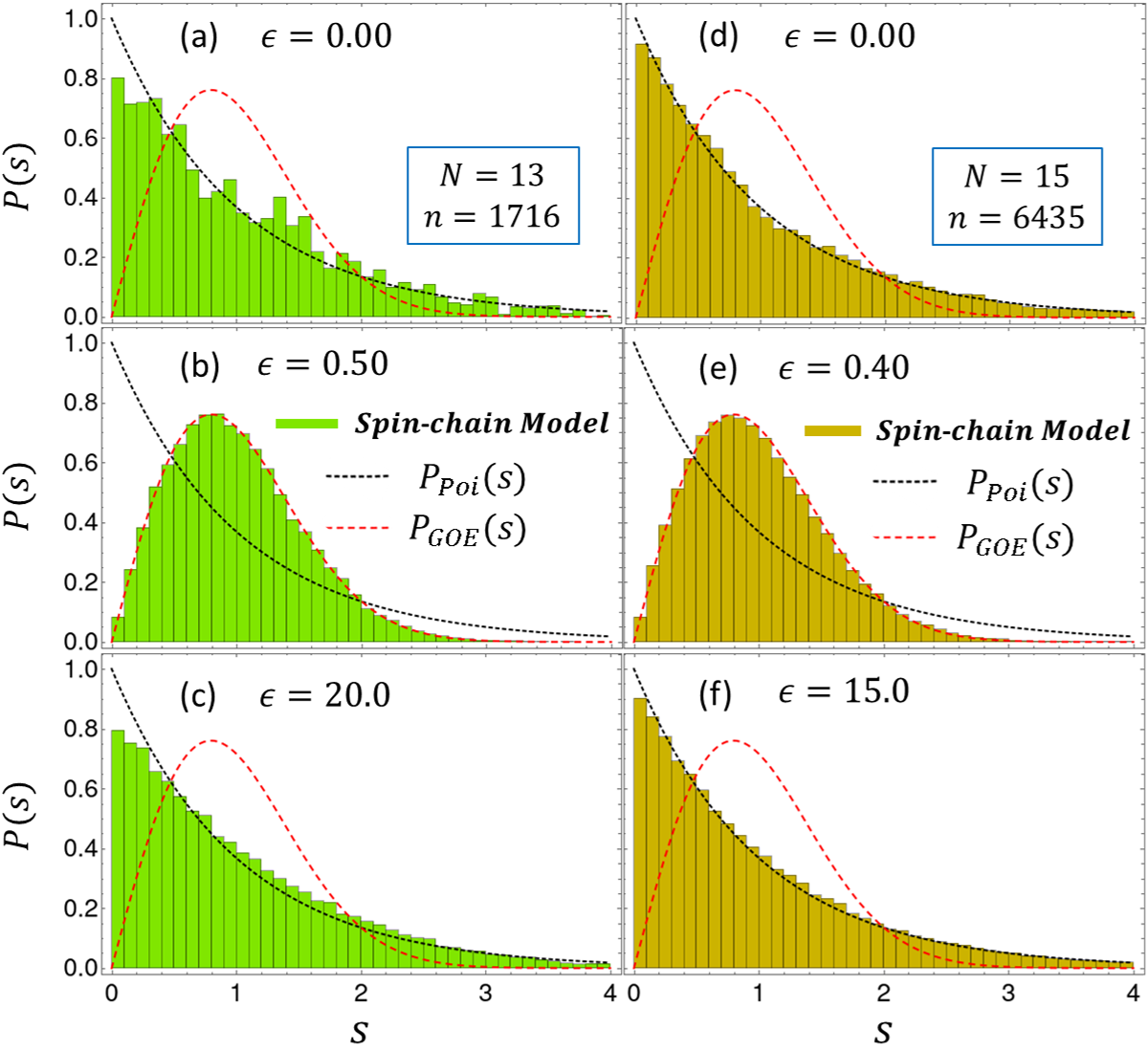}
\par\end{centering}
\caption{NNSD for the Poissonian-to-GOE-to-Poissonian {\em re-entrant} crossovers with increasing $\epsilon$,
within the spin-chain model $H_{1}$, for the lattice sizes $N=13$
($\mathcal{M}=25$) {[}(a)-(c){]} and $N=15$ ($\mathcal{M}=15$)
{[}(d)-(f){]}. (a) shows that the NNSD follows the Poissonian distribution
when $\epsilon$ is zero, (b) shows that the NNSD follows the GOE
distribution for $\epsilon=0.5$, and (c) shows how Poisson distribution
is recovered due to eigenvector localization for a typical large disorder
($\epsilon=20.0$). Similarly, for $N=15$, NNSD is plotted in (d)-(f)
for Possonian-to-GOE-to-Poissonian crossover. One observes that the GOE
and the Poissonian in the localized limit are obtained for lower values
of $\epsilon$ in the $N=15$ lattice, compared to the $N=13$ lattice. Here the localized Poissonian limit is already achieved for $\epsilon=15$.
\label{fig:8_Spin_model_Poi_GOE_NNSD}}
\end{figure}

Now, we are interested to study the Poissonian-GOE-Poissonian (\textit{non-chaotic}
$\rightarrow$ \textit{chaotic} $\rightarrow$ \textit{localized}
transition) re-entrant crossover, in the spin-chain model $H_{1}$.
We previously discussed in Sec. \ref{sec:Spin-Hamiltonians}, how
the spin-chain model $H_{1}$ preserves both (the $T_{0}$ and the
$T$) time-reversal symmetries and has a diagonal disorder (random
Ising interaction along the z-axis). In addition, $H_{1}$ possesses
full rotational symmetry about the $\mathrm{z}$-axis and also has
the discrete rotational symmetries (rotation by $\pi$) about the
$\mathrm{x}$- or the $\mathrm{y}$-axes (or for any axis in the $\mathrm{x}$-$\mathrm{y}$
plane, for that matter), which is in consonance with condition II.2
of Table \ref{tab:Conditions_for_RMT_ensembles}. As $H_{1}$ commutes
with $\mathrm{S^{z}}$, we need to restrict any calculation for $H_{1}$,
to a fixed $\mathrm{S^{z}}$ subspace (see Table \ref{tab:Crossovers_between_various_ Symmetry_ Classes_and_their_criteria}
for full details). Here, for an \textit{odd} $N$, we consider the
$\mathrm{S^{z}}=\frac{1}{2}$ subspace ($n=1716$ for $N=13$ and
$n=6435$ for $N=15$ systems), which is the lowest $\mathrm{S}^{\mathrm{z}}$
subspace and hence has representations from all possible total $\mathrm{S}$
sectors. At $\epsilon=0.0$, there is no disorder in the system, and
we achieve the Poissonian limit for the NNSD studies {[}Fig. \ref{fig:8_Spin_model_Poi_GOE_NNSD}(a)
for the $N=13$ and Fig. \ref{fig:8_Spin_model_Poi_GOE_NNSD}(d) for the $N=15$
lattice sizes{]}. Due to the \textit{self-averaging} or the \textit{spectral
ergodicity} property of RMT \cite{Haake-Chaos_book,Pandey1979_ergodicity,Debojyoti_paper_1_2022},
we observe much smoother Histogram plots for the $N=15$ system, compared
to the $N=13$ system. The GOE distribution is achieved at $\epsilon=0.5$
{[}Fig. \ref{fig:8_Spin_model_Poi_GOE_NNSD}(b){]} and $\epsilon=0.4$ {[}Fig. \ref{fig:8_Spin_model_Poi_GOE_NNSD}(e){]},
for the $N=13$ and the $N=15$ systems, respectively. Further increase
of diagonal disorder, results in the onset of localization in the eigenstates.
This recovers the Poissonian distribution in the strongly disordered
limit, viz. $\epsilon=20.0$ and $\epsilon=15.0$ for the $N=13$
{[}Fig. \ref{fig:8_Spin_model_Poi_GOE_NNSD}(c){]} and the $N=15$ {[}Fig. \ref{fig:8_Spin_model_Poi_GOE_NNSD}(f){]}
systems, respectively. At this stage we should take note of the similarities in the trends and the associated physics, with the re-entrant Poissonian-to-GSE-to-Poissonian crossover, discussed at the beginning of this section. This is also strongly reminiscent of a similar Poissonian
$\rightarrow$ GOE $\rightarrow$ Poissonian re-entrant crossover
encountered in our earlier paper \cite{Debojyoti_paper_1_2022},
where a random inhomogeneous magnetic field term competes with the
Heisenberg term to bring about the many-body localization, in the
large diagonal disorder limit.

\subsection{Spectral Rigidity and Number Variance\label{subsec:long_range_results}}

Till now, our studies seem to indicate that the short-range eigenvalue
correlation statistics of the spin-chain models, follow RMT predictions,
quite accurately. This is also borne out by earlier studies of short-range
spectral correlations in other 1D spin models \cite{Debojyoti_paper_1_2022,Modak's-paper_2014,Hamazaki_DM_RMT_paper,Santos_Gubin_Inverted_spin,Avishai-Richert-PRB}.
We now want to study the long-range eigenvalue correlation properties
and related spectral crossovers, in our above spin model, in the presence
of tunable symmetry-breaking physical parameters. We also compare
the values of the symmetry-breaking physical crossover parameters
between the short-range (NNSD) and long-range (spectral rigidity and
number variance) spectral fluctuation studies. As already mentioned, in Table \ref{tab:Crossovers_between_various_ Symmetry_ Classes_and_their_criteria},
we have listed the physical crossover parameter values for the NNSD.

\begin{figure}
\begin{centering}
\includegraphics[scale=0.4]{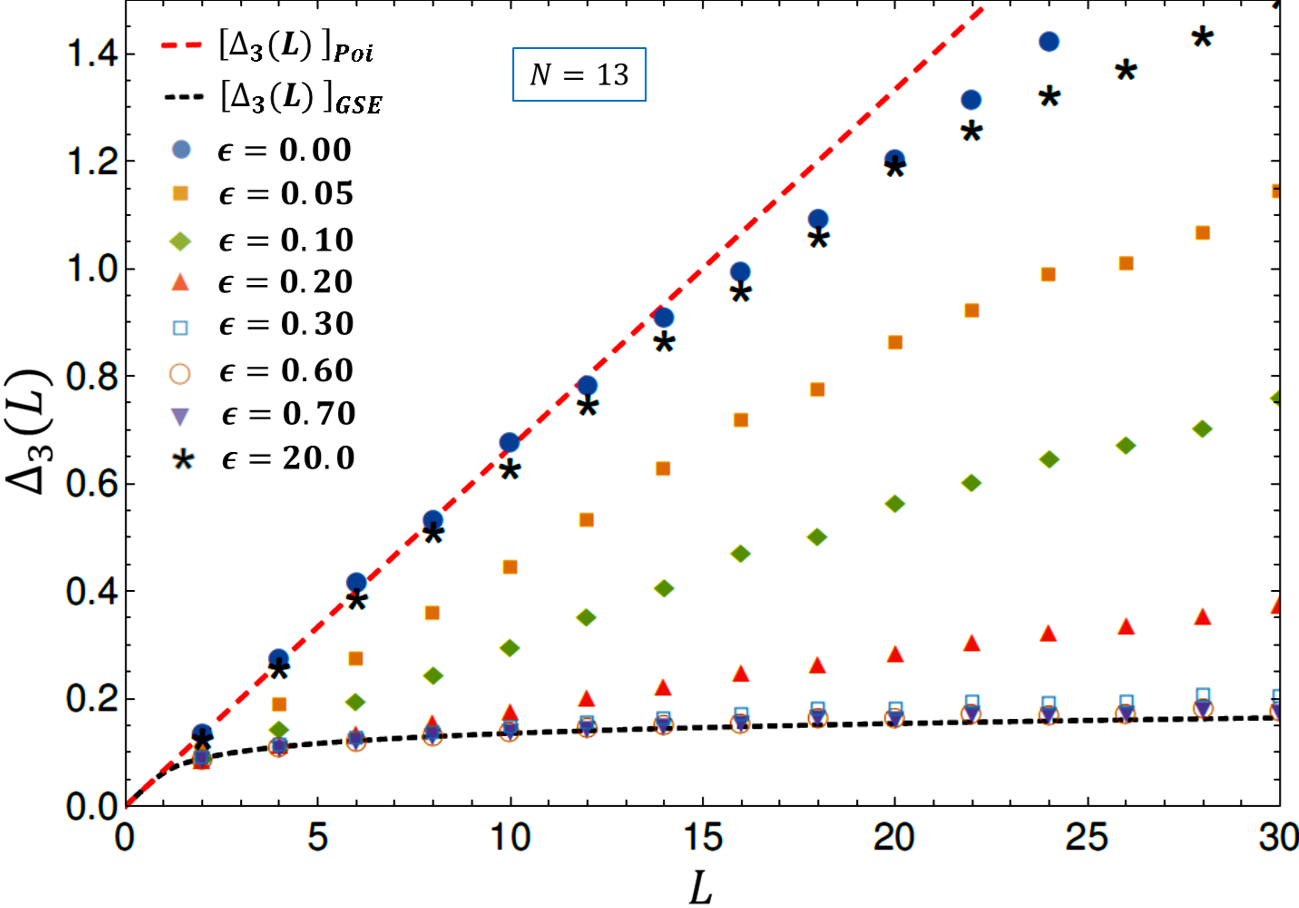}
\par\end{centering}
\caption{Spectral Rigidity ($\Delta_{3}$-statistic) for the $N=13$ spin-chain
($\mathcal{M}=15$) across the Poissonian-to-GSE crossover, with increasing
$\epsilon$, and fixed $D=0.2$ (symbols). For the extremal cases,
where one expects the physical system to follow either of the pure
statistics (Poissonian or GSE), we have also plotted the RMT exact
analytical predictions (broken lines), for comparison. It also shows the $\Delta_{3}$-statistc
for the many-body localized phase, brought about by the large disorder
($\epsilon=20.0$) (black asteriks), which is again expected to follow
the Poissonian result approximately (red broken line). \label{fig:9_Delta_3_Poi_GSE}}
\end{figure}

Using Eq. (\ref{eq:spectral_rigidity}), we calculate the $\Delta_{3}(L)$-statistic
for the lattice size $N=13$, in the various possible spectral
crossovers listed in Table \ref{tab:Crossovers_between_various_ Symmetry_ Classes_and_their_criteria}, as described below. In Fig. \ref{fig:9_Delta_3_Poi_GSE},
we show the $\Delta_3(L)$ for Poissonian-to-GSE-to-Poissonian re-entrant crossover plots with increasing $\epsilon$. This simply introduces randomness without breaking any time-reversal symmetry. Here, we study the long-range
correlations amongst eigenvalues with interval length ranging from
$L=2$ to $L=30$, in steps of $2$. In this calculation, one of the Kramers
degenerate partners is discarded from each doublet. We observe that
$\Delta_{3}(L)$ follows the analytical Poissonian result, $\left[\Delta_{3}(L)\right]_{Poi}$,
up to $L\sim14$ and then deviates from it. With increasing value
of the random parameter $\epsilon$, Poissonian-to-GSE crossover is
achieved at $\epsilon=0.6$ and $\Delta_{3}(L)$ follows the analytical
GSE result, $\left[\Delta_{3}(L)\right]_{GSE}$, quite well, up to
large eigenvalue interval lengths. We have earlier achieved the GSE
crossover in NNSD at same same value of the random parameter $\epsilon$.
Even an increased value of $\epsilon=0.7$ does not reflect any better
agreement, as is evident from the figure. This shows that, at $\epsilon=0.6$,
the eigenvalue fluctuation properties of the spin-chain system ($H_{3}$),
follow GSE statistics at both the \textit{local} (or short-range)
and the \textit{global} (or long-range) scales.

\begin{figure}
\begin{centering}
\includegraphics[scale=0.4]{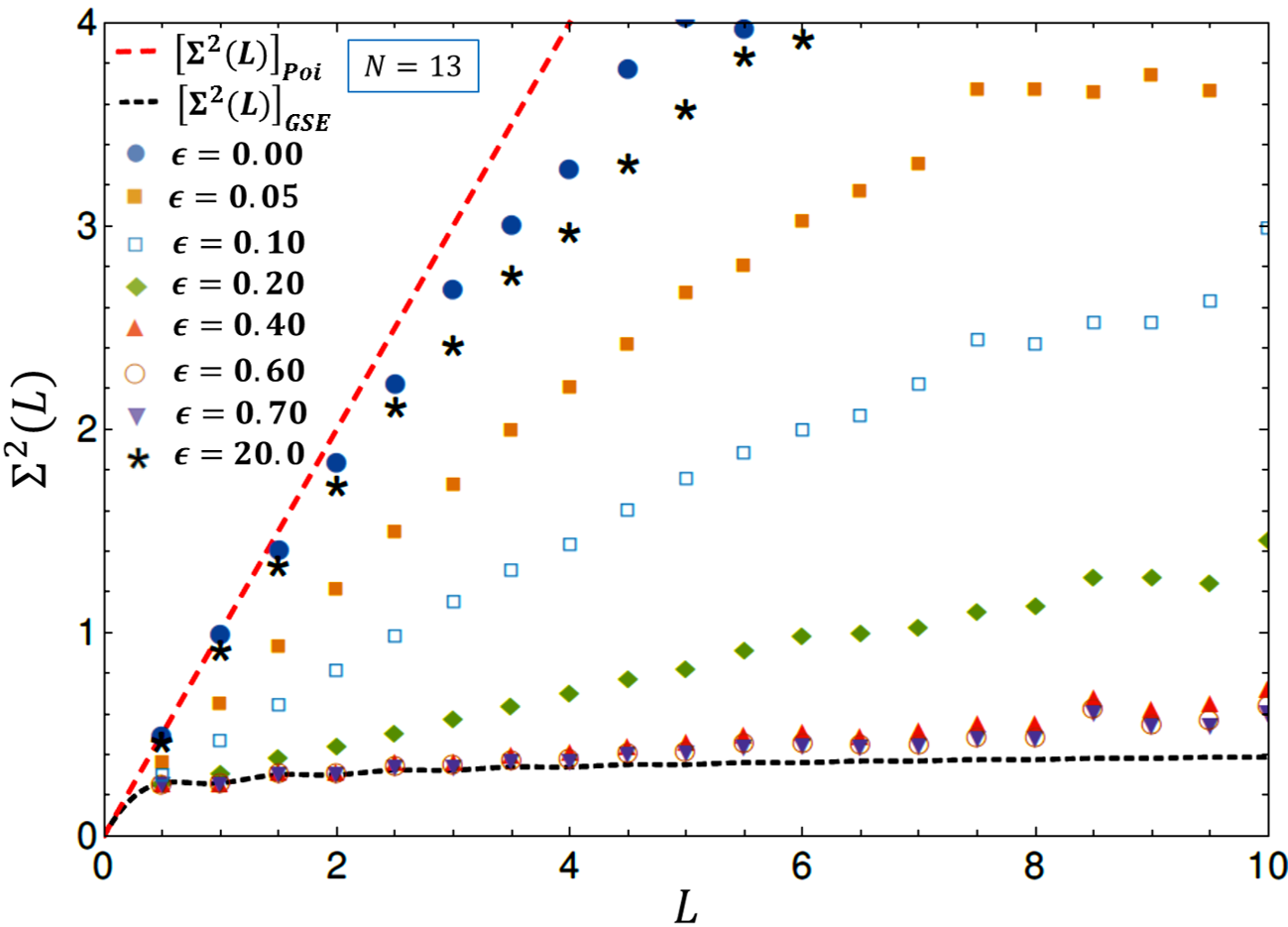}
\par\end{centering}
\caption{Number variance ($\Sigma^{2}$-statistc) for the $N=13$ spin-chain
($\mathcal{M}=15$) across the Poissonian-to-GSE crossover with increasing
$\epsilon$, and fixed $D=0.2$ (symbols). For the extremal cases,
where one expects the physical system to follow either of the pure
statistics (Poissonian or GSE), we have also plotted the RMT exact
analytical predictions (broken lines), for comparison. It also shows the $\Sigma^{2}$-statistc
for the many-body localized phase, brought about by the large disorder
($\epsilon=20.0$) (black asteriks), which is again expected to follow
the Poissonian result approximately (red broken line). \label{fig:10_Sigma_Poi_GSE}}
\end{figure}

Now, using Eq. (\ref{eq:number_variance}), we calculate the number variance $\Sigma^{2}(L)$ for the spin-chain systems with lattice
size $N=13$, considering the eigenvalue interval length starting
from $L=0.5$ to $L=10$, in steps of $0.5$. In Fig. \ref{fig:10_Sigma_Poi_GSE},
we show the Poissonian-to-GSE-to-Poissonian crossover in $\Sigma^{2}(L)$-statistic
for the spin-chain model $H_{3}.$ We observe that, in the absence
of the random Ising interaction, $\Sigma^{2}(L)$ follows the Poissonian
analytical prediction, $\left[\Sigma^{2}(L)\right]_{Poi}$, only up
to an eigenvalue interval length $L\sim2$, beyond that it deviates
from the ideal result. In the GSE regime ($\epsilon=0.6$), the $\Sigma^{2}(L)$-statistic
of the spin-chain model almost perfectly overlaps with the oscillatory
statistical function $\left[\Sigma^{2}(L)\right]_{GSE}$ (the oscillatory
nature comes from the full integral expressions discussed in the Appendix
\ref{sec:Full-Integral-Expressions-for-Delta-and-Sigma}), up to an
eigenvalue interval length $L\sim4$, beyond that it deviates. As
the figure shows, even an increased value of $\epsilon=0.7$ does
not result in better agreement.

The two Poissonian limits, $\epsilon=0.0$ and $\epsilon=20.0$, coincide with the ideal Poissonian result $\left[\Delta_{3}(L)\right]_{Poi}$ ($\left[\Sigma^{2}(L)\right]_{Poi}$) (broken red line), till up to $L\sim14$ and 8 ($L\sim2$ and 1), respectively. Though both the ordered ($\epsilon=0.0$) and the highly disordered ($\epsilon=20.0$) limits tend to agree with the ideal Poissonian prediction till a finite $L$ value, it seems that the disordered/localized Poissonian limit starts to deviate sooner (for a lower $L$) than the ordered Poissonian limit from the ideal Poissonian result, for both the $\Delta_{3}$ and the $\Sigma^{2}$ statistics. This seems unlike the short-range correlations [see Fig. \ref{fig:1_Poi_GSE_NNSD}, especially panels (a) and (f)] where both these limits seem to follow the Poissonian distribution equally faithfully (for similar behavior in a different spin model, see Ref. \cite{Debojyoti_paper_1_2022}). {\em This property of long-range correlations may serve to distinguish between the two different Poissonian regimes, which NNSD is not able to distinguish.}

\begin{figure}
\begin{centering}
\includegraphics[scale=0.4]{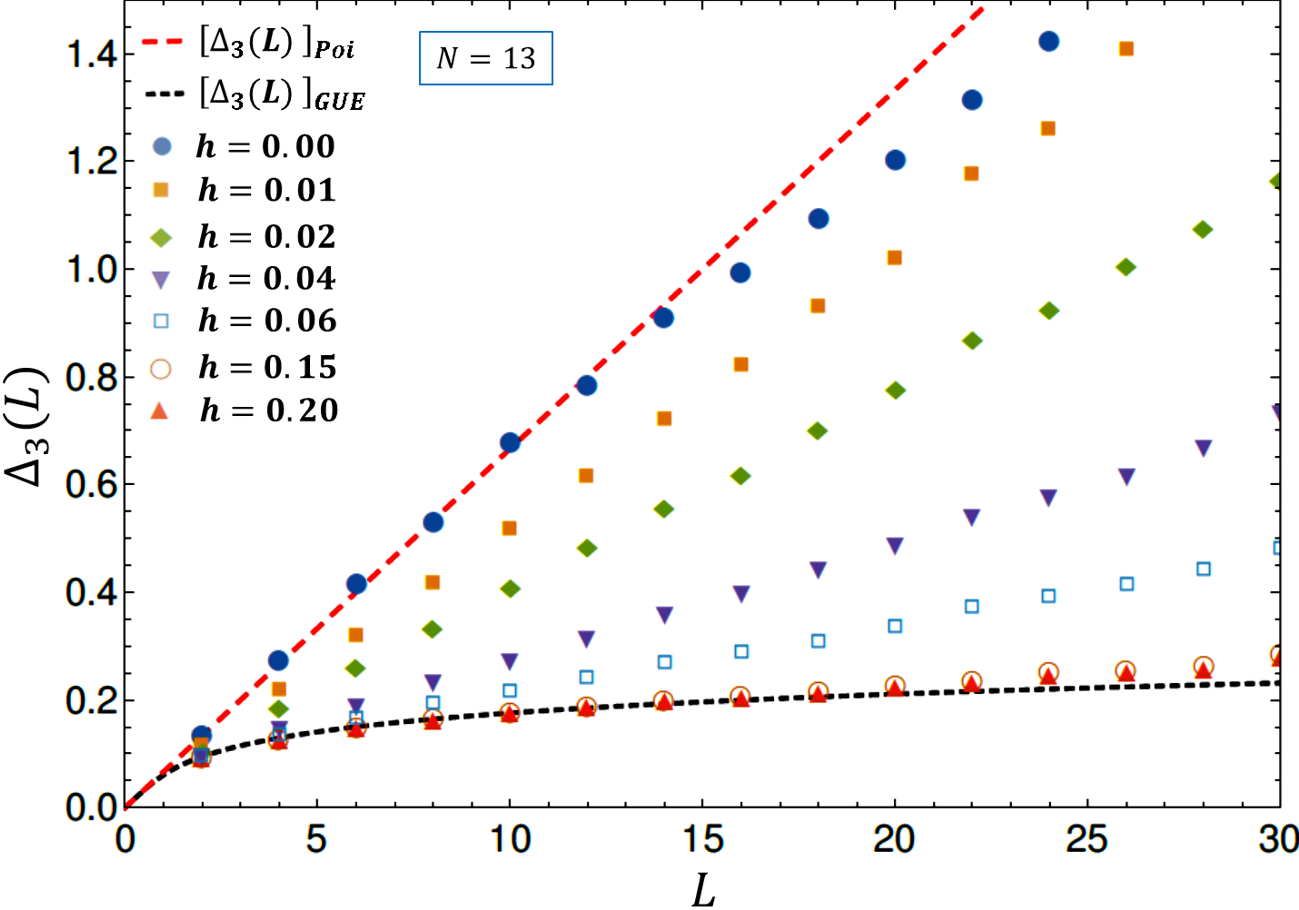}
\par\end{centering}
\caption{Spectral Rigidity ($\Delta_{3}$-statistic) for the $N=13$ spin-chain
($\mathcal{M}=15$) across the Poissonian-to-GUE crossover with increasing
$h$, and fixed $D=0.2$ (symbols). For the extremal cases, where
one expects the physical system to follow either of the pure statistics
(Poissonian or GUE), we have also plotted the RMT exact analytical
predictions (broken lines), for comparison.\label{fig:11_Delta_3_Poi_GUE}}
\end{figure}

Now, we study the spectral rigidity of the spin-chain system in model
$H_{2}$ with $N=13$. The system undergoes a Poissonian-to-GUE crossover
(see Table \ref{tab:Crossovers_between_various_ Symmetry_ Classes_and_their_criteria})
with increasing $h$, keeping $D$ fixed at $0.2$. Fig. \ref{fig:11_Delta_3_Poi_GUE}
shows the $\Delta_{3}(L)$-statistic with interval length starting
from $L=2$ to $L=30$. We observe that at $h=0.0$ (KD is present,
one of Kramers degenerate partners is discarded from each doublet),
$\Delta_{3}(L)$-statistic coincides with the $\left[\Delta_{3}(L)\right]_{Poi}$
for $L$ values up to $14$, just like for the Poissonian-to-GSE transition,
discussed above. At the \textit{local} (or short-range) GUE limit
($h=0.15$), we notice that $\Delta_{3}(L)$-statistic follows the
standard GUE result, $\left[\Delta_{3}(L)\right]_{GUE}$, till $L\sim14$.
If we further increase $h$ to $0.2$, we do notice some improvement
in agreement with $\left[\Delta_{3}(L)\right]_{GUE}$, up to $L\sim20$.
So, for this case, $h=0.2$ serves as a better crossover point for
the \textit{global} GUE crossover, and $h=0.15$ can be regarded as
a more of a \textit{local} crossover point. We have also plotted the
$\Delta_{3}(L)$-statistic for the intermediate cases in Fig. \ref{fig:11_Delta_3_Poi_GUE}.

\begin{figure}
\begin{centering}
\includegraphics[scale=0.4]{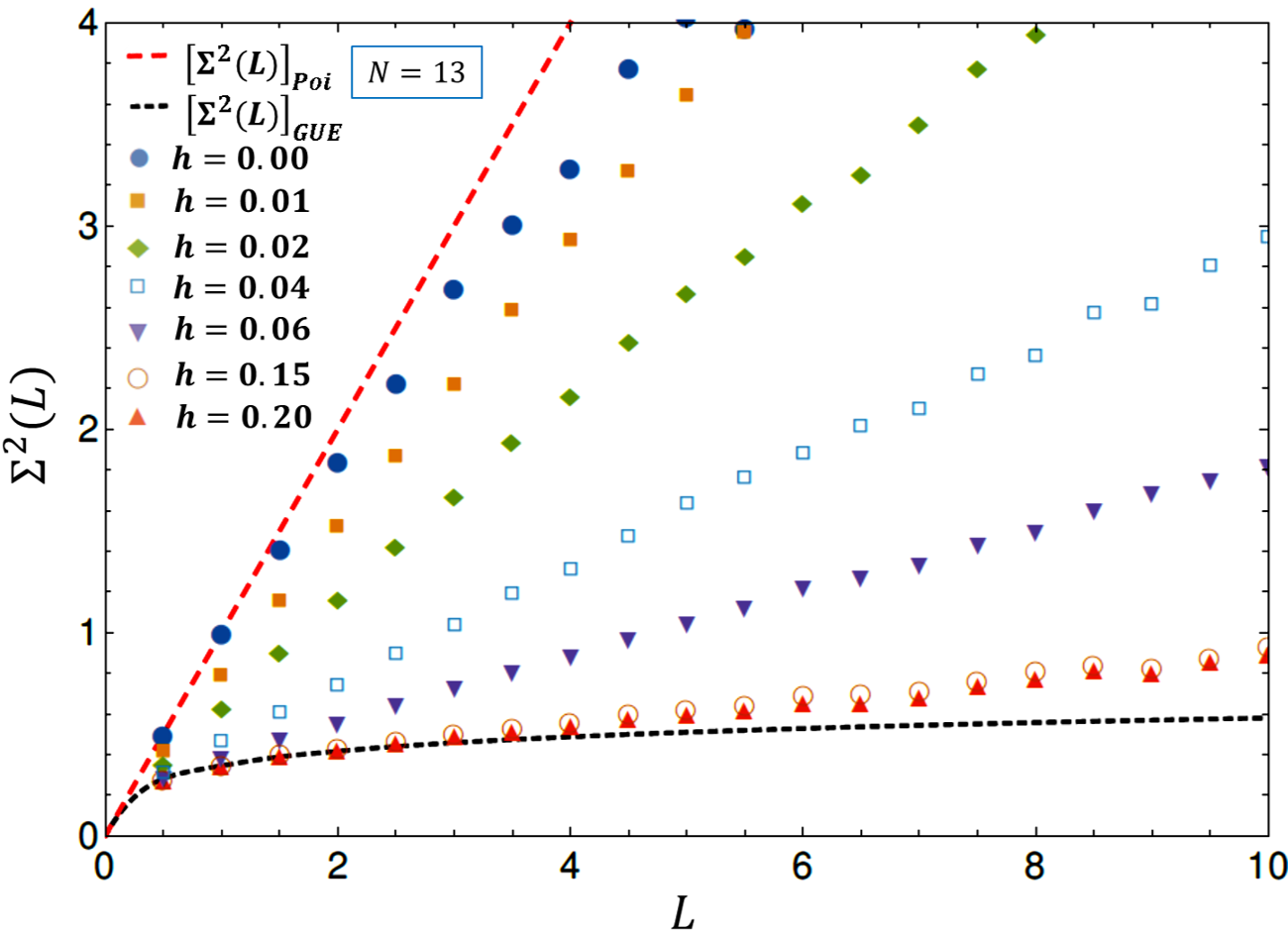}
\par\end{centering}
\caption{Number variance ($\Sigma^{2}$-statistc) for the $N=13$ spin-chain
($\mathcal{M}=15$) across the Poissonian-to-GUE crossover with increasing
$h$, and fixed $D=0.2$ (symbols). For the extremal cases, where
one expects the physical system to follow either of the pure statistics
(Poissonian or GUE), we have also plotted the RMT exact analytical
predictions (broken lines), for comparison. \label{fig:12_Sigma_Poi_GUE}}
\end{figure}

In Fig. \ref{fig:12_Sigma_Poi_GUE}, we plot the level number variance
$\Sigma^{2}(L)$ of the spin-chain model $H_{2}$, having $N=13$
sites, for the eigenvalue interval length starting from $L=0.5$ to
$L=10$, in steps of $0.5$. We observe that, $\Sigma^{2}(L)$ follows
the analytical Poissonian result (at $h=0.0$) only up to $L\sim2$,
beyond that it deviates from this ideal value, just like in the Poissonian-to-GSE
crossover. The Poissonian-to-GUE crossover is achieved at $h=0.2$,
but coincides with the $\left[\Sigma^{2}(L)\right]_{GUE}$ only till
the interval length $L\sim4$. Comparing this with the lower value
of the NNSD crossover point, $h=0.15$, we conclude that, a relatively
higher symmetry-breaking field is required, in this case, to achieve the crossover
in the long-range eigenvalue correlation studies.

\begin{figure}
\begin{centering}
\includegraphics[scale=0.4]{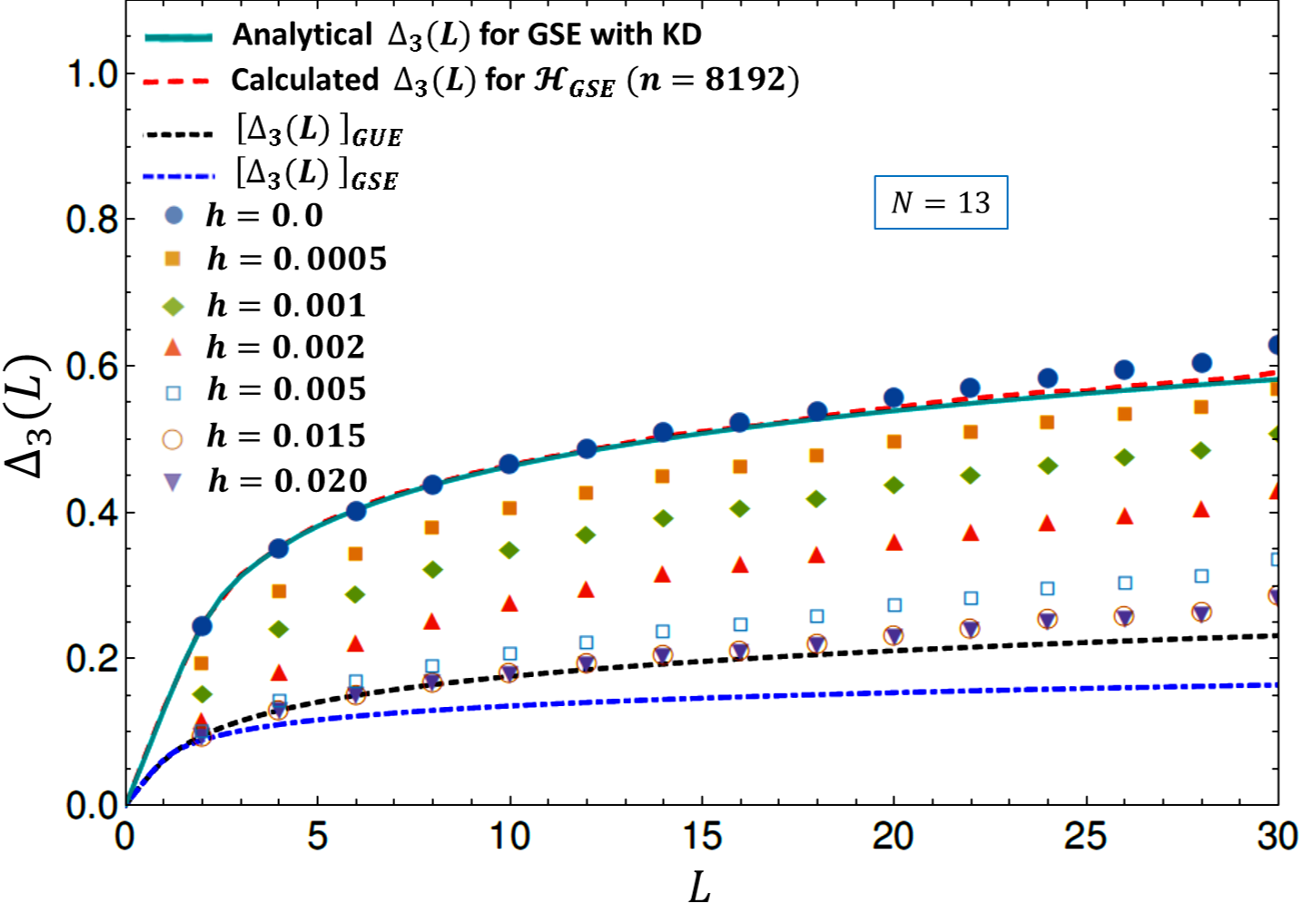}
\par\end{centering}
\caption{Calculated Spectral Rigidity ($\Delta_{3}$-statistic) of the \textit{consecutive}
eigenvalues (keeping all eigenvalues, including the Kramers degenerate
ones) for the $N=13$ spin-chain ($\mathcal{M}=15$) across the GSE-to-GUE
crossover with increasing $h$, and fixed $D=0.2$ and $\epsilon=0.6$
(symbols). To compare their extent of agreement with RMT predictions, the analytical expression for the $\Delta_{3}$-statistic of the non-standard GSE using the consecutive eigenvalues is plotted (cyan line) for a very small value of the crossover parameter (see Appendix~\ref{sec:Full-Integral-Expressions-for-Delta-and-Sigma}). For comparison, here we
also plot the $\Delta_{3}$-statistic of the RMT matrix model, $\mathcal{H}_{GSE}$
($n=8192$ and $\mathcal{M}=10$) (broken lines). One should note
that the non-standard \textit{diluted} GSE plot (red broken line)
lies above the standard GUE plot (black broken line), indicating lower
correlation or level-repulsion on the average. The standard GSE result
(blue broken line) is also plotted for comparison, and indeed lies
below the standard GUE, confirming a stronger correlation, as expeced.
\label{fig:13_Delta_3_GSE_GUE_consecutive_eigenvalues}}
\end{figure}

\begin{figure}
\begin{centering}
\includegraphics[scale=0.4]{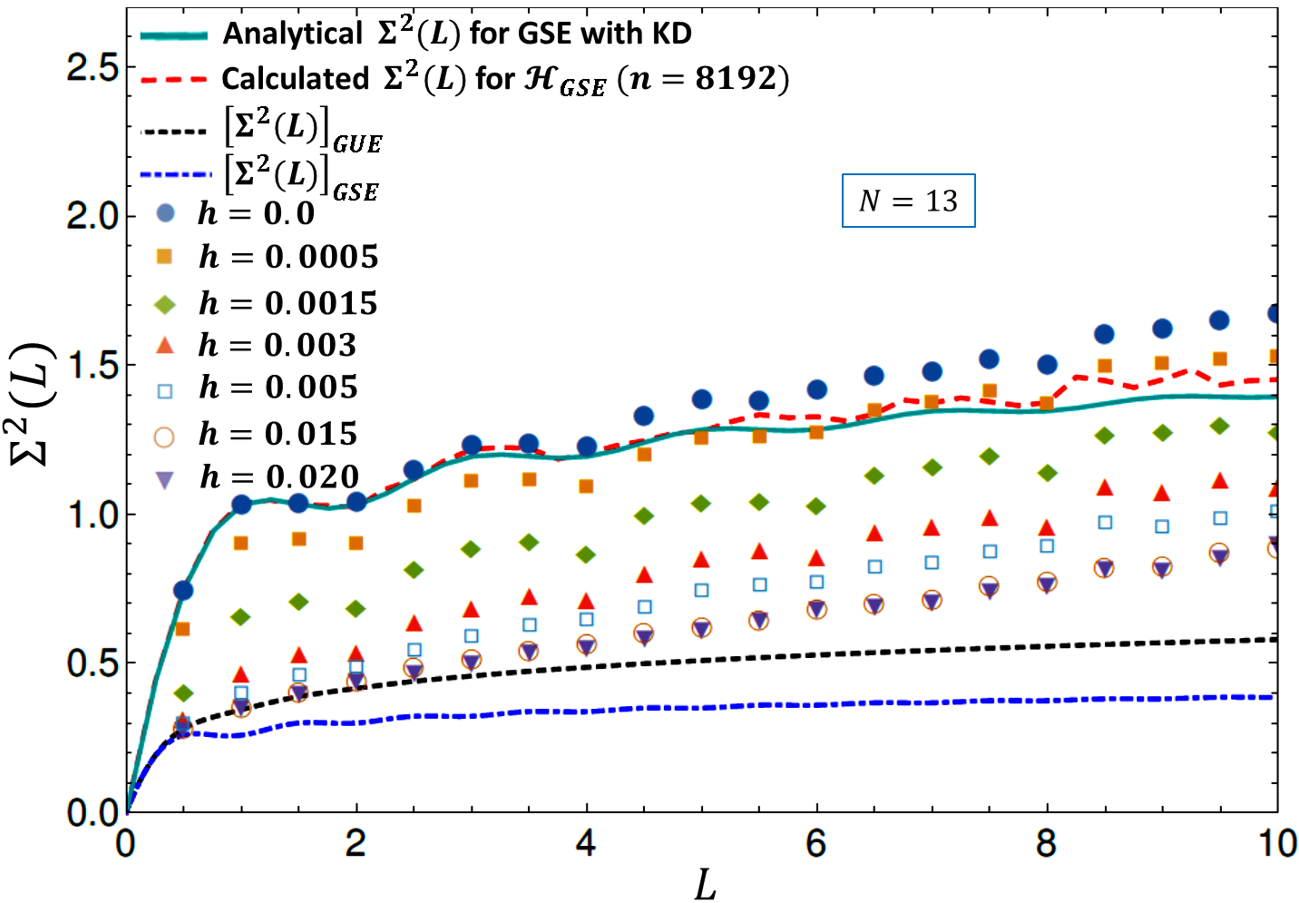}
\par\end{centering}
\caption{Number variance ($\Sigma^{2}$-statistc) of the \textit{consecutive}
eigenvalues for the $N=13$ spin-chain system ($\mathcal{M}=15$),
it shows the GSE-to-GUE crossover with increasing $h$, and fixed
$D=0.2$ and $\epsilon=0.6$ (symbols). To compare their extent of agreement with RMT predictions, the analytical expression for the $\Sigma^{2}$-statistic of the non-standard GSE using the consecutive eigenvalues is plotted (cyan line) for a very small value of the crossover parameter (see Appendix~\ref{sec:Full-Integral-Expressions-for-Delta-and-Sigma}). For comparison, here we plot the $\Sigma^{2}(L)$-statistic
from the RMT matrix model, $\mathcal{H}_{GSE}$ ($n=8192$ and $\mathcal{M}=15$)
(broken lines). One should note that the non-standard \textit{diluted}
GSE plot (red broken line) lies above the standard GUE plot (black
broken line), indicating lower correlation or level-repulsion on the
average. The standard GSE result (blue broken line) is also plotted
for comparison, and indeed lies below the standard GUE, confirming
a stronger correlation, as expeced. \label{fig:14_Sigma_GSE_GUE_consecutive_eigenvalues}}
\end{figure}

In Sec. \ref{subsec:NNSD_results}, we have studied the short-range
GSE-to-GUE (see Table \ref{tab:Crossovers_between_various_ Symmetry_ Classes_and_their_criteria})
crossovers for both the cases, where the KD was removed and retained
in the eigenvalue spectrum. Here, we study the $\Delta_{3}$-statistic
and the number variance of the spin-chain model $H$ for the GSE-to-GUE
crossover, in the Figs. \ref{fig:13_Delta_3_GSE_GUE_consecutive_eigenvalues}
and \ref{fig:14_Sigma_GSE_GUE_consecutive_eigenvalues}, respectively,
by considering the \textit{consecutive} eigenvalues (i.e., retaining
all eigenvalues including the Kramers degenerate ones). The analytical
form of the $\Delta_{3}$-statistic as well as that of $\Sigma^{2}$-statistic can be obtained from
 the two-level cluster function for the GSE-to-GUE crossover~\cite{MehtaPandey1983}; see Appendix~\ref{sec:Full-Integral-Expressions-for-Delta-and-Sigma}. The GSE limit of this crossover is of special interest due to the consideration of Kramers degenerate eigenvalues. For this limit, we also show the results using the RMT matrix model, $\mathcal{H}_{GSE}$,
having a dimension $n=8192$ (similar to our physical $N=13$ spin-chain
model), and calculate the long-range statistics of this model for
comparing with that of the physical spin model. In Fig. \ref{fig:13_Delta_3_GSE_GUE_consecutive_eigenvalues},
we observe that $\Delta_{3}(L)$ plot based on the analytical result for a very small value of the crossover parameter (see Appendix~\ref{sec:Full-Integral-Expressions-for-Delta-and-Sigma}) agrees very well with the numerical $\Delta_{3}(L)$ plot for $\mathcal{H}_{GSE}$, in the GSE limit. Furthermore, for $h=0.0$, the $\Delta_{3}$-statistic
of $H$ follows the RMT $\Delta_{3}(L)$ till a large $L\thicksim20$. We notice that this $\Delta_{3}$-statistic
of the \textit{non-standard} GSE, obtained by retaining all Kramers
degeneracies, is always higher in value than the standard GUE class
statistics ($\left[\Delta_{3}(L)\right]_{GUE}$), implying that it
is \textit{less correlated} than the GUE on the average. This is contrary
to the standard GSE result (after removing the KD by hand), $\left[\Delta_{3}(L)\right]_{GSE}$,
plotted for reference, which is \textit{more correlated} than the
standard GUE class (always lower in value). This is a rather interesting
result, and may be understood along the lines that retaining all Kramers doublets
(no level repulsion between Kramers doublets) amounts to reducing
the average correlation or level-repulsion compared to the standard
GSE case. So it may be looked upon as a \textit{diluted} GSE limit.
At the GUE (NNSD) limit, obtained at about $h=0.015$, the calculated
spectral rigidity from the physical model follows the analytical result
$\left[\Delta_{3}(L)\right]_{GUE}$ till about $L\thicksim12$ and
any further increase in $h$ $(=0.02)$ does not display a better
agreement. So, in this study, the limiting case $h=0.015$ can be
designated as a\textit{ global} crossover point. In Fig. \ref{fig:14_Sigma_GSE_GUE_consecutive_eigenvalues},
we observe that, at GSE limit, the calculated $\Sigma^{2}(L)$ of
the spin-chain model, follows the oscillating numerical plot of the RMT analytical as well as the
matrix model $\mathcal{H}_{GSE}$, up to $L\thicksim4$. This again
corresponds to the \textit{diluted} GSE limit, as discussed above,
and is \textit{less correlated} compared to the standard GUE limit.
On the other hand, just as in the $\Delta_{3}$ case, the plot for
the standard GSE (blue broken lines) is \textit{more correlated} and
lies below the standard GUE plot. Beyond $h=0.005$, the oscillations
in the $\Sigma^{2}(L)$-statistic reduce significantly as we increase
$h$. The calculated $\Sigma^{2}(L)$ for the physical system at the
GUE limit ($h=0.015$ and $0.02$), follows the analytical form of
the GUE class, $\left[\Sigma^{2}(L)\right]_{GUE}$, only till $L\thicksim2$.
The standard GSE result (after removing the KD by hand) for the $\Sigma^{2}(L)$-statistic,
$\left[\Sigma^{2}(L)\right]_{GSE}$, is also plotted in Fig. \ref{fig:14_Sigma_GSE_GUE_consecutive_eigenvalues},
and is consistent with its more correlated nature compared to $\left[\Sigma^{2}(L)\right]_{GUE}$.
Both studies ($\Delta_{3}$ and $\Sigma^{2}$ statistics) show that
all the intermediate cases of the GSE-to-GUE crossovers tend to follow
trends that lie between the two limiting cases, resulting in a smooth
and continuous transition between the two limits.

\begin{figure}
\begin{centering}
\includegraphics[scale=0.4]{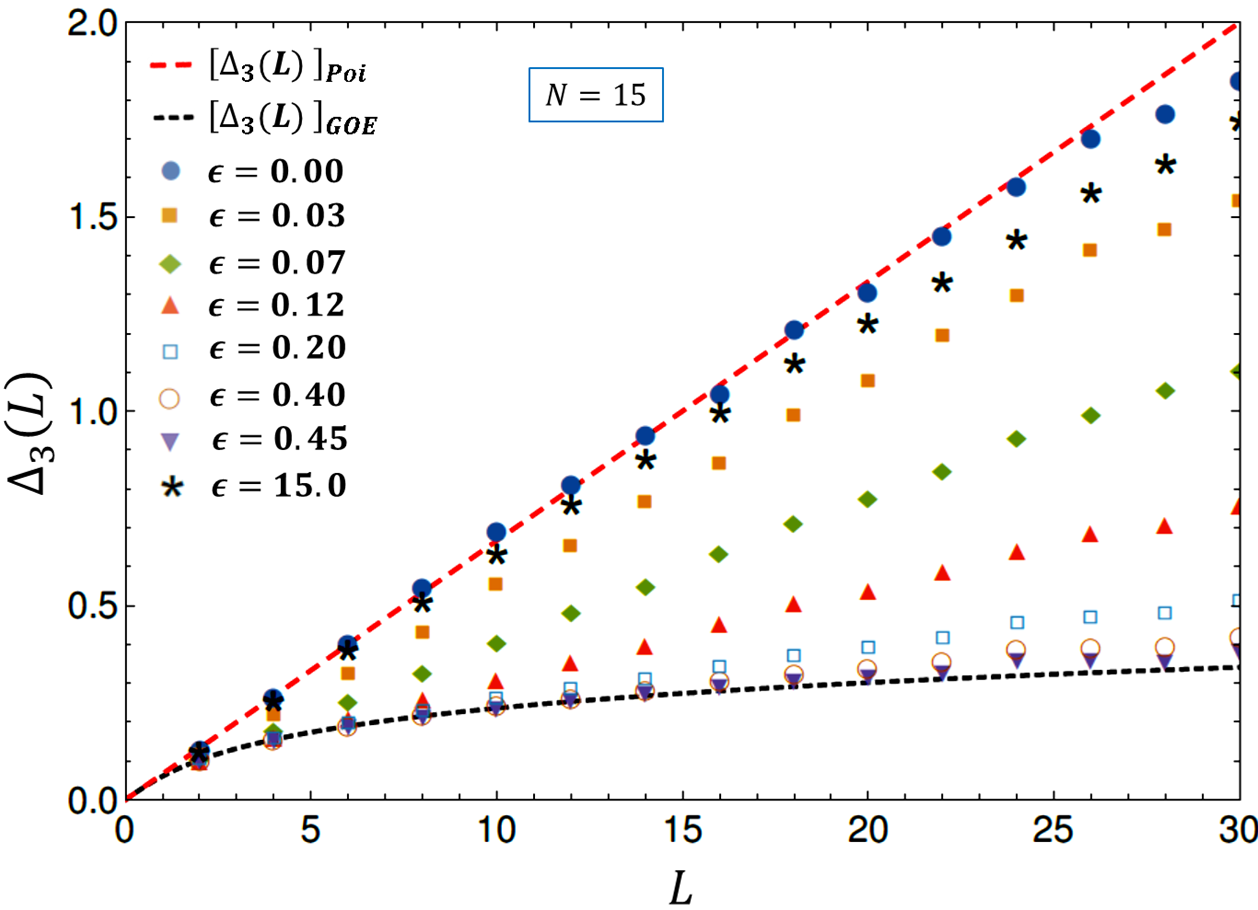}
\par\end{centering}
\caption{Spectral Rigidity ($\Delta_{3}$-statistic) for the $N=15$ spin-chain
($\mathcal{M}=15$) across the Poissonian-to-GOE crossover with increasing
$\epsilon$, and fixed $D=0.2$ (symbols). It also shows the $\Delta_{3}$-statistic
in the many-body localized limit, brought about by the large disorder
($\epsilon=15.0$) (black asteriks), which is again expected to follow
the Poissonian result approximately (red broken line). The GOE limit
exact result is also plotted (black broken line) to show the extent
of agreement with the physical spin system. \label{fig:15_Delta_3_Poi_GOE}}
\end{figure}

\begin{figure}
\begin{centering}
\includegraphics[scale=0.4]{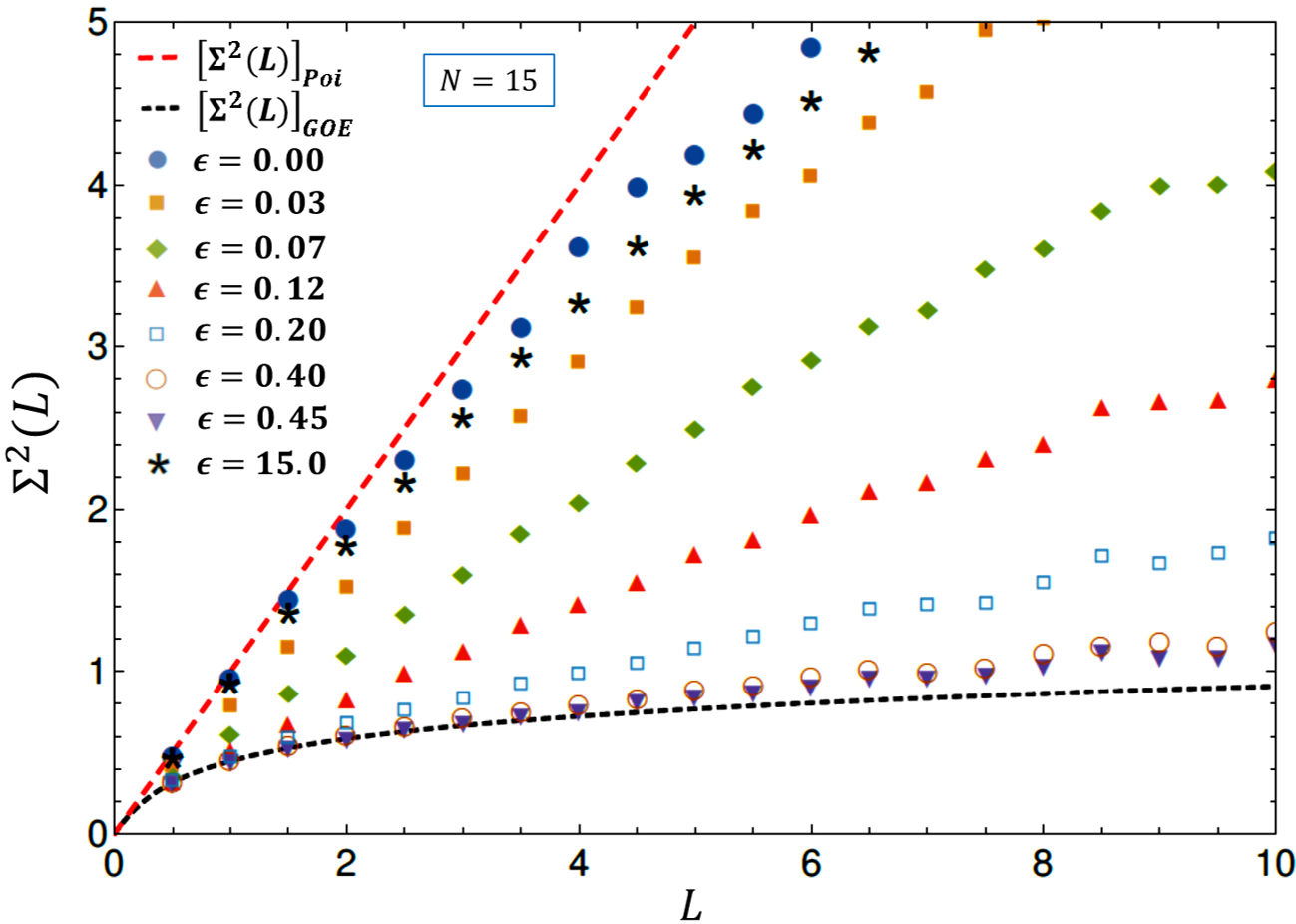}
\par\end{centering}
\caption{Number variance ($\Sigma^{2}$-statistc) for the $N=15$ spin-chain
($\mathcal{M}=15$) across the Poissonian-to-GOE crossover with increasing
$\epsilon$, and fixed $D=0.2$ (symbols). It also shows the $\Sigma^{2}$-statistc
for the many-body localized phase, brought about by the large disorder
($\epsilon=15.0$) (black asteriks), which is again expected to follow
the Poissonian result approximately (red broken line). The GOE limit
exact result is also plotted (black broken line) to show the extent
of agreement with the physical spin system. \label{fig:16_Sigma_Poi_GOE}}
\end{figure}

Next, we investigate the long-range correlation studies of the re-entrant
Poissonian-GOE-Poissonian crossover (see Table \ref{tab:Crossovers_between_various_ Symmetry_ Classes_and_their_criteria})
for the spin-chain model $H_{1}$ with $N=15$ lattice sites. While
the \textit{ordered} limit for $\epsilon=0.0$ is the integrable 1D
Heisenberg model, the \textit{highly disordered} ($\epsilon=15.0$)
limit includes a very large Ising anisotropy that completely suppresses
the quantum fluctuations from the Heisenberg term and results in a
many-body localized state that spans only a few relevant basis states.
This again is expected to follow the Poissonian distribution \cite{Debojyoti_paper_1_2022}.
Due to a low dimension of the $\mathrm{S^{z}=1/2}$ subspace in this
case, where the total $\mathrm{S^{z}}$ is conserved, we discarded
the $N=13$ system in favor of the $N=15$ system, for the long-range
correlation calculations of the Poissonian-to-GOE crossover. The corresponding
$\Delta_{3}(L)$-statistic is plotted in Fig. \ref{fig:15_Delta_3_Poi_GOE}
and the $\Sigma^{2}(L)$-statistic is plotted in Fig. \ref{fig:16_Sigma_Poi_GOE}. Just as we encountered in the Poissonian-to-GSE-to-Poissonian re-entrant crossover (Fig. \ref{fig:1_Poi_GSE_NNSD}), the two Poissonian limits, $\epsilon=0.0$ and $\epsilon=15.0$, coincide
with the ideal Poissonian result $\left[\Delta_{3}(L)\right]_{Poi}$
($\left[\Sigma^{2}(L)\right]_{Poi}$) (broken red line), till up to
$L\sim24$ and $8$ ($L\sim2$ and $1$), respectively. Again both
the \textit{ordered} ($\epsilon=0$) and the \textit{highly disordered}
($\epsilon=15.0$) limits tend to agree with the ideal Poissonian
prediction till a finite $L$ value, but it seems that the \textit{disordered/localized} Poissonian
limit starts to deviate sooner (for a lower $L$) than the \textit{ordered}
Poissonian limit from the ideal Poissonian result, for both the $\Delta_{3}$
and the $\Sigma^{2}$ statistics. The long-range spectral correlations again seem to capture the subtle differences between the two Poissonian regimes, that the short-range correlations fail to capture {[}see Figs. \ref{fig:8_Spin_model_Poi_GOE_NNSD}(d)-\ref{fig:8_Spin_model_Poi_GOE_NNSD}(f){]}.

At GOE limit, $\Delta_{3}(L)$-statistic obeys $\left[\Delta_{3}(L)\right]_{GOE}$
till $L\sim16$ for $\epsilon=0.4$, at which we observed NNSD. If
we increase $\epsilon$ to $0.45$, we do notice improvement in agreement
with the standard GOE result, up to till $L\sim28$. At $\epsilon=0.45$,
the $\Sigma^{2}(L)$-statistic follow the $\left[\Sigma^{2}(L)\right]_{GOE}$,
only up to an eigenvalue interval $L\sim5$, and even an increased
$\epsilon$ does not perform any better. 

So, in some cases, we encountered that relatively strong symmetry-breaking
interactions are required to achieve a \textit{global} RMT crossover
compared to the corresponding \textit{local} crossover. However, in
the majority of cases, the long-range fluctuations of physical systems
deviate from the standard RMT results after a certain $L$ value,
despite perfectly following the short-range NNSD results. We conclude
that our physical system is ``sufficiently random'' for correlations
amongst neighboring eigenvalues, but in many cases fail to fully follow
the RMT results for correlations amongst distant eigenvalues. Large
$L$ deviations in long-range studies suggest a possible breakdown
of universality in the level fluctuations of a physical system, implying
that spin models are not as correlated as random matrices. Furthermore,
for the spin-chain system, while quantifying the same level of long-range correlation, the $\Sigma^{2}$-statistic deviates from
the standard RMT results at a much faster rate than the $\Delta_{3}$-statistic. This could be
due to the fact that the $\Delta_{3}$-statistic is an integral transform of
the number variance (see Eq. (\ref{eq:appendix_C_integral_transform_between_Sigma_and_Delta3}) of Appendix \ref{sec:Full-Integral-Expressions-for-Delta-and-Sigma}), resulting in an agreement with the spin-system results till a much larger value of $L$ \cite{Vyas_long_range_comparison,T.GuhrReview}. 

\section{Conclusions\label{sec:Conclusion}}

In this work, we have studied the spectral correlation properties of an interacting quantum spin-chain system with various competing terms, including a coupling to an external stochastic magnetic field. By tuning the relative amplitudes of these terms, we can alter the unitary and anti-unitary symmetries associated with the Hamiltonian. This, in turn, enables us to realize spectral crossovers amongst various Poissonian and Wigner-Dyson classes (GOE, GUE, GSE) of random matrix theory. To quantify these spectral crossovers, we have employed both short-range and long-range random matrix observables, viz. the nearest neighbor spacing distribution for the former case, and spectral rigidity and number variance for the latter. The key findings from our extensive study are listed below for both short-range and long-range statistics.
\vspace{3mm}

\noindent
\textbf{\textit{Short-range Statistics:}}

\begin{itemize}

\item	The short-range statistic, NNSD, in all the crossovers, viz., Poissonian-to-GSE, Poissonian-to-GUE, GSE-to-GUE, and Poissonian-to-GOE, exhibit extremely good agreement with the RMT predictions in the two extremes of the crossovers and hence establishes the universality of local spectral fluctuations for our spin-chain system. Moreover, in the intermediate regime, it nicely captures the gradual change in the level-repulsion behavior. 

\item	A crucial aspect of our exploration pertaining to the GSE-to-GUE crossover is retention of the Kramers doubly degenerate eigenvalues in the GSE limit, which is typically not done while studying the spectral correlation properties of the GSE class. In this limit, the NNSD exhibits a Dirac-delta peak at the origin along with a broad Wigner-Dyson like hump at a finite spacing value of $s$. For this case, we have also presented an analytical expression for the NNSD of the renormalized GSE class, which matches well with the observations from the RMT matrix model and the spin-chain system.

\item	Within the GSE-to-GUE crossover, as the Kramers degeneracy is lifted via the breaking of an appropriate anti-unitary symmetry, an interesting double peak structure is observed in the NNSD, which eventually turns into a single peak Wigner-Dyson curve of the standard  GUE NNSD. The evolution of the eigenspectra undergoing the GSE-to-GUE transition is presented through the study of the Marginal Spectral Densities (MSD) and the Densities of States (DOS) for both the spin model as well as the relevant RMT crossover matrix model.

\item The re-entrant Poissonian-to-GOE-to-Poissonian (non-random$\to$chaotic$\to$localized) crossover is observed by gradually increasing the disorder in the system via an increase in the random Ising coupling, $\epsilon$. The appearance of the Poissonian statistics in the two extreme limits around the Wigner-Dyson distribution, one delocalized and the other localized in the many-body Slater basis, is successfully demonstrated in this study.

\end{itemize}

\noindent
\textbf{\textit{Long-range Statistics:}}

\begin{itemize}

\item  The long-range correlation statistics match with RMT predictions in the extreme limits up to a certain spectral length interval, $L$ and deviate for larger $L$ values, which indicates that the spectral correlations for the physical spin system are indeed different from RMT long-range correlations as one examines intervals of larger lengths. This may, in turn, be related to the relatively short spatial range of the physical interactions in our spin models.

\item In the GSE limit of the GSE-to-GUE crossover, the consequence of keeping all the eigenvalues (including the Kramers doublets) is that a less correlated behavior, viz. higher values of $\Delta_3(L)$ and $\Sigma^2(L)$ compared to the standard GSE, and also the standard GUE, is observed for our spin system. This seems to nicely agree with similar trends shown by the analytical RMT results for the standard GUE and GSE extremes as well as the GSE limit with Kramers degeneracy (referred to as the {\em diluted} GSE limit) as derived in Appendix \ref{sec:Full-Integral-Expressions-for-Delta-and-Sigma}, in that the {\em diluted} GSE plots lie not only above the stanard GSE, but also above the plots for the standard GUE, for both $\Delta_3$ and $\Sigma^2$, indicating lower correlation.

\item Another key finding of our study is that the integrable Poissonian limit, which is realized in the two extreme regimes, one in the absence and the other in the presence of many-body localization, while not being distinguishable at the level of the NNSD, does show a discernable behavior with respect to the long-range correlations. The latter, therefore, can serve as indicators to distinguish between these two regimes.

\end{itemize}

From the above, it is evident that by tuning the various competing terms in the spin-chain Hamiltonian, which control various unitary and anti-unitary symmetries, a rich variety of crossovers is observed in the short-range as well as in the long-range statistics of the eigenvalues. One particularly interesting aspect of our study has been examining the consequences of retaining the Kramers degenerate eigenvalues in the GSE limit. It would be of interest to explore this facet in other many-body physical systems as well. One would also like to go beyond the eigenvalues and quantify the behaviour of eigenvectors in such symmetry crossovers in physical systems using statistics such as inverse participation ratio, generalized information entropies and multifractal dimensions. Some of these investigations are already under way.

\begin{acknowledgments} D.K. and S.S.G. acknowledge the Science and Engineering Research Board (SERB),
Department of Science and Technology (DST), Government of India, for financial support via Project No. ECR/2016/002054. S.K. acknowledges financial support provided by SERB, DST, Government of India, via Project No. CRG/2022/001751. We would like to dedicate this article to the memory of Prof. Akhilesh Pandey, a doyen in the field of Random Matrix Theory. \end{acknowledgments}

\appendix

\section{Simplification of the DM Term}\label{sec:simplification-of-DM-term}

This appendix presents a simplification of the DM interaction term
($H_{DM}$) mentioned in Eq. (1). We have, 
\begin{equation}
H_{DM}=\sum_{j=1}^{N-1}\boldsymbol{\boldsymbol{D}}\cdot[\mathbf{S}_{j}\times\mathbf{S}_{j+1}].\label{eq:DM_appendix_A_1}
\end{equation}

\begin{flushleft}
After expanding the term $\boldsymbol{\boldsymbol{D}}\cdot[\mathbf{S}_{j}\times\mathbf{S}_{j+1}]$
in terms of the spin components ($\mathrm{S^{x}},\mathrm{S^{y}},\mathrm{S^{z}}$)
we get, 
\begin{align}
\nonumber
\boldsymbol{\boldsymbol{D}}\cdot[\mathbf{S}_{j}\times\mathbf{S}_{j+1}] & =D_{x}\left(\mathrm{S}_{\mathit{j}}^{\mathit{\mathrm{y}}}\mathrm{S}_{j+1}^{\mathit{\mathrm{z}}}-\mathrm{S}_{\mathit{j}}^{\mathit{\mathrm{z}}}\mathrm{S}_{\mathit{j}+1}^{\mathit{\mathrm{y}}}\right)\\
 & 
 \nonumber
 ~~ +D_{\mathrm{y}}\left(-\mathrm{S}_{\mathit{j}}^{\mathit{\mathrm{x}}}\mathrm{S}_{\mathit{j}+1}^{\mathit{\mathrm{z}}}+\mathrm{S}_{j}^{\mathit{\mathrm{z}}}\mathrm{S}_{j+1}^{\mathit{\mathrm{x}}}\right)\\
 &~~ +D_{\mathrm{z}}\left(\mathrm{S}_{j}^{\mathit{\mathrm{x}}}\mathrm{S}_{j+1}^{\mathit{\mathrm{y}}}-\mathrm{S}_{\mathit{j}}^{\mathit{\mathrm{y}}}\mathrm{S}_{j+1}^{\mathit{\mathrm{x}}}\right).
\label{eq:DM_appendix_A_2}
\end{align}

Now, in terms of the site spin $\mathit{raising}$ ($\mathrm{S}_{j}^{+}$)
and $\mathit{lowering}$ ($\mathrm{\mathrm{S}_{\mathit{j}}^{-}}$)
(or spin $\mathit{ladder}$) operators, 
\begin{equation}
\mathrm{S}_{j}^{\mathit{\mathrm{x}}}=\frac{1}{2}\left(\mathrm{S_{\mathit{j}}^{+}}+\mathrm{S_{\mathit{j}}^{-}}\right),\:\mathrm{S}_{\mathit{j}}^{\mathit{\mathrm{y}}}=\frac{1}{2i}\left(\mathrm{S_{\mathit{j}}^{+}}-\mathrm{S_{\mathit{j}}^{-}}\right),\label{eq:DM_appendix_A_3}
\end{equation}
we have after simplification: 
\par\end{flushleft}
\begin{align}
\nonumber
&\boldsymbol{\boldsymbol{D}}\cdot[\mathbf{S}_{j}\times\mathbf{S}_{j+1}]\\
\nonumber
&~ =\frac{iD_{\mathrm{x}}}{2}\left(-\mathrm{S_{\mathit{j}}^{+}}\mathrm{S}_{j+1}^{\mathrm{z}}+\mathrm{S_{\mathit{j}}^{-}}\mathrm{S}_{j+1}^{\mathrm{z}}+\mathrm{S_{\mathit{j}}^{z}}\mathrm{S}_{j+1}^{+}-\mathrm{S_{\mathit{j}}^{\mathit{\mathrm{z}}}}\mathrm{S}_{j+1}^{-}\right)\\
\nonumber
&~ +\frac{D_{\mathrm{y}}}{2}\left(-\mathrm{S_{\mathit{j}}^{+}}\mathrm{S}_{j+1}^{\mathrm{z}}-\mathrm{S_{\mathit{j}}^{-}}\mathrm{S}_{j+1}^{\mathrm{z}}+\mathrm{S_{\mathit{j}}^{\mathit{\mathrm{z}}}}\mathrm{S}_{j+1}^{+}+\mathrm{S_{\mathit{j}}^{\mathit{\mathrm{z}}}}\mathrm{S}_{j+1}^{-}\right)\\
&~ +\frac{iD_{\mathrm{z}}}{2}\left(\mathrm{S_{\mathit{j}}^{+}}\mathrm{S}_{j+1}^{-}-\mathrm{S_{\mathit{j}}^{-}}\mathrm{S}_{j+1}^{+}\right).
\label{eq:DM_appendix_A_4}
\end{align}
The terms which involve a single spin ladder operator, like $\mathrm{S_{\mathit{j}}^{+}}\mathrm{S}_{j+1}^{\mathrm{z}}$
($\mathrm{S_{\mathit{j}}^{-}}\mathrm{S}_{j+1}^{\mathrm{z}}$), increase
(decrease) the total $\mathrm{S^{z}}$ of the system by $1$. On the
other hand, the two-site terms like $\mathrm{S_{\mathit{j}}^{+}}\mathrm{S}_{j+1}^{\mathrm{-}}$
and $\mathrm{S_{\mathit{j}}^{-}}\mathrm{S}_{j+1}^{\mathrm{+}}$, conserve
total $\mathrm{S^{z}}$. The terms with two similar spin ladder operators
(like $\mathrm{S}_{j}^{+}\mathrm{S}_{j+1}^{+}$ and $\mathrm{S}_{j}^{-}\mathrm{S}_{j+1}^{-}$),
cancel out during the simplification. From Eq. (\ref{eq:DM_appendix_A_4}),
we observe that the spin operators associated with the $D_{\mathrm{x}}$
and $D_{\mathrm{y}}$ terms, connect nearest-neighbor $\mathrm{S^{z}}$
sectors, but the spin operators associated with the $D_{\mathrm{z}}$
term do not connect different $\mathrm{S^{z}}$ sectors. Thus, for
non-zero $D_{\mathrm{x}}$ and/or $D_{\mathrm{y}}$, we cannot carry
out calculations within a fixed $\mathrm{S^{z}}$ sector.

\section{Conservation of the Spin Component along the Direction of the Vector Coupling under the DM Interaction Term}\label{sec:Commutation_Between_H_DM_and_S_dot_D_cap}

In this appendix, we show that $H_{DM}$ commutes with $\ensuremath{{\bf S}\cdot\hat{D}}$,
resulting in $H_{DM}$ having cylindrical symmetry only about the
direction of $\hat{D}$. We consider two spins, $\mathbf{S}_{1}$
and $\mathbf{S}_{2}$, for which $H_{DM}=\boldsymbol{\boldsymbol{D}}\cdot[\mathbf{S}_{1}\times\mathbf{S}_{2}]$
and $\mathbf{S}=\mathbf{S}_{1}+\mathbf{S}_{2}$. Using a tensorial notation along with the summation convention, we write the vector $\boldsymbol{\boldsymbol{D}}=D_{\alpha}\hat{e}_{\alpha}$,
where $\hat{e}_{\alpha}$ is the unit vector along the three cartesian axes ($\alpha=1,2,3$ corresponds
to the $\mathrm{x,y,z}$ axes, respectively). The unit vector along
$\boldsymbol{\boldsymbol{D}}$ is $\hat{D}=d_{\alpha}\hat{e}_{\alpha}$,
where $d_{\alpha}=D_{\alpha}/D$ ($d_{\alpha}d_{\alpha}=1$).

Now, we have
\begin{equation}
\left[H_{DM},\ensuremath{{\bf S}\cdot\hat{D}}\right]=H_{DM}(\ensuremath{{\bf S}\cdot\hat{D}})-(\ensuremath{{\bf S}\cdot\hat{D}})H_{DM}.\label{eq:appendix_commutation_eq_1}
\end{equation}
 The first term in the RHS can be written as,
\begin{align}
\nonumber
&H_{DM}(\ensuremath{{\bf S}\cdot\hat{D}})\\
\nonumber
&=(\varepsilon_{\alpha\beta\gamma}D_{\alpha}\mathrm{S}_{1\beta}\mathrm{S}_{2\gamma})(\mathrm{S}_{1\mu}+\mathrm{S}_{2\mu})d_{\mu}\\
\nonumber
&=d_{\mu}(\varepsilon_{\alpha\beta\gamma}D_{\alpha}\mathrm{S}_{1\beta}\mathrm{S}_{2\gamma}\mathrm{S}_{1\mu}+\varepsilon_{\alpha\beta\gamma}D_{\alpha}\mathrm{S}_{1\beta}\mathrm{S}_{2\gamma}\mathrm{S}_{2\mu})\\
\nonumber
&=d_{\mu}(\varepsilon_{\alpha\beta\gamma}D_{\alpha}\mathrm{S}_{1\beta}\mathrm{S}_{1\mu}\mathrm{S}_{2\gamma}+i\varepsilon_{\alpha\beta\gamma}\varepsilon_{\gamma\mu\eta}D_{\alpha}\mathrm{S}_{1\beta}\mathrm{S}_{2\eta}\\
&+\varepsilon_{\alpha\beta\gamma}D_{\alpha}\mathrm{S}_{1\beta}\mathrm{S}_{2\mu}\mathrm{S}_{2\gamma}),\label{eq:appendix_commutation_eq_2}
\end{align}
where in the last step we have used the fact that, spin operators
from different sites commute and the commutation relation for the same-site spin operators is $\left[\mathrm{S}_{2\gamma},\mathrm{S}_{2\mu}\right]=i\varepsilon_{\gamma\mu\eta}\mathrm{S}_{2\eta}$, which may be combined into: $\left[\mathrm{S}_{p\gamma},\mathrm{S}_{q\mu}\right]=i\delta_{pq}\varepsilon_{\gamma\mu\eta}\mathrm{S}_{p\eta}$.
Similarly, we get the further simplified form as, 
\begin{align}
\nonumber
&H_{DM}(\ensuremath{{\bf S}\cdot\hat{D}})\\
\nonumber
&=d_{\mu}(i\varepsilon_{\alpha\beta\gamma}\varepsilon_{\beta\mu\eta}D_{\alpha}\mathrm{S}_{1\eta}\mathrm{S}_{2\gamma}+\mathrm{S}_{1\mu}\varepsilon_{\alpha\beta\gamma}D_{\alpha}\mathrm{S}_{1\beta}\mathrm{S}_{2\gamma}\\
&+i\varepsilon_{\alpha\beta\gamma}\varepsilon_{\gamma\mu\eta}D_{\alpha}\mathrm{S}_{1\beta}\mathrm{S}_{2\eta}+\mathrm{S}_{2\mu}\varepsilon_{\alpha\beta\gamma}D_{\alpha}\mathrm{S}_{1\beta}\mathrm{S}_{2\gamma}).\label{eq:appendix_commutation_eq_3}
\end{align}

Now, we use the identity: $\varepsilon_{ijk}\varepsilon_{imn}=\delta_{jm}\delta_{kn}-\delta_{jn}\delta_{km}$,
and the anti-symmetry of the Levi-Civita tensor, whence the {\em first} term of Eq. (\ref{eq:appendix_commutation_eq_3}) becomes,
\begin{align}
\nonumber
&i\varepsilon_{\alpha\beta\gamma}\varepsilon_{\beta\mu\eta}D_{\alpha}\mathrm{S}_{1\eta}\mathrm{S}_{2\gamma}d_{\mu}\\
\nonumber
&=-i\varepsilon_{\beta\alpha\gamma}\varepsilon_{\beta\mu\eta}D_{\alpha}\mathrm{S}_{1\eta}\mathrm{S}_{2\gamma}d_{\mu}\\
\nonumber
&=-i(D_{\mu}d_{\mu})(\mathrm{S}_{1\gamma}\mathrm{S}_{2\gamma})+i(D_{\alpha}\mathrm{S}_{1\alpha})(\mathrm{S}_{2\mu}d_{\mu})\\
\nonumber
&=-i(\hat{D}\cdot\boldsymbol{\boldsymbol{D}})(\mathbf{S}_{1}\cdot\mathbf{S}_{2})+i(\mathbf{S}_{1}\cdot\boldsymbol{\boldsymbol{D}})(\mathbf{S}_{2}\cdot\hat{D})\\
&=-iD\left[\mathbf{S}_{1}\cdot\mathbf{S}_{2}-(\mathbf{S}_{1}\cdot\hat{D})(\mathbf{S}_{2}\cdot\hat{D})\right].\label{eq:appendix_commutation_eq_4}
\end{align}
Similarly, the {\em third} term of Eq. (\ref{eq:appendix_commutation_eq_3})
becomes, 
\begin{equation}
i\varepsilon_{\alpha\beta\gamma}\varepsilon_{\gamma\mu\eta}D_{\alpha}\mathrm{S}_{1\beta}\mathrm{S}_{2\eta}d_{\mu}=iD\left[\mathbf{S}_{1}\cdot\mathbf{S}_{2}-(\mathbf{S}_{1}\cdot\hat{D})(\mathbf{S}_{2}\cdot\hat{D})\right].\label{eq:appendix_commutation_eq_5}
\end{equation}
 From the Eqs. (\ref{eq:appendix_commutation_eq_4}) and (\ref{eq:appendix_commutation_eq_5}),
we see that the {\em first} and the {\em third} terms of Eq. (\ref{eq:appendix_commutation_eq_3})
cancel out each other. From the {\em second} and the {\em fourth} terms of Eq.
(\ref{eq:appendix_commutation_eq_3}) we get,
\begin{align}
\nonumber
H_{DM}(\ensuremath{{\bf S}\cdot\hat{D}})&=(\mathrm{S}_{1\mu}d_{\mu})(\varepsilon_{\alpha\beta\gamma}D_{\alpha}\mathrm{S}_{1\beta}\mathrm{S}_{2\gamma})\\
\nonumber
&+(\mathrm{S}_{2\mu}d_{\mu})(\varepsilon_{\alpha\beta\gamma}D_{\alpha}\mathrm{S}_{1\beta}\mathrm{S}_{2\gamma})\\
\nonumber
&=(\mathrm{S}_{1\mu}+\mathrm{S}_{2\mu})d_{\mu}(\varepsilon_{\alpha\beta\gamma}D_{\alpha}\mathrm{S}_{1\beta}\mathrm{S}_{2\gamma})\\
&=(\ensuremath{{\bf S}\cdot\hat{D}})H_{DM}.\label{eq:appendix_commutation_eq_6}
\end{align}
So we get, $\left[H_{DM},\ensuremath{{\bf S}\cdot\hat{D}}\right]=0$, that establishes the said conservation law and resultant cylindrical symmetry about the $\boldsymbol{D}$-axis.

\section{Calculation of the NNSD of the GSE Class with KD}\label{sec:NNSD-of-the-GSE-Class-with-KD}

Consider the ordered spectrum, with Kramers degeneracy, having $n$
levels, $\varepsilon_{1}=\varepsilon_{2}<\varepsilon_{3}=\varepsilon_{4}<\cdots<\varepsilon_{n-1}=\varepsilon_{n}$,
where $n$ is even. In finding the nearest neighbor spacings, we have
($n-1$) level-spacings in all. The degenerate pairs will lead to
$\ensuremath{\frac{n}{2}}$ zero level-spacings, while the nondegenerate
ones will lead to $\ensuremath{(\frac{n}{2}-1)}$ nonzero (positive)
level-spacings. Denoting the unnormalized level-spacings by the variable
$x$, the zero-spacings give rise to a Dirac delta {[}$\delta(x)${]}
peak at $x=0$, and the distribution of non-zero part is expected
to follow the standard NNSD of GSE, $P_{GSE}(x)$. Including both
these contributions, the NNSD in the case where the Kramers degeneracy
is retained, should be given by a function of the form:
\begin{align}
f^{n}(x)  =f_{GSE}^{n}(x)+f_{0}^{n}(x)
  =\mu P_{GSE}(x)+\nu\delta(x),
\label{eq:NNSD_GSE_Without_removing_KD-1}
\end{align}
 where $\mu$ and $\nu$ are determined by making use of the
fractional contributions from the two types of spacings (\textit{zero}
and \textit{finite}) to the overall normalized distribution $f^{n}(x)$,
also using the individual normalization properties of $P_{GSE}(x)$
and $\delta(x)$, as discussed below. The normalization condition
demands that the integrated weight of $f^{n}(x)$ should be 1. So,
the fractional weight of the non-zero spacings becomes, 
\begin{equation}
\intop_{0}^{\infty}f_{GSE}^{n}(x)\mathrm{d}x=\frac{(\ensuremath{\frac{n}{2}}-1)}{(\ensuremath{n}-1)}=\frac{1}{2}\left(\frac{n-2}{n-1}\right),\label{eq:NNSD_GSE_Without_removing_KD-2}
\end{equation}
 and of the zero spacings becomes, 
\begin{equation}
\intop_{0}^{\infty}f_{0}^{n}(x)\mathrm{d}x=\frac{\ensuremath{\frac{n}{2}}}{(n-1)}=\frac{1}{2}\left(\frac{n}{n-1}\right).\label{eq:NNSD_GSE_Without_removing_KD-3}
\end{equation}
 From Eq. (\ref{eq:NNSD_GSE_Without_removing_KD-1}) and Eq. (\ref{eq:NNSD_GSE_Without_removing_KD-2})
we get,
\begin{equation}
\mu\intop_{0}^{\infty}P_{GSE}(x)\mathrm{d}x=\frac{1}{2}\left(\frac{n-2}{n-1}\right),\label{eq:NNSD_GSE_Without_removing_KD-4}
\end{equation}
and using the normalization condition $\intop_{0}^{\infty}P_{GSE}(x)\mathrm{d}x=1$,
we find, $\mu=\frac{1}{2}\left(\frac{n-2}{n-1}\right).$ Again,
from Eq. (\ref{eq:NNSD_GSE_Without_removing_KD-1}) and Eq. (\ref{eq:NNSD_GSE_Without_removing_KD-3})
we have, 
\begin{equation}
\nu\intop_{0}^{\infty}\delta(x)\mathrm{d}x=\frac{1}{2}\left(\frac{n}{n-1}\right).\label{eq:NNSD_GSE_Without_removing_KD-5}
\end{equation}
 Using the definition of the Dirac delta function, $\intop_{0}^{\infty}\delta(x)\mathrm{d}x:=\frac{1}{2}$,
we have $\nu=\frac{n}{n-1},$ and $f^{n}(x)$ now becomes, 
\begin{equation}
f^{n}(x)=\frac{1}{2}\left(\frac{n-2}{n-1}\right)P_{GSE}(x)+\left(\frac{n}{n-1}\right)\delta(x).\label{eq:NNSD_GSE_Without_removing_KD-6}
\end{equation}

We need to now calculate the average spacings for the distribution
$f^{n}(x)$. We consider, 
\begin{align}
\nonumber
\mathcal{D} & =\intop_{0}^{\infty}f^{n}(x)\mathrm{d}x\\
\nonumber
& =\frac{1}{2}\left(\frac{n-2}{n-1}\right)\int_{0}^{\infty}xP_{GSE}(x)+\left(\frac{n}{n-1}\right)\int_{0}^{\infty}x\delta(x)\\
& =\frac{1}{2}\left(\frac{n-2}{n-1}\right),
\label{eq:NNSD_GSE_Without_removing_KD-7}
\end{align}
 here we have used the unfolding condition $\int_{0}^{\infty}xP_{GSE}(x)=1$
and the relation for the Dirac delta function, $\int_{0}^{\infty}x\delta(x):=0$.
In order to again make the average spacing equal to $1$, we define
the normalized variable, $s=x/\mathcal{D}$, and rewrite the distribution
in terms of this new variable. Considering the Jacobian of transformation
$\left|\mathrm{d}x/\mathrm{d}s\right|=\frac{1}{2}\left(\frac{n-2}{n-1}\right)$,
we get the rescaled distribution as,
\begin{align}
\nonumber
\mathcal{P}_{GSE}^{n}(s) & =\frac{1}{2}\left[\frac{n-2}{n-1}\right]\left\{ f^{n}\left(\frac{1}{2}\left[\frac{n-2}{n-1}\right]s\right)\right\} \\
\nonumber
& =\frac{1}{2}\left[\frac{n-2}{n-1}\right]\left\{ \frac{1}{2}\left[\frac{n-2}{n-1}\right]P_{GSE}\left(\frac{1}{2}\left[\frac{n-2}{n-1}\right]s\right)\right\} \\
& +\frac{1}{2}\left[\frac{n-2}{n-1}\right]\left\{ \left[\frac{n}{n-1}\right]\delta\left(\frac{1}{2}\left[\frac{n-2}{n-1}\right]s\right)\right\} 
\label{eq:NNSD_GSE_Without_removing_KD-8}
\end{align}
 Using the standard analytical expression for $P_{GSE}(s)$, based
on the Wigner surmise, and the scaling property of the Dirac Delta
function, $\delta(ks)=\delta(s)/\left|k\right|$, we simplify the
Eq. (\ref{eq:NNSD_GSE_Without_removing_KD-8}) and get the modified
analytical form of the NNSD, for the GSE class {[}$\mathcal{P}_{GSE}^{n}(s)${]},
where the Kramers degeneracy is not removed from the eigenspectrum.
For dimension $n$, it is presented as,

\begin{align}
\nonumber
\mathcal{P}_{GSE}^{n}(s) & =\ensuremath{\mathop{\left[\frac{2^{12}}{3^{6}\pi^{3}}\left(\frac{n-2}{n-1}\right)^{6}s^{4}\right]\exp\left[-\frac{16}{9\pi}\left(\frac{n-2}{n-1}\right)^{2}s^{2}\right]}}\\
& +\left(\frac{n}{n-1}\right)\delta(s).
\label{eq:NNSD_GSE_Without_removing_KD-9}
\end{align}
 For a large $n$ (like for our $n=8192$ calculation, but not the
$n=4$ calculation), Eq. (\ref{eq:NNSD_GSE_Without_removing_KD-9})
assumes the asymptotic form 
\begin{equation}
\mathcal{P}_{GSE}(s)=\ensuremath{\mathop{\left(\frac{2^{12}}{3^{6}\pi^{3}}s^{4}\right)\exp\left(-\frac{16}{9\pi}s^{2}\right)}+\delta(s)}.\label{eq:NNSD_GSE_Without_removing_KD-10}
\end{equation}

\section{Full Integral Expressions for $\Delta_{3}(L)$ and $\Sigma^{2}(L)$}\label{sec:Full-Integral-Expressions-for-Delta-and-Sigma}

In this appendix, we present the full integral expressions of the
$\Delta_{3}(L)$ and $\Sigma^{2}(L)$ statistics in terms of the two-level
$\mathit{cluster}$ $\mathit{function}s$ \cite{Mehta-Book}. The
correlation characteristics of a single cluster of $n$-levels are
described by the $\mathit{cluster}$ $\mathit{function}$, which is
separate from the lower order correlations \cite{Mehta-Book}.
It vanishes when any one (or more) of the level separations ($\left|\tilde{\varepsilon}_{i}-\tilde{\varepsilon}_{j}\right|$)
increases relative to the local mean level spacing (which is unity
for our case). The two-level $\mathit{cluster}$ $\mathit{function}s$
for various cases are listed below \cite{Mehta-Book}, 
\begin{align}
&Y_{Poi}(r)=0,\\
&Y_{GOE}(r)=\left(\frac{\sin(\pi r)}{\pi r}\right)^{2}\nonumber\\
&~~~~~~~~~~~ +\left(\frac{\cos(\pi r)}{r}-\frac{\sin(\pi r)}{\pi r^{2}}\right)\left(\frac{1}{2}-\frac{\mathrm{Si}(\pi r)}{\pi}\right),\\
&Y_{GUE}(r)=\left(\frac{\sin(\pi r)}{\pi r}\right)^{2},\\
&Y_{GSE}(r)=\left(\frac{\sin(2\pi r)}{2\pi r}\right)^{2}\nonumber\\
&~~~~~~~~~~~ -\left(\frac{\cos(2\pi r)}{r}-\frac{\sin(2\pi r)}{2\pi r^{2}}\right)\left(\frac{\mathrm{Si}(2\pi r)}{2\pi}\right),\label{eq:appendix_C_1}
\end{align}
 where $r=\left|\tilde{\varepsilon}_{1}-\tilde{\varepsilon}_{2}\right|$
and $\mathrm{Si}(z)=\int_{0}^{z}\sin(t)/t\,dt$ is the standard sine
integral. 

The full integral expressions of the averaged $\Delta_{3}$-statistic
and $\Sigma^{2}$-statistic are given by \cite{Mehta-Book}, 
\begin{equation}
\Delta_{3}(L)=\frac{L}{15}-\frac{1}{15L^{4}}\int_{0}^{L}\left(L-r\right)^{3}\left(2L^{2}-9Lr-3r^{2}\right)Y(r)dr,\label{eq:appendix_C_2}
\end{equation}
 and 
\begin{equation}
\Sigma^{2}(L)=L-2\int_{0}^{L}(L-r)Y(r)dr.\label{eq:appendix_C_3}
\end{equation}
It is also known that $\Delta_{3}(L)$ is an integral transform of $\Sigma^2(L)$ \cite{T.GuhrReview,Vyas_long_range_comparison,Brody_Random_Matrix_Physics_review},
\begin{equation}
\Delta _{3}(L)=\frac{2}{L^{4}}\int_{0}^{L}(L^{3}-2L^{2}r+r^{3})\Sigma ^{2}(r)dr.
\label{eq:appendix_C_integral_transform_between_Sigma_and_Delta3}
\end{equation}

 The above integrals for $\Delta_{3}(L)$ and $\Sigma^{2}(L)$ for
the Wigner-Dyson ensembles can be performed in terms of the sine and
cosine integrals, but are lengthy. In Sec. \ref{subsec:long_range_results},
we used numerical evaluations of the aforementioned integral formulas
for the specific ranges of $L$, and presented the results in the
relevant Figs. \ref{fig:9_Delta_3_Poi_GSE}-\ref{fig:16_Sigma_Poi_GOE}.

For the GSE-to-GUE crossover, the exact as well as asymptotic expressions for correlation functions and cluster functions are known~\cite{PandeyMehta1983}. The evaluation of spectral rigidity and number variance requires the two-level cluster function for the unfolded eigenvalues, which is given as \cite{PandeyMehta1983,KumarPandey2011a},
\begin{equation}
Y_{GSE-GUE}(\lambda,r)=\left(\frac{\sin(\pi r)}{\pi r}\right)^{2}-I(\lambda,r)K(\lambda,r),\label{eq:appendix_C_4}
\end{equation}
 where 
 \begin{align}
 I(\lambda,r)&=-\frac{1}{\pi}\int_{0}^{\pi}\frac{\sin(kr)}{k}e^{2\lambda^{2}k^{2}}dk,\\
K(\lambda,r)&=-\frac{1}{\pi}\int_{\pi}^{\infty}k\sin(kr)e^{-2\lambda^{2}k^{2}}dk.
\end{align}
The parameter $\lambda\sim\sqrt{n}\,\alpha$ in the above expressions is the rescaled-crossover
parameter. We use the above $\lambda$-dependent
cluster function expression in Eqs. (\ref{eq:appendix_C_2}) and (\ref{eq:appendix_C_3})
to obtain the spectral rigidity and number variance. The GSE limit
is obtained for $\lambda\to0$, whereas the GUE limit is achieved
for $\lambda\to\infty$, for which the product $I(\lambda,r)K(\lambda,r)$
goes to zero. It should be noted that the crossover is almost complete
for $\lambda\sim1$, i.e., $\alpha\sim 1/\sqrt{n}$. 

The above integrals need to be numerically evaluated to obtain the number variance
and spectral rigidity in the GSE-to-GUE crossover. The limit $\lambda\to0$ poses difficulty in
numerical evaluation as $I(\lambda,r)$ and $K(\lambda,r)$ approach
$-\text{sgn}(r)/2$ and $\delta(r)/r$, respectively; the latter signifying
the Kramers degeneracy. Therefore, we perform the evaluation of the
above integrals for nonzero, but small $\lambda$ values using very high precision. For instance, for the GSE curves in Figs. 12 and 13, we have used $\lambda=1/600$.

\bibliographystyle{apsrev4-2}
\end{document}